\documentclass[11pt]{article}
\usepackage{jheppub}
\usepackage{bm}
\usepackage{amsmath,amssymb}
\usepackage{graphics}
\usepackage{braket}
\usepackage{mathtools}
\usepackage{mathrsfs}
\usepackage{comment}
\usepackage{autobreak}

\usepackage[dvipsnames]{xcolor}

\edef\restoreparindent{\parindent=\the\parindent\relax}
\usepackage{parskip}
\restoreparindent

\usepackage{tikz}
\usetikzlibrary{arrows,arrows.meta,intersections, calc,positioning,decorations.pathreplacing,decorations.pathmorphing,shapes}
\usetikzlibrary{patterns}
\usetikzlibrary{decorations.markings}
\usetikzlibrary{knots}

\def\tr{{\rm tr}}
\def\Tr{{\rm Tr}}
\def\d{{\rm d}}
\def\i{{\rm i}}

\def\CC{{\cal C}}
\def\CD{{\cal D}}

\def\CH{{\cal H}}

\def\CI{{\cal I}}

\def\CM{{\cal M}}
\def\CN{{\cal N}}

\def\CS{{\cal S}}

\def\CV{{\cal V}}

\def\BB{\mathbb{B}}

\def\BD{\mathbb{D}}

\def\BH{\mathbb{H}}
\def\BR{\mathbb{R}}
\def\BS{\mathbb{S}}
\def\BT{\mathbb{T}}
\def\BZ{\mathbb{Z}}

\def\d{\mathrm{d}}

\def\SU{\mathrm{SU}}

\def\U{\mathrm{U}}

\makeatletter
\newsavebox{\@brx}
\newcommand{\llangle}[1][]{\savebox{\@brx}{\(\m@th{#1\langle}\)}%
  \mathopen{\copy\@brx\kern-0.5\wd\@brx\usebox{\@brx}}}
\newcommand{\rrangle}[1][]{\savebox{\@brx}{\(\m@th{#1\rangle}\)}%
  \mathclose{\copy\@brx\kern-0.5\wd\@brx\usebox{\@brx}}}
\makeatother

\newcommand{\Z}{\mathbb{Z}}

\newcommand{\HS}{\mathcal{H}}
\newcommand{\tm}[2]{\tau^{#1|#2}}
\newcommand{\ttm}[2]{\tilde{\tau}^{#1|#2}}
\newcommand{\Smat}[2]{{\mathcal{S}_{#1}}^{#2}}

\newcommand{\de}{\partial}
\newcommand{\be}{\begin{equation}}
\newcommand{\ba}{\begin{eqnarray}}
\newcommand{\ea}{\end{eqnarray}}
\newcommand{\ee}{\end{equation}}

\newcommand{\s}{\sqrt}
\newcommand{\vp}{\varphi}

\newcommand{\ddd}{\cdot\cdot\cdot}

\newcommand{\la}{\langle}
\newcommand{\lb}{\rangle}
\newcommand{\bea}{\begin{eqnarray}}
\newcommand{\eea}{\end{eqnarray}}
\newcommand{\bes}{\begin{equation*}}
\newcommand{\beas}{\begin{eqnarray*}}
\newcommand{\eeas}{\end{eqnarray*}}
\newcommand{\bas}{\begin{array*}}
\newcommand{\eas}{\end{array*}}
\newcommand{\ees}{\end{equation*}}

\def\mod{\ {\rm mod}\ }

\preprint{YITP-21-69, IPMU21-0043}
\title{\boldmath Topological pseudo entropy}

\author[a]{Tatsuma Nishioka,}
\author[a,b,c]{Tadashi Takayanagi}
\author[a]{and Yusuke Taki}

\affiliation[a]{Yukawa Institute for Theoretical Physics, Kyoto University,\\
Kitashirakawa Oiwakecho, Sakyo-ku, Kyoto 606-8502, Japan}

\affiliation[b]{Inamori Research Institute for Science,\\
620 Suiginya-cho, Shimogyo-ku,
Kyoto 600-8411, Japan}

\affiliation[c]{Kavli Institute for the Physics and Mathematics
 of the Universe (WPI),\\
University of Tokyo, Kashiwa, Chiba 277-8582, Japan}

\abstract{
We introduce a pseudo entropy extension of topological entanglement entropy called topological pseudo entropy.
Various examples of the topological pseudo entropies are examined in three-dimensional Chern-Simons gauge theory with Wilson loop insertions. Partition functions with knotted Wilson loops are directly related to topological pseudo (R\'enyi) entropies.
We also show that the pseudo entropy in a certain setup is equivalent to the interface entropy in two-dimensional conformal field theories (CFTs), and leverage the equivalence to calculate the pseudo entropies in particular examples.
Furthermore, we define a pseudo entropy extension of the left-right entanglement entropy in two-dimensional boundary CFTs and derive a universal formula for a pair of arbitrary boundary states.
As a byproduct, we find that the topological interface entropy for rational CFTs has a contribution identical to the topological entanglement entropy on a torus.
}

\begin{document} 
\maketitle

\section{Introduction}

Entanglement entropy has played an important role as a useful quantum order parameter in various quantum many-body systems \cite{Vidal:2002rm,Calabrese:2004eu,Eisert:2008ur,Kitaev:2005dm,Levin:2006zz}.
In particular, the topological entanglement entropy \cite{Kitaev:2005dm,Levin:2006zz}  
can characterize topologically ordered phases.
A prominent example of topological 
field theory is a three-dimensional Chern-Simons gauge theory, where the topological entanglement entropy can be computed by the famous surgery method \cite{Witten:1988hf} as first shown in \cite{Dong:2008ft}. 
Refer to \cite{Fendley:2006gr,2012PhRvL108s6402Q,Balasubramanian:2016sro,Wen:2016snr,Wen:2016bla,Wong:2017pdm,Balasubramanian:2018por,Fliss:2020cos,Berthiere:2020ihq} for further developments.

Recently, a quantity called the pseudo entropy was introduced in \cite{Nakata:2021ubr}, mainly motivated by finding a counterpart to a generalization of holographic entanglement 
entropy \cite{Ryu:2006bv,Ryu:2006ef,Hubeny:2007xt,Nishioka:2009un,Rangamani:2016dms} to Euclidean time-dependent backgrounds.
The pseudo entropy itself is a generalization of 
entanglement entropy that depends on both the initial state $|\psi\lb$ and the final state $|\vp\lb$, defined as follows.
Let $\ket{\psi},\ket{\varphi}\in\HS_A\otimes\HS_B$ be unnormalized states satisfying $\braket{\varphi|\psi}\neq0$. Define the transition matrix as
\begin{align}
  \tau^{\psi|\varphi}\equiv\frac{\ket{\psi}\bra{\varphi}}{\braket{\varphi|\psi}}\ ,
\end{align}
and its reduced version as
\begin{align}
  \tau_A^{\psi|\varphi}\equiv\Tr_B\left[\tau^{\psi|\varphi}\right]\ .
\end{align}
The pseudo R\'{e}nyi entropy is
\begin{align}\label{Pseudo_Renyi}
  S^{(n)}\left(\tau_A^{\psi|\varphi}\right)
    \equiv
        \frac{1}{1-n}\log\,\Tr_A\left[\left(\tau_A^{\psi|\varphi}\right)^n\right]\ ,
\end{align}
and we define the pseudo entropy by taking a limit $n\to1$:
\begin{align}
  S\left(\tau_A^{\psi|\varphi}\right)
    \equiv
        \lim_{n\to1}S^{(n)}\left(\tau_A^{\psi|\varphi}\right)
    =
        -\Tr_A\left[\tau_A^{\psi|\varphi}\log\tau_A^{\psi|\varphi}\right]\ .
\end{align}
Since the transition matrix is not Hermitian in general, the pseudo entropy can take complex values.
This guides us to define the following real-valued quantity 
\begin{align}\label{diff_pseudo_entropy}
    \Delta S^{(n)}\left(\tau_A^{\psi|\varphi}\right)
        \equiv         
            \frac{1}{2}\left[S^{(n)}\left(\tau_A^{\psi|\varphi}\right)+ S^{(n)}\left(\tau_A^{\varphi|\psi}\right)-S^{(n)}\left(\tau_A^{\psi|\psi}\right)- S^{(n)}\left(\tau_A^{\vp|\varphi}\right)\right]\ ,
\end{align}
where note the relation $S^{(n)}\left(\tau_A^{\varphi|\psi}\right)=S^{(n)}\left(\tau_A^{\psi|\varphi}\right)^*$
and the fact that the latter two terms are the standard entanglement R\'{e}nyi entropy for $|\psi\lb$
and $|\vp\lb$, respectively.
In other words, 
\begin{align}\label{diff_pseudo_entropy_limit}
\Delta S\left(\tau_A^{\psi|\varphi}\right)\equiv \lim_{n\to1}\Delta S^{(n)}\left(\tau_A^{\psi|\varphi}\right)
\end{align}
is the difference between the real part of the pseudo entropy and the averaged entanglement entropy.

In \cite{Mollabashi:2020yie,Mollabashi:2021xsd}, 
the pseudo entropy was numerically evaluated for 
the Lifshitz free scalar field and for Ising and XY spin models. 
These calculations showed that the difference 
$\Delta S\left(\tau_A^{\psi|\varphi}\right)$ always takes non-positive values when $|\psi\lb$
and $|\vp\lb$ belong to the same phase.
However, it turns out that when the two states are in different quantum phases, the difference typically takes positive values.
This implies that the pseudo entropy can distinguish two different quantum phases. 
A heuristic explanation of this interesting behavior was given in \cite{Mollabashi:2021xsd} based on holography, where an anti-de Sitter space emerges in the gravity dual along the interface between two quantum phases, which enhances the pseudo entropy.

Motivated by these, the purpose of the present paper is to introduce a pseudo entropy 
extension of topological entanglement entropy, which we call  topological pseudo entropy.
We will explicitly evaluate the topological pseudo entropy in various examples in three-dimensional Chern-Simons gauge theory.
We will also point out that the pseudo entropy in a class of specific setups is equivalent to the interface entropy \cite{Sakai:2008tt,Gutperle:2015kmw,Gutperle:2017enx,Brehm:2015lja,Brehm:2015plf,Wen:2017smb,Chen:2018pda,Lou:2019heg,Brehm:2020agd} in conformal field theories (CFTs).
We will also provide and evaluate a pseudo entropy extension of 
the left-right entanglement entropy \cite{PandoZayas:2014wsa,Das:2015oha,Berthiere:2020ihq} in CFTs.

This paper is organized as follows. In section \ref{sec:CS} we calculate the topological pseudo entropy in various setups of a three-dimensional Chern-Simons gauge theory and provide its interpretations in the light of quantum entanglement and geometry.
In section \ref{sec:CHM}, we explain how to calculate the pseudo entropy in CFTs via conformal transformations and show that the pseudo entropy in a special case of CFTs is equivalent to the interface entropy.
In section \ref{sec:LRPE}, we introduce the pseudo entropy extension of the left-right entanglement entropy.
In section \ref{sec:conclusion}, we summarize our conclusions.
In the appendix \ref{ap:sutwo}, we provide explicit values for the $\SU(2)$ Chern-Simons gauge theory.
In the appendix \ref{ap:multi_bdy}, we evaluate the pseudo entropy for multi-boundary states in Chern-Simons gauge theory.


\section{Topological pseudo entropy in Chern-Simons gauge theory}\label{sec:CS}
Consider the three-dimensional Chern-Simons gauge theory with the gauge group $\SU(N)$ at level $k$.
The partition functions of the Chern-Simons theory with Wilson lines can be calculated from the knowledge of two-dimensional (2d) conformal field theory of $\widehat{\SU(N)}_k$ Wess-Zumino-Witten (WZW) model \cite{Witten:1988hf} as quantum states in the Chern-Simons theory correspond to the conformal blocks of the 2d CFT. First we explain how to calculate pseudo entropy in Chern-Simons theory from section \ref{subsec:replica} to section \ref{subsec:CS}. 
Next, we calculate the entanglement entropy or the pseudo entropy for states on $\BS^2$ with two excitations in section \ref{subsec:twoexcitation} and four excitations in section \ref{subsec:excitation}, and states on $\BT^2$ in section \ref{subsec:torus}. 
In section \ref{ss:geometric_interpretation}, we consider the geometric interpretation for pseudo entropy from the above calculations. 
Finally in section \ref{subsec:boundary}, we consider the definition of boundary states in Chern-Simons theory by analogy with boundary conformal field theory (BCFT) for comparison with the results in later sections. We investigate another example of multi-boundary states in Chern-Simons theory in appendix \ref{ap:multi_bdy}.

\subsection{Replica trick}\label{subsec:replica}

Before considering the Chern-Simons theory, we review how to compute the pseudo entropy in quantum field theory.
We can compute the pseudo entropy on a spatial region $\Sigma=A\cup B$ as well as the entanglement entropy by using the replica trick.
We consider a Euclidean field theory with an action $I[\Phi]$, where $\Phi$ is the collection of fields. We prepare the two states $\ket{\psi}$ and $\ket{\varphi}$ by inserting operators $\mathcal{O}_\psi$ and $\mathcal{O}_\varphi$ respectively to the path integral on the past of $\Sigma$:
\begin{align}
  \braket{\Phi_0|\psi}
    &=
        \int_{\Phi|_{\Sigma}=\Phi_0}\CD\Phi\ \mathcal{O}_\psi[\Phi]\,e^{-I[\Phi]}
        =
            \begin{tikzpicture}[thick,scale=1.5,baseline={([yshift=-.5ex]current bounding box.center)}]
                    \draw[fill=lightgray!20!white] (0.9,0) arc (0:-180:0.9 and 0.8);
                    \draw[dotted] (0.9,0) -- (-0.9,0);
                    \node at (0,-0.4) {$\mathcal{O}_\psi$};
                    \node at (-0.5,0.15) {$\small\Phi_0$};
                    \node at (1.05,0) {$\Sigma$};
            \end{tikzpicture}
        \ , \label{eq:psipi}\\
  \braket{\Phi_0|\varphi}
    &=
        \int_{\Phi|_{\Sigma}=\Phi_0}\CD\Phi\ \mathcal{O}_\varphi[\Phi]\,e^{-I[\Phi]}
        =
        \begin{tikzpicture}[thick,scale=1.5,baseline={([yshift=-.5ex]current bounding box.center)}]
                \draw[fill=lightgray!20!white] (0.9,0) arc (0:-180:0.9 and 0.8);
                \draw[dotted] (0.9,0) -- (-0.9,0);
                \node at (0,-0.4) {$\mathcal{O}_\varphi$};
                \node at (-0.5,0.15) {\small$\Phi_0$};
                \node at (1.05,0) {$\Sigma$};
        \end{tikzpicture}
        \label{eq:varphipi}\ ,
\end{align}
where $\Phi_0$ is a boundary condition of $\Phi$ on $\Sigma$ and $\ket{\Phi_0}$ is a state on $\Sigma$ defined by $\hat{\Phi}|_\Sigma\ket{\Phi_0}=\Phi_0\ket{\Phi_0}$.
The vertical direction in the figure is the imaginary time.
The inserted operators $\mathcal{O}_\psi$ and $\mathcal{O}_\varphi$ may be collections of line operators like Wilson loops as well as local operators. 
The inner product is given by gluing the manifolds for $\ket{\psi}$ and $\ket{\varphi}$ along $\Sigma$ and integrating over the boundary condition:
\begin{align}\label{eq:innprod}
  \braket{\varphi|\psi}=\int\CD\Phi_0\braket{\varphi|\Phi_0}\braket{\Phi_0|\psi}=
  \begin{tikzpicture}[thick,scale=1.5,baseline={([yshift=-.5ex]current bounding box.center)}]
                \draw[fill=lightgray!20!white] (0,0) ellipse (0.9 and 0.8);
                \draw[dotted] (0.9,0) -- (-0.9,0);
                \node at (0,-0.4) {$\mathcal{O}_\psi$};
                \node at (0,0.4) {$\mathcal{O}^\dagger_\varphi$};
    \end{tikzpicture}
  \ .
\end{align}
We call the resulting manifold $\mathcal{M}_1$.
Then we can interpret $\braket{\varphi|\psi}$ as a partition function on $\mathcal{M}_1$ in the presence of $\mathcal{O}_\psi$ and $\mathcal{O}_\varphi^{\dagger}$, so we denote it by $Z\left[\mathcal{M}_1;\mathcal{O}_\psi,\mathcal{O}_\varphi^{\dagger}\right]$.

Next, we evaluate $\Tr_A\left[(\Tr_B\ket{\psi}\bra{\varphi})^n\right]$.
A partial trace over $B$ corresponds to the gluing only over $B$, thus the unnormalized version of the reduced transition matrix is 
\begin{align}\label{eq:redtmfig}
  \ttm{\psi}{\varphi}_A\equiv\Tr_B\left[\ket{\psi}\bra{\varphi}\right]
    =
        \int\CD[\Phi_0|_B]\braket{\varphi|\Phi_0}\braket{\Phi_0|\psi}=
    \begin{tikzpicture}[thick,scale=1.5,baseline={([yshift=-.5ex]current bounding box.center)}]
                \draw[fill=lightgray!20!white] (0,0) ellipse (0.9 and 0.8);
                \draw[dotted,fill=white] (0,0) arc (180:0:0.3 and 0.05);
                \draw[dotted,fill=white] (0,0) arc (180:360:0.3 and 0.05);
                \draw[dotted] (0,0) -- (-0.9,0) ;
                \draw[dotted] (0.6,0) -- (0.9,0) ;
                \node at (0,-0.4) {$\mathcal{O}_\psi$};
                \node at (0,0.4) {$\mathcal{O}^\dagger_\varphi$};
                \node at (0.35,0.18) {\small$A$};
                \node at (-0.45,0.1) {\small$B$};
                \node at (0.75,0.1) {\small$B$} ;
    \end{tikzpicture}
        \ .
\end{align}
To compute the $n^{\text{th}}$ power of $\ttm{\psi}{\varphi}_A$, we prepare $n$ copies of the manifold in \eqref{eq:redtmfig} and glue them along the subregion $A$ cyclically:
\begin{align}\label{eq:replica}
  \Tr_A\left[\left(\ttm{\psi}{\varphi}_A\right)^n\right]=
  \begin{tikzpicture}[thick,scale=1.5,baseline={([yshift=-.5ex]current bounding box.center)}]
            \begin{scope}
                \draw[fill=lightgray!20!white] (0,0) ellipse (0.9 and 0.8);
                \draw[dotted,fill=white] (0,0) arc (180:0:0.3 and 0.05);
                \draw[dotted,fill=white] (0,0) arc (180:360:0.3 and 0.05);
                \draw[dotted] (0,0) -- (-0.9,0) ;
                \draw[dotted] (0.6,0) -- (0.9,0) ;
                \node at (0,-0.4) {$\mathcal{O}_\psi$};
                \node at (0,0.4) {$\mathcal{O}^\dagger_\varphi$};
            \end{scope}
            \begin{scope}[shift={(2,0)}]
                \draw[fill=lightgray!20!white] (0,0) ellipse (0.9 and 0.8);
                \draw[dotted,fill=white] (0,0) arc (180:0:0.3 and 0.05);
                \draw[dotted,fill=white] (0,0) arc (180:360:0.3 and 0.05);
                \draw[dotted] (0,0) -- (-0.9,0) ;
                \draw[dotted] (0.6,0) -- (0.9,0) ;
                \node at (0,-0.4) {$\mathcal{O}_\psi$};
            \end{scope}
            \node at (3.4,0) {\Large $\cdots$};
            \begin{scope}[shift={(4.8,0)}]
                \draw[fill=lightgray!20!white] (0,0) ellipse (0.9 and 0.8);
                \draw[dotted,fill=white] (0,0) arc (180:0:0.3 and 0.05);
                \draw[dotted,fill=white] (0,0) arc (180:360:0.3 and 0.05);
                \draw[dotted] (0,0) -- (-0.9,0) ;
                \draw[dotted] (0.6,0) -- (0.9,0) ;
                \node at (0,-0.4) {$\mathcal{O}_\psi$};
            \end{scope}
            \begin{scope}
                \begin{knot}[background color=lightgray!20!white]
                    \strand[->,OliveGreen] (0.4,0.05) .. controls (0.75,0.3) ..(1.3,0) .. controls (1.75,-0.3) .. (2.2,-0.05) ;
                    \strand[OliveGreen] (2.4,0.05) .. controls (2.75,0.3)..(3.1,0.1);
                    \strand[->,OliveGreen] (3.8,0.1).. controls (4.5,-0.3) .. (5,-0.05);
                    \strand[->,OliveGreen] (5.1,0.05) .. controls (4,1) and (2.5,0.7) .. (1.5,0) .. controls (0.7,-0.5) .. (0.2,-0.05);
                \end{knot}
            \end{scope}
            \node[fill=lightgray!20!white] at (2,0.4) {$\mathcal{O}^\dagger_\varphi$};
            \node[fill=lightgray!20!white] at (4.8,0.4) {$\mathcal{O}^\dagger_\varphi$};
    \end{tikzpicture}
  \ .
\end{align}
We denote the glued manifold in \eqref{eq:replica} by $\mathcal{M}_n$ and the partition function on $\mathcal{M}_n$ by $Z \left[\mathcal{M}_n;\mathcal{O}_\psi,\mathcal{O}_\varphi^{\dagger}\right]$.
Finally we obtain the pseudo entropy 
\begin{align}
    \begin{aligned}
    S\left(\tm{\psi}{\varphi}_A\right)
        &=
            \lim_{n\to1}\frac{1}{1-n}\log\Tr_A\left[
            \left(  
            \frac{\ttm{\psi}{\varphi}_A}{
            \Tr_A\left[\ttm{\psi}{\varphi}_A\right]}
            \right)^n\right] \\
        &=
            -\left.\frac{\partial}{\partial n}\log\frac{Z \left[\mathcal{M}_n;\mathcal{O}_\psi,\mathcal{O}_\varphi^{\dagger}\right]}{Z\left[\mathcal{M}_1;\mathcal{O}_\psi,\mathcal{O}_\varphi^{\dagger}\right]^n}\right|_{n=1}\ .
    \end{aligned}
\end{align}

\subsection{Chern-Simons theory and modular \texorpdfstring{$\mathcal{S}$}{S}-matrix}

The Chern-Simons theory on a 3d manifold $\mathcal{M}$ with gauge group $\SU(N)$ is defined by the action 
\begin{align}
  I_\text{CS}[A]= -\i\,\frac{k}{4\pi}\int_{\mathcal{M}}\tr\left[A\wedge \d A+\frac{2}{3}\, A\wedge A\wedge A\right]\ ,
\end{align}
where $A$ is a connection one-form and the trace is taken over the Lie algebra associated with $\SU(N)$.
A prefactor $k$, which has to take an integer value for gauge invariance, is called the level of the Chern-Simons theory.
Since the action does not depend on the metric, Chern-Simons theory is a topological field theory. 
Topological invariance is such a strong property that we can obtain a lot of information from the invariance.
We will focus on observables that are also topologically invariant, i.e., Wilson loops, defined by 
\begin{align}
  W_R[A]=\tr_R\,\mathcal{P}\exp\left(\int_CA\right)\ ,
\end{align}
where the trace is taken over the representation space of a representation $R$ of $\SU(N)$, $\mathcal{P}$ means the path ordered integral along a closed loop $C$.

We can evaluate the partition function of a Chern-Simons theory by using the fact that there is a duality between Chern-Simons theories and WZW models \cite{Witten:1988hf}.
Before describing the duality, we recapitulate several notions about WZW models.
Let $\chi_i(\tau)$ be a character of a WZW model on a torus with a complex structure $\tau$, where $i$ denotes an integrable representation of an affine Lie algebra $\widehat{\SU(N)}_k$. The modular invariance of the theory amounts to the transformation law for the character:
\begin{align}
  \chi_i(-1/\tau) = \sum_{j}\Smat{i}{j}\,\chi_j(\tau)\ ,
\end{align}  
where $\Smat{i}{j}$ is called modular $\mathcal{S}$-matrix, which is a unitary and  symmetric matrix, 
\begin{align} \label{eq:unitary}
  \sum_{l}\Smat{i}{l}\,{\left(\mathcal{S}^{\dagger}\right)_l}^j
    = {\delta_{i}}^j\ , \qquad \Smat{i}{j} = \Smat{j}{i}\ . 
\end{align} 
Moreover, the square of the modular $\mathcal{S}$-matrix is identical to the charge conjugation $\mathcal{C}$:
\begin{align}
   \sum_{l}\Smat{i}{l}\,\Smat{l}{j}={\mathcal{C}_i}^j={\delta_i}^{\bar{j}}\ ,
\end{align}  
where $\bar{j}$ denotes the charge conjugate representation of $j$. This leads to 
the identity $\Smat{i}{j}=\left(\Smat{i}{\bar{j}}\right)^*$.
 In particular, we find that the matrix element 
 $\Smat{0}{i}=\Smat{i}{0}$ is
real valued for any $i$.

For an example of $\widehat{\SU(2)}_k$ WZW theory, the modular $\mathcal{S}$-matrix can be written as 
\begin{align}\label{eq:SmatrixSU2}
  \Smat{i}{j}=\sqrt{\frac{2}{k+2}}\,\sin\left[\frac{\pi (2i+1)(2j+1)}{k+2}\right]\ ,
\end{align}
where the subscripts $i,j=0,\ldots,k/2$ label the integrable representations of $\widehat{\SU(2)}_k$ and $0$ denotes the identity representation. Note that $\CS$-matrix for $\widehat{\SU(2)}_k$ is real. We summarize the properties and several explicit values of $\SU(2)$ $\CS$-matrix in appendix \ref{ap:sutwo}. 

There is another important relation between the modular $\mathcal{S}$-matrix and the fusion coefficients ${N_{ij}}^k$, known as the Verlinde formula \cite{Verlinde:1988}:
\begin{align}
  {N_{ij}}^k=\sum_l\frac{\Smat{i}{l}\,\Smat{j}{l}\,{\left(\CS^\dagger\right)_l}^k}{\Smat{0}{l}}\ ,
\end{align}
or equivalently
\begin{align}\label{eq:Verlinde2}
  \sum_k {N_{ij}}^k\,\frac{\Smat{k}{m}}{\Smat{0}{m}}=\frac{\Smat{i}{m}}{\Smat{0}{m}}\,\frac{\Smat{j}{m}}{\Smat{0}{m}}\ .
\end{align}
Regarding ${N_{ij}}^k$ as the $(j,k)$-component of the matrix $\bm{\mathrm{N}}_i$, $\Smat{i}{m}/\Smat{0}{m}$ in \eqref{eq:Verlinde2} is an eigenvalue of $\bm{\mathrm{N}}_i$. In particular, $m=0$ yields the largest eigenvalue 
\begin{align}\label{quantum_dim}
  d_i=\frac{\Smat{i}{0}}{\Smat{0}{0}}\ ,
\end{align}
called quantum dimension for the representation $i$.
Note that $\Smat{0}{0}$ and $\Smat{i}{0}$ are real, so that the quantum dimensions are also real.
The total quantum dimension is defined by
\begin{align}\label{eq:totalqd}
  \CD=\sqrt{\sum_{i}|d_i|^2}=\frac{1}{\Smat{0}{0}}\ .
\end{align}
The second equality in \eqref{eq:totalqd} follows from the unitarity condition \eqref{eq:unitary}. 

Finally let us describe the duality between Chern-Simons and WZW theories.
Consider a Chern-Simons theory with Wilson loops and take a spatial submanifold $\Sigma\simeq\BS^2$.
When $\Sigma$ has some intersections with Wilson loops $W_{R_i}[A]$, the Hilbert space on $\Sigma$ is given by
\begin{align}\label{eq:invsp}
    \mathcal{H}_\Sigma
        =
        \mathrm{Inv}\left(\bigotimes_i\mathscr{R}_i\right)\ ,
\end{align}
where $\mathscr{R}_i$ denotes the representation space of an integrable representation $R_i$, and ``$\mathrm{Inv}$'' means that it takes only the invariant subspace.
The subscript $i$ in \eqref{eq:invsp} runs over all the intersections of Wilson lines and $\Sigma$.
In particular, if there are no intersections, then the Hilbert space is one-dimensional.

\subsection{Computation of partition functions in Chern-Simons gauge theory}\label{subsec:CS}

With the input of the modular properties of 2d CFTs, we can evaluate the partition functions in Chern-Simons theory by Witten's method \cite{Witten:1988hf}. 

We cut a manifold $\CM$ along a submanifold $\Sigma\simeq\BS^2$ into two parts $\CM'$ and $\CM''$. When $\Sigma$ has no intersections with any Wilson loops, the Hilbert space on $\Sigma$ is one-dimensional by \eqref{eq:invsp}. Therefore we can attach a hemisphere to each of the cross-sections, then we have
\begin{align}\label{eq:cutting}
    Z\left[\CM\right]
        =
        \frac{Z\left[\CM'\right]\, Z\left[\CM''\right]}{Z\left[\BS^3\right]}\ .
\end{align}
\begin{figure}
    \centering
    \begin{tikzpicture}
        \begin{scope}
        \begin{scope}[xshift=0.5cm]
            \draw (-3, -0.25) ellipse (1 and 0.5);
            \draw (-3, 0.25) arc (90:270:0.1 and 0.5);
            \draw[draw=none, fill=lightgray!20!white] (-3, -0.25) ellipse (0.1 and 0.5);
            \draw[dotted] (-3, -0.75) arc (-90:90:0.1 and 0.5);
            \node at (-3.5, -0.25) {\Large $\CM'$};
            \node at (-2.5, -0.25) {\Large $\CM''$};
        \end{scope}
        
        \node at (-0.8, -0.25) {\Large $=$};
        
        \draw (1, 1) arc (90:270:1 and 0.5);
        \draw[draw=none, fill=lightgray!50] (1, 0.5) ellipse (0.1 and 0.5);
        \draw[fill=RoyalBlue, fill opacity=0.5] (1, 1) arc (90:270:0.1 and 0.5);
        \draw[dotted] (1, 0) arc (-90:90:0.1 and 0.5);
        \draw[fill=RoyalBlue, fill opacity=0.5] (1, 0) arc (-90:90:0.5 and 0.5);
        \node at (0.5, 0.5) {\Large $\CM'$};
        
        \node at (2, 0.5) {\Large $\times$};
        
        \draw (3, 0) arc (-90:90:1 and 0.5);
        \draw[fill=RoyalBlue, fill opacity=0.5] (3, 1) arc (90:270:0.5 and 0.5);
        \draw[draw=none, fill=lightgray!20!white]
        (3, 0.5) ellipse (0.1 and 0.5);
        \draw (3, 1) arc (90:270:0.1 and 0.5);
        \draw[dotted] (3, 0) arc (-90:90:0.1 and 0.5);
        \node at (3.5, 0.5) {\Large $\CM''$};
        
        \draw (0, -0.25) --+ (4, 0);
        
        \draw[fill=RoyalBlue, fill opacity=0.5] (2, -1) circle (0.5);
        \draw (2, -0.5) arc (90:270:0.1 and 0.5);
        \draw[dotted] (2, -1.5) arc (-90:90:0.1 and 0.5);
        \node at (1.1, -1) {\Large $\BS^3$};
    \end{scope}
    
    \end{tikzpicture}
    \caption{A manifold can be decomposed into two by cutting it a half and attaching hemispheres to each of them. }
    \label{fig:cut}
\end{figure}
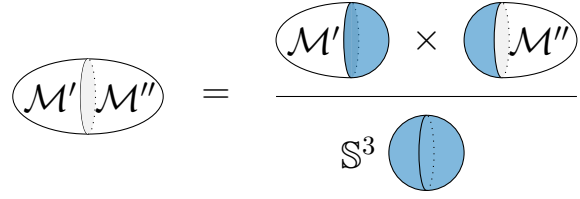
Figure \ref{fig:cut} shows this relation graphically. 
We can apply this method also to the case where $\Sigma$ has two punctures $R_i$ and $\overline{R}_i$ because the Hilbert space is one-dimensional. We consider the case $\CM=\BS^2\times\BS^1$ including two Wilson loops wrapping along $\BS^1$. Applying the above method, we obtain 
\begin{align}
    Z\left[\BS^2\times\BS^1;R_i,\overline{R}_j\right]={\delta_{i}}^{j}\ .
\end{align} 

Next, we would like to evaluate a partition function on a sphere $\BS^3$. This can be obtained by gluing two solid tori along their common boundary $\BT^2$.
When we glue the two, we perform the modular transformation for one of the tori as depicted in figure \ref{fig:CS1}.
Thus the partition function on $\BS^3$ without any Wilson loops becomes
\begin{align}
    Z\left[\BS^3\right]=\sum_i\Smat{0}{i}\,Z\left[\BS^2\times\BS^1;R_i\right]=\Smat{0}{0}\ .
\end{align}
Moreover, the $\BS^3$ partition function with a single Wilson loop in a representation 
$R_i$ and that with two linked Wilson loops in representations $R_i$ and $R_j$ are given by
\begin{align} \label{eq:pf1}
    \begin{aligned}
    Z\left[\BS^3; R_i\right] 
        &=
            \Smat{0}{i}\ ,\\
    Z\left[\BS^3; L(R_i,R_j)\right] 
        &=
            \Smat{i}{j}\ .
    \end{aligned}
\end{align}
Figure \ref{fig:CS1} shows the calculations for these results.

\begin{figure}[t]\label{fig:glue}
  \centering
    \begin{tikzpicture}[thick]
         \begin{scope}[scale=0.4]
            \draw[fill=lightgray!20!white] (0,0) ellipse (3 and 1.5);
            \begin{scope}
            \clip (0,-1.8) ellipse (3 and 2.5);
            \draw (0,2.2) ellipse (3 and 2.5);
            \end{scope}
            \begin{scope}
            \clip (0,2.2) ellipse (3 and 2.5);
            \draw[fill=white] (0,-2.2) ellipse (3 and 2.5);
            \end{scope}
            \draw[very thick, OliveGreen] (0, 0) ellipse (2.3 and 1);
        \end{scope}
        \begin{scope}[scale=0.4, xshift=8cm, rotate=90]
            \draw[fill=lightgray!20!white] (0,0) ellipse (3 and 1.5);
            \begin{scope}
            \clip (0,-1.8) ellipse (3 and 2.5);
            \draw (0,2.2) ellipse (3 and 2.5);
            \end{scope}
            \begin{scope}
            \clip (0,2.2) ellipse (3 and 2.5);
            \draw[fill=white] (0,-2.2) ellipse (3 and 2.5);
            \end{scope}
            \draw[very thick, OliveGreen] (0, 0) ellipse (2.3 and 1);
        \end{scope}
        \node at (-2.2, -0.1) {\Large $\displaystyle \sum_i \Smat{i}{j}$};
        \node[OliveGreen] at (0, 1) {\Large $R_i$};
        \node[OliveGreen] at (2.4, -0.8) {\Large $R_j$};
        \node at (1.8, 0) {\Large $=$};
        \node at (0, -1) {\large $\tau$};
        \node at (4.3, -1) {\large $\displaystyle -\frac{1}{\tau}$};
    \end{tikzpicture}
    
    \begin{tikzpicture}[thick]
         \begin{scope}[scale=0.4]
            \draw[fill=lightgray!20!white] (0,0) ellipse (3 and 1.5);
            \begin{scope}
            \clip (0,-1.8) ellipse (3 and 2.5);
            \draw (0,2.2) ellipse (3 and 2.5);
            \end{scope}
            \begin{scope}
            \clip (0,2.2) ellipse (3 and 2.5);
            \draw[fill=white] (0,-2.2) ellipse (3 and 2.5);
            \end{scope}
            \draw[very thick, BrickRed] (0, 0) ellipse (2.3 and 1);
        \end{scope}
        \begin{scope}[scale=0.4, xshift=7.3cm, rotate=90]
            \draw[fill=lightgray!20!white] (0,0) ellipse (3 and 1.5);
            \begin{scope}
            \clip (0,-1.8) ellipse (3 and 2.5);
            \draw (0,2.2) ellipse (3 and 2.5);
            \end{scope}
            \begin{scope}
            \clip (0,2.2) ellipse (3 and 2.5);
            \draw[fill=white] (0,-2.2) ellipse (3 and 2.5);
            \end{scope}
        \end{scope}
        \draw (-3.5, 0) circle (1);
        \draw[dotted] (-3.5, 0) ellipse (1 and 0.3);
        \draw[very thick, BrickRed] (-3.5, 0.2) ellipse (0.7 and 0.35);
        \node[BrickRed] at (-3.5, -0.5) {\Large $R_i$};
        \node at (-1.9, 0) {\Large $=$};
        \node[BrickRed] at (0, 1) {\Large $R_i$};
        \node at (1.7, 0) {\Large $\cdot$};
        \node at (4.5, 0) {\Large $=~ \Smat{0}{i}$};
    \end{tikzpicture}
    
    \begin{tikzpicture}[thick]
         \begin{scope}[scale=0.4]
            \draw[fill=lightgray!20!white] (0,0) ellipse (3 and 1.5);
            \begin{scope}
            \clip (0,-1.8) ellipse (3 and 2.5);
            \draw (0,2.2) ellipse (3 and 2.5);
            \end{scope}
            \begin{scope}
            \clip (0,2.2) ellipse (3 and 2.5);
            \draw[fill=white] (0,-2.2) ellipse (3 and 2.5);
            \end{scope}
            \draw[very thick, BrickRed] (0, 0) ellipse (2.3 and 1);
        \end{scope}
        \begin{scope}[scale=0.4, xshift=7.3cm, rotate=90]
            \draw[fill=lightgray!20!white] (0,0) ellipse (3 and 1.5);
            \begin{scope}
            \clip (0,-1.8) ellipse (3 and 2.5);
            \draw (0,2.2) ellipse (3 and 2.5);
            \end{scope}
            \begin{scope}
            \clip (0,2.2) ellipse (3 and 2.5);
            \draw[fill=white] (0,-2.2) ellipse (3 and 2.5);
            \end{scope}
            \draw[very thick, OliveGreen] (0, 0) ellipse (2.3 and 1);
        \end{scope}
        \draw (-3.5, 0) circle (1);
        \draw[dotted] (-3.5, 0) ellipse (1 and 0.3);
        \begin{knot}[flip crossing=1]
            \strand[very thick, BrickRed] (-3.7, 0) ellipse (0.6 and 0.3);
            \strand[very thick, OliveGreen] (-3, 0) ellipse (0.3 and 0.6);
        \end{knot}
        \node[BrickRed] at (-3.7, -0.6) {\Large $R_i$};
        \node[OliveGreen] at (-3.6, 0.6) {\Large $R_j$};
        \node at (-1.9, 0) {\Large $=$};
        \node[BrickRed] at (0, 1) {\Large $R_i$};
        \node[OliveGreen] at (2.1, -0.8) {\Large $R_j$};
        \node at (1.7, 0) {\Large $\cdot$};
        \node at (4.5, 0) {\Large $=~ \Smat{i}{j}$};
    \end{tikzpicture}    
    \caption{The modular transformation in Chern-Simons gauge theory and the evaluations of the partition functions with Wilson loops. The horizontal solid tori have a complex structure $\tau$ while the vertical ones have $-1/\tau$. The dot means the gluing along the torus on the boundaries of two solid tori.}
    \label{fig:CS1}
\end{figure}

By using these results and \eqref{eq:cutting}, we can calculate the partition functions with multiple disconnected Wilson loops.
For example, the $\BS^3$ partition function with two disconnected Wilson loops in representations $R_i$ and $R_j$ (see figure \ref{fig:CS2}) is computed as 
\begin{align}
    Z\left[\BS^3; R_i,R_j\right] = \frac{\Smat{0}{i}\, \Smat{0}{j}}{\Smat{0}{0}}\ .
\end{align}

\begin{figure}[t]
  \centering
    \begin{tikzpicture}[thick]
    \begin{scope}[yshift=-3.5cm]
        \begin{scope}[xshift=0.5cm]
            \draw (-3, -0.25) circle (1);
            \draw[dotted] (-3, -0.25) ellipse (1 and 0.3); 
            \draw[BrickRed, very thick] (-3.5, -0.25) ellipse (0.3 and 0.15);
            \draw[OliveGreen, very thick] (-2.5, -0.25) ellipse (0.3 and 0.15);
            \node[BrickRed] at (-3.5, -0.8) {$R_i$};
            \node[OliveGreen] at (-2.5, -0.8) {$R_j$};
         \end{scope} 
     
        \node at (-0.7, -0.25) {\Large $=$};
        
        \draw (1, 0.5) circle (0.5);
        \draw[dotted] (1, 0.5) ellipse (0.5 and 0.15);
        \draw[BrickRed, very thick] (1, 0.65) ellipse (0.3 and 0.15);
        \node[BrickRed] at (1, 0.25) {$R_i$};
        
        \node at (2, 0.5) {\Large $\times$};
        
        \draw (3, 0.5) circle (0.5);
        \draw[dotted] (3, 0.5) ellipse (0.5 and 0.15);
        \draw[OliveGreen, very thick] (3, 0.65) ellipse (0.3 and 0.15);
        \node[OliveGreen] at (3, 0.25) {$R_j$};
        
        \draw (0, -0.25) --+ (4, 0);
        
        \draw (2, -1) circle (0.5);
        \draw[dotted] (2, -1) ellipse (0.5 and 0.15);

        \node at (5.3, -0.25) {\Large $\displaystyle = ~ \frac{\Smat{0}{i}\, \Smat{0}{j}}{\Smat{0}{0}}$};
    \end{scope}
    \end{tikzpicture}
    \caption{We can calculate $Z\left[\BS^3;R_i,R_j\right]$ by applying \eqref{eq:cutting} and \eqref{eq:pf1}.}
    \label{fig:CS2}
\end{figure}

\subsection{Topological entanglement entropy on \texorpdfstring{$\BS^2$}{S2} with two excitations} \label{subsec:twoexcitation}

Before we go to our main target of topological pseudo entropy, we would like to 
start with the calculation of topological entanglement entropy in a simple setup.
Refer to \cite{Dong:2008ft} for more extensive analysis.
We consider a setup where a state $|\psi\lb$
is defined by a path integral on a hemisphere $\BB^3$ such that on its boundary $\BS^2$,
there are two quasi-particle (i.e.\,anyon) excitations one of which is in a representation $R_i$ and the other is in $\overline{R}_i$ of $\widehat{\SU(N)}_k$, so that they form a singlet. 
We choose the subsystem $A$ on the sphere $\BS^2$, such that $A$ includes the excitation in $R_i$ and its complement $B$ does that in $\overline{R}_i$.

The entanglement entropy 
\begin{align}
S(\rho_A)=-\Tr_A\left[\rho_A\log\rho_A\right]\ ,
\end{align}
of the reduced density matrix 
\begin{align}
\rho_A=\mbox{Tr}_B\left[\frac{|\psi\lb\la\psi|}{\braket{\psi|\psi}}\right]
\end{align}
can be computed via the replica trick we reviewed in section \ref{subsec:replica}. 

We can construct the state $\ket{\psi}$ by a path integral over $\BB^3$ inserting a Wilson line operator ending on the excitations ($\mathcal{O}_\psi=W_{R_i}$ in \eqref{eq:psipi}).
The partial trace over $B$ can be performed by gluing only the subregion $B$ of $\BS^2$ and the product of two $\rho_A$'s
can be done by gluing the subregion $A$, then $\Tr_A\left[(\Tr_B\ket{\psi}\bra{\psi})^n\right]$ becomes a partition function on $\BS^3$ with a Wilson loop.
Figure \ref{fig:twoexcitations} shows the calculation of $n=2$ case.
Divided by the normalization factor, we obtain 
\begin{figure}[t]
  \begin{center}
    \begin{tikzpicture}[thick]
      \begin{scope}
        \draw[fill=lightgray!20!white] (0,0) circle (1);
        \draw[dotted] (0,1) arc (90:-90:0.2 and 1);
        \node at (-0.9,0.9) {$A$};
        \node at (0.9,0.9) {$B$};
        \begin{scope}
          \draw[BrickRed] (-0.8,0)--(0.8,0) ;
          \draw[BrickRed] (-0.85,-0.05)--(-0.75,0.05);
          \draw[BrickRed] (-0.85,0.05)--(-0.75,-0.05);
          \draw[BrickRed] (0.85,-0.05)--(0.75,0.05);
          \draw[BrickRed] (0.85,0.05)--(0.75,-0.05);    
          \draw[BrickRed] (0.05,0)--(-0.05,0.1);   
          \draw[BrickRed] (0.05,0)--(-0.05,-0.1);   
        \end{scope}
        \draw (0,1) arc (90:270:0.2 and 1);
        \node at (0,-1.4) {$\ket{\psi}$};
      \end{scope}
      
      \begin{scope}[xshift=2.5cm]
        \draw[fill=lightgray!20!white] (0,0) circle (1);       
        \draw[dotted] (0,1) arc (90:-90:0.2 and 1);
        \node at (-0.9,0.9) {$\bar{B}$};
        \node at (0.9,0.9) {$\bar{A}$};
        \begin{scope}
          \draw[BrickRed] (-0.8,0)--(0.8,0) ;
          \draw[BrickRed] (-0.85,-0.05)--(-0.75,0.05);
          \draw[BrickRed] (-0.85,0.05)--(-0.75,-0.05);
          \draw[BrickRed] (0.85,-0.05)--(0.75,0.05);
          \draw[BrickRed] (0.85,0.05)--(0.75,-0.05);
          \draw[BrickRed] (0.05,0)--(-0.05,0.1);   
          \draw[BrickRed] (0.05,0)--(-0.05,-0.1);   
        \end{scope}
        \draw (0,1) arc (90:270:0.2 and 1);
        \node at (0,-1.4) {$\bra{\psi}$};
      \end{scope}

      \begin{scope}[xshift=5cm]
        \draw[fill=lightgray!20!white] (0,0) circle (1);       
        \draw[dotted] (0,1) arc (90:-90:0.2 and 1);
        \node at (-0.9,0.9) {$A$};
        \node at (0.9,0.9) {$B$};
        \begin{scope}
          \draw[BrickRed] (-0.8,0)--(0.8,0) ;
          \draw[BrickRed] (-0.85,-0.05)--(-0.75,0.05);
          \draw[BrickRed] (-0.85,0.05)--(-0.75,-0.05);
          \draw[BrickRed] (0.85,-0.05)--(0.75,0.05);
          \draw[BrickRed] (0.85,0.05)--(0.75,-0.05);
          \draw[BrickRed] (0.05,0)--(-0.05,0.1);   
          \draw[BrickRed] (0.05,0)--(-0.05,-0.1);   
        \end{scope}
        \draw (0,1) arc (90:270:0.2 and 1);
        \node at (0,-1.4) {$\ket{\psi}$};
      \end{scope}

      \begin{scope}[xshift=7.5cm]
        \draw[fill=lightgray!20!white] (0,0) circle (1);        
        \draw[dotted] (0,1) arc (90:-90:0.2 and 1);
        \node at (-0.9,0.9) {$\bar{B}$};
        \node at (0.9,0.9) {$\bar{A}$};
        \begin{scope}
          \draw[BrickRed] (-0.8,0)--(0.8,0) ;
          \draw[BrickRed] (-0.85,-0.05)--(-0.75,0.05);
          \draw[BrickRed] (-0.85,0.05)--(-0.75,-0.05);
          \draw[BrickRed] (0.85,-0.05)--(0.75,0.05);
          \draw[BrickRed] (0.85,0.05)--(0.75,-0.05);
          \draw[BrickRed] (0.05,0)--(-0.05,0.1);   
          \draw[BrickRed] (0.05,0)--(-0.05,-0.1);   
        \end{scope}
        \draw (0,1) arc (90:270:0.2 and 1);
        \node at (0,-1.4) {$\bra{\psi}$};
      \end{scope}
      
      \begin{scope}[xshift=9.5cm]
        \node at (0,0) {\Large $\to$};
      \end{scope}

      \begin{scope}[xshift=11.5cm]
        \draw (0,0) circle (1);
        \draw[dotted] (0,0) ellipse (1 and 0.3);
        \draw[BrickRed,very thick] (0,0) ellipse (0.4 and 0.2);
        \node[BrickRed] at (0,0.5) {$R_i$};
        \node at (0,-1.4) {$\Tr_A\left[(\Tr_B\ket{\psi}\bra{\psi})^2\right]$};
      \end{scope}

    \end{tikzpicture}
    \caption{We can calculate $\Tr_A\left[(\Tr_B\ket{\psi}\bra{\psi})^2\right]$ by gluing $B$ with the neighboring $\bar{B}$, corresponding to taking the partial trace over $B$, and $A$ with the neighboring $\bar{A}$, corresponding to the product of $\rho_A$. The last $\bar{A}$ is glued to the first $A$, corresponding to the trace over $A$.}
    \label{fig:twoexcitations}
  \end{center}
\end{figure}
\begin{align}
    \Tr_A\left[\rho_A^n\right]
        =
        \frac{Z\left[\BS^3; R_i\right]}{Z\left[\BS^3; R_i\right]^n}\ .
\end{align}
Thus, the topological entanglement entropy is given by 
\begin{align}\label{qdimin}
    \begin{aligned}
    S(\rho_A)
        &=
        \log Z\left[\BS^3; R_i\right]\\
        &=
        \log \Smat{0}{i}\\
        &=
        -\log\, \CD+\log d_i\ . 
    \end{aligned}
\end{align}

If we do not insert any excitation, we have by simply setting $d_j=0$,\footnote{In $d=3$ dimensions, the pseudo entropy can have an area law UV divergent term. In the Chern-Simons theory calculation, however, the partition function is a topological invariant, i.e., independent of any scale, after renormalizing the UV divergence in an appropriate scheme \cite{Witten:1988hf}. Hence in this case the pseudo entropy is free from the area law term and becomes scale independent.} 
\begin{align}
    \begin{aligned}
    S(\rho_A)
        &=
        \log Z\left[\BS^3\right] \\
        &=
        \log \Smat{0}{0}\\
        &=
        -\log \CD\ . 
    \end{aligned}
\end{align}
This vacuum topological entanglement entropy is related to the total quantum dimension \cite{Kitaev:2005dm,Levin:2006zz} and is expected to measure the degrees of freedom of edge modes, which is analogous to the area term in the holographic entanglement entropy. When we add an anyon, the topological entanglement entropy increases by the amount of log of the quantum dimension as in \eqref{qdimin}.

\subsection{Topological pseudo entropy on \texorpdfstring{$\BS^2$}{S2} with four excitations} \label{subsec:excitation}
We consider the case that the spatial region is $\BS^2$, which is divided to two subregions $A$ and $B$ as the figures show in \eqref{eq:exc1} and there are four excitations.
For simplicity, we only consider fundamental (called $j$) or anti-fundamental (called $\bar{j}$) excitations. For the total charge to vanish, two of the four excitations must be fundamental and the others must be anti-fundamental.
There are then two possible cases: 1) a pair of $j$ and $\bar{j}$ in $A$ and the other pair in $B$, and 2) two $j$'s in $A$ and two $\bar{j}$'s in $B$.
We prepare these states by the path integral.
The excitations will be the edges of Wilson lines.
There are many ways to connect the excitations so that the Wilson lines make some knots.
In what follows we will show they give rise to nontrivial contributions to the pseudo or entanglement entropies.

\subsubsection{Case 1: \texorpdfstring{$j$}{j} and \texorpdfstring{$\bar{j}$}{j} in \texorpdfstring{$A$}{A}, the others in \texorpdfstring{$B$}{B}}
In this case, there are two configurations of Wilson lines which end on one $j$ and one $\bar{j}$.
We set $\ket{\psi}$ and $\ket{\varphi}$ as 
\begin{align}\label{eq:exc1}
  \ket{\psi}=
    \begin{tikzpicture}[thick,scale=1.5,baseline={([yshift=-.5ex]current bounding box.center)}]
        \begin{scope}[decoration={markings, mark=at position 0.5 with {\arrow{>}}}]
        \draw[fill=lightgray!20!white] (0,0) circle (1);
        \draw[dotted] (0,1) arc (90:-90:0.2 and 1);
        \node at (-0.9,0.9) {$A$};
        \node at (0.9,0.9) {$B$};
        \begin{scope}[shift={(-0.7,0.4)}] 
          \draw[BrickRed] (-0.05,-0.05)--(0.05,0.05);
          \draw[BrickRed] (-0.05,0.05)--(0.05,-0.05);
        \end{scope}
        \begin{scope}[shift={(-0.7,-0.4)}] 
          \draw[BrickRed] (-0.05,-0.05)--(0.05,0.05);
          \draw[BrickRed] (-0.05,0.05)--(0.05,-0.05);
        \end{scope}
        \begin{scope}[shift={(0.7,0.4)}] 
          \draw[BrickRed] (-0.05,-0.05)--(0.05,0.05);
          \draw[BrickRed] (-0.05,0.05)--(0.05,-0.05);
        \end{scope}
        \begin{scope}[shift={(0.7,-0.4)}] 
          \draw[BrickRed] (-0.05,-0.05)--(0.05,0.05);
          \draw[BrickRed] (-0.05,0.05)--(0.05,-0.05);
        \end{scope}
         \draw[BrickRed, postaction={decorate}] (-0.7,-0.4) to [out=60,in=300] (-0.7,0.4);
         \draw[BrickRed,postaction={decorate}] (0.7,0.4) to [out=240,in=120] (0.7,-0.4);
        \node[BrickRed] at (-0.55,0.6) {\footnotesize$j$} ;
        \node[BrickRed] at (-0.55,-0.6) {\footnotesize$\bar{j}$} ;
        \node[BrickRed] at (0.55,0.6) {\footnotesize$\bar{j}$} ;
        \node[BrickRed] at (0.55,-0.6) {\footnotesize$j$} ;
        
        \draw (0,1) arc (90:270:0.2 and 1);
      \end{scope}
    \end{tikzpicture}
    , \qquad
  \ket{\varphi}=
    \begin{tikzpicture}[thick,scale=1.5,baseline={([yshift=-.5ex]current bounding box.center)}]
        \begin{scope}[decoration={markings, mark=at position 0.5 with {\arrow{>}}}]
        \draw[fill=lightgray!20!white] (0,0) circle (1);
        \draw[dotted] (0,1) arc (90:-90:0.2 and 1);
        \node at (-0.9,0.9) {$A$};
        \node at (0.9,0.9) {$B$};
        \begin{scope}[shift={(-0.7,0.4)}] 
          \draw[BrickRed] (-0.05,-0.05)--(0.05,0.05);
          \draw[BrickRed] (-0.05,0.05)--(0.05,-0.05);
        \end{scope}
        \begin{scope}[shift={(-0.7,-0.4)}] 
          \draw[BrickRed] (-0.05,-0.05)--(0.05,0.05);
          \draw[BrickRed] (-0.05,0.05)--(0.05,-0.05);
        \end{scope}
        \begin{scope}[shift={(0.7,0.4)}] 
          \draw[BrickRed] (-0.05,-0.05)--(0.05,0.05);
          \draw[BrickRed] (-0.05,0.05)--(0.05,-0.05);
        \end{scope}
        \begin{scope}[shift={(0.7,-0.4)}] 
          \draw[BrickRed] (-0.05,-0.05)--(0.05,0.05);
          \draw[BrickRed] (-0.05,0.05)--(0.05,-0.05);
        \end{scope}
        \draw[BrickRed, postaction={decorate}] (0.7,0.4) to [out=210,in=330] (-0.7,0.4);
        \draw[BrickRed,postaction={decorate}] (-0.7,-0.4) to [out=30,in=150] (0.7,-0.4);
        \node[BrickRed] at (-0.55,0.6) {\footnotesize$j$} ;
        \node[BrickRed] at (-0.55,-0.6) {\footnotesize$\bar{j}$} ;
        \node[BrickRed] at (0.55,0.6) {\footnotesize$\bar{j}$} ;
        \node[BrickRed] at (0.55,-0.6) {\footnotesize$j$} ;
        
        \draw (0,1) arc (90:270:0.2 and 1);
      \end{scope}
    \end{tikzpicture}
\end{align}

We first calculate the entanglement entropies of $\ket{\psi}$ and $\ket{\varphi}$.
For $\ket{\psi}$, $\Tr_A\left[\left(\tilde{\rho}^{\psi}_A\right)^n\right]$ equals to the partition function on $\BS^3$ that includes $2n$ Wilson loops in the representation $R_j$.
Thus 
\begin{align}
    \begin{aligned}
    \Tr_A\left[\left(\rho^{\psi}_A\right)^n\right]
        &=
        \frac{Z\left[\BS^3;R_j\right]^{2n}/Z\left[\BS^3\right]^{2n-1}}{\left(Z\left[\BS^3;R_j\right]^2/Z\left[\BS^3\right]\right)^n} \\
        &=
        Z\left[\BS^3\right]^{1-n}\\
        &=
        \left(\Smat{0}{0}\right)^{1-n}\ .
    \end{aligned}
\end{align}
Since $\Smat{0}{0}=\CD^{-1}$, we have
\begin{align}\label{eq:exc1psi}
  S\left(\rho^{\psi}_A\right) = -\log \CD\ .
\end{align}
For $\ket{\varphi}$, $\Tr_A\left[\left(\tilde{\rho}^{\varphi}_A\right)^n\right]$ equals to the partition function on $\BS^3$ that includes two Wilson loops:
\begin{align}
    \begin{aligned}
    \Tr_A\left[\left(\rho^{\varphi}_A\right)^n\right]
        &=
            \frac{Z\left[\BS^3;R_j\right]^2/Z\left[\BS^3\right]}{\left(Z\left[\BS^3;R_j\right]^2/Z\left[\BS^3\right]\right)^{n}} \\
        &=
            \left[\frac{(\Smat{0}{j})^2}{\Smat{0}{0}}\right]^{1-n}\ .
    \end{aligned}
\end{align}
Therefore, we have
\begin{align}\label{eq:exc1phi}
  S\left(\rho^{\varphi}_A\right)
   =
   -\log \CD+2\log d_j\ .
\end{align}

Next, we calculate the pseudo entropy of the reduced transition matrix:
\begin{align}
  \tau_A^{\psi|\varphi}=\Tr_B\left[\frac{\ket{\psi}\bra{\varphi}}{\braket{\varphi|\psi}}\right]\ .
\end{align}
$\Tr_A\left[\left(\tilde{\tau}_A^{\psi|\varphi}\right)^n\right]$ equals to the partition function on $\BS^3$ with $n$ Wilson loop, so
\begin{align}
    \begin{aligned}
    \Tr_A\left[\left(\tau_A^{\psi|\varphi}\right)^n\right]
        &=
            \frac{Z\left[\BS^3;R_j\right]^n/Z\left[\BS^3\right]^{n-1}}{Z\left[\BS^3;R_j\right]^n} \\
        &=\left(\Smat{0}{0}\right)^{1-n}\ .
    \end{aligned}
\end{align}
Therefore, the pseudo entropy is 
\begin{align}\label{exc1pseudo}
  S\left(\tau_A^{\psi|\varphi}\right)=-\log \CD\ .
\end{align}
In this case the difference of the pseudo entropy from the entanglement entropy is negative:
\begin{align}
  \Delta S=-\log d_j<0\ .
\end{align}

The results \eqref{eq:exc1psi}, \eqref{eq:exc1phi}, and \eqref{exc1pseudo} are easily interpreted as follows.
$\ket{\psi}$ is not entangled since no Wilson lines connect $A$ and $B$, so that $S\left(\rho^{\psi}_A\right)$ has no non-topological contributions. On the other hand, $S\left(\rho^{\varphi}_A\right)$ has the term $2\log d_j$ because $\ket{\varphi}$ is entangled due to the Wilson lines connecting the two points $A$ and $B$.
As shown in \cite{Nakata:2021ubr}, the pseudo entropy is zero when either state has no entanglement.
Now $\ket{\psi}$ has no entanglement, so $S\left(\tm{\psi}{\varphi}_A\right)$ has no terms other than the topological term.

\subsubsection{Case 2: two \texorpdfstring{$j$}{j}'s in \texorpdfstring{$A$}{A}, the others in \texorpdfstring{$B$}{B}}\label{sec:CS_case2}
We define states $\ket{\psi_a} (a\in\Z)$ as follows. First we define at $a=0$
\begin{align}\label{eq:psizero}
  \ket{\psi_0}=
  \begin{tikzpicture}[thick,scale=1.5,baseline={([yshift=-.5ex]current bounding box.center)}]
        \begin{scope}[decoration={markings, mark=at position 0.5 with {\arrow{>}}}]
        \draw[fill=lightgray!20!white] (0,0) circle (1);
        \draw[dotted] (0,1) arc (90:-90:0.2 and 1);
        \node at (-0.9,0.9) {$A$};
        \node at (0.9,0.9) {$B$};
        \begin{scope}[shift={(-0.7,0.4)}] 
          \draw[BrickRed] (-0.05,-0.05)--(0.05,0.05);
          \draw[BrickRed] (-0.05,0.05)--(0.05,-0.05);
        \end{scope}
        \begin{scope}[shift={(-0.7,-0.4)}] 
          \draw[BrickRed] (-0.05,-0.05)--(0.05,0.05);
          \draw[BrickRed] (-0.05,0.05)--(0.05,-0.05);
        \end{scope}
        \begin{scope}[shift={(0.7,0.4)}] 
          \draw[BrickRed] (-0.05,-0.05)--(0.05,0.05);
          \draw[BrickRed] (-0.05,0.05)--(0.05,-0.05);
        \end{scope}
        \begin{scope}[shift={(0.7,-0.4)}] 
          \draw[BrickRed] (-0.05,-0.05)--(0.05,0.05);
          \draw[BrickRed] (-0.05,0.05)--(0.05,-0.05);
        \end{scope}
        \draw[BrickRed, postaction={decorate}] (0.7,0.4) to [out=210,in=330] (-0.7,0.4);
        \draw[BrickRed,postaction={decorate}] (0.7,-0.4) to [out=150,in=30] (-0.7,-0.4);
        \node[BrickRed] at (-0.55,0.6) {\footnotesize$j$} ;
        \node[BrickRed] at (-0.55,-0.6) {\footnotesize$j$} ;
        \node[BrickRed] at (0.55,0.6) {\footnotesize$\bar{j}$} ;
        \node[BrickRed] at (0.55,-0.6) {\footnotesize$\bar{j}$} ;
        
        \draw (0,1) arc (90:270:0.2 and 1);
      \end{scope}
    \end{tikzpicture}
    \ .
\end{align}
Then we define $\ket{\psi_a}(a\in\Z_+)$ by twisting the region $B$ $a$ times:
\begin{align}\label{eq:psipos}
  \ket{\psi_1}=
  \begin{tikzpicture}[thick,scale=1.5,baseline={([yshift=-.5ex]current bounding box.center)}]
        \begin{scope}[decoration={markings, mark=at position 0.3 with {\arrow{>}}}]
        \draw[fill=lightgray!20!white] (0,0) circle (1);
        \draw[dotted] (0,1) arc (90:-90:0.2 and 1);
        \node at (-0.9,0.9) {$A$};
        \node at (0.9,0.9) {$B$};
        \begin{scope}[shift={(-0.7,0.4)}] 
          \draw[BrickRed] (-0.05,-0.05)--(0.05,0.05);
          \draw[BrickRed] (-0.05,0.05)--(0.05,-0.05);
        \end{scope}
        \begin{scope}[shift={(-0.7,-0.4)}] 
          \draw[BrickRed] (-0.05,-0.05)--(0.05,0.05);
          \draw[BrickRed] (-0.05,0.05)--(0.05,-0.05);
        \end{scope}
        \begin{scope}[shift={(0.7,0.4)}] 
          \draw[BrickRed] (-0.05,-0.05)--(0.05,0.05);
          \draw[BrickRed] (-0.05,0.05)--(0.05,-0.05);
        \end{scope}
        \begin{scope}[shift={(0.7,-0.4)}] 
          \draw[BrickRed] (-0.05,-0.05)--(0.05,0.05);
          \draw[BrickRed] (-0.05,0.05)--(0.05,-0.05);
        \end{scope}
        \begin{knot}[background color=lightgray!20!white]
        \strand[BrickRed, postaction={decorate}] (0.7,0.4) -- (-0.7,-0.4);
        \strand[BrickRed, postaction={decorate}] (0.7,-0.4) -- (-0.7,0.4);
        \end{knot}
        \node[BrickRed] at (-0.55,0.6) {\footnotesize$j$} ;
        \node[BrickRed] at (-0.55,-0.6) {\footnotesize$j$} ;
        \node[BrickRed] at (0.55,0.6) {\footnotesize$\bar{j}$} ;
        \node[BrickRed] at (0.55,-0.6) {\footnotesize$\bar{j}$} ;
        
        \draw (0,1) arc (90:270:0.2 and 1);
      \end{scope}
    \end{tikzpicture}
  
  \ ,\qquad\ket{\psi_2}=
  \begin{tikzpicture}[thick,scale=1.5,baseline={([yshift=-.5ex]current bounding box.center)}]
        \begin{scope}[decoration={markings, mark=at position 0.5 with {\arrow{>}}}]
        \draw[fill=lightgray!20!white] (0,0) circle (1);
        \draw[dotted] (0,1) arc (90:-90:0.2 and 1);
        \node at (-0.9,0.9) {$A$};
        \node at (0.9,0.9) {$B$};
        \begin{scope}[shift={(-0.7,0.4)}] 
          \draw[BrickRed] (-0.05,-0.05)--(0.05,0.05);
          \draw[BrickRed] (-0.05,0.05)--(0.05,-0.05);
        \end{scope}
        \begin{scope}[shift={(-0.7,-0.4)}] 
          \draw[BrickRed] (-0.05,-0.05)--(0.05,0.05);
          \draw[BrickRed] (-0.05,0.05)--(0.05,-0.05);
        \end{scope}
        \begin{scope}[shift={(0.7,0.4)}] 
          \draw[BrickRed] (-0.05,-0.05)--(0.05,0.05);
          \draw[BrickRed] (-0.05,0.05)--(0.05,-0.05);
        \end{scope}
        \begin{scope}[shift={(0.7,-0.4)}] 
          \draw[BrickRed] (-0.05,-0.05)--(0.05,0.05);
          \draw[BrickRed] (-0.05,0.05)--(0.05,-0.05);
        \end{scope}
        \begin{knot}[background color=lightgray!20!white,flip crossing=2]
        \strand[BrickRed, postaction={decorate}] (0.7,0.4) .. controls (0,-0.5) .. (-0.7,0.4);
        \strand[BrickRed, postaction={decorate}] (0.7,-0.4) .. controls (0,0.5) .. (-0.7,-0.4);
        \end{knot}
        \node[BrickRed] at (-0.55,0.6) {\footnotesize$j$} ;
        \node[BrickRed] at (-0.55,-0.6) {\footnotesize$j$} ;
        \node[BrickRed] at (0.55,0.6) {\footnotesize$\bar{j}$} ;
        \node[BrickRed] at (0.55,-0.6) {\footnotesize$\bar{j}$} ;
        
        \draw (0,1) arc (90:270:0.2 and 1);
      \end{scope}
    \end{tikzpicture}
  \ ,\qquad\ket{\psi_3}=\cdots
\end{align}
On the other hand, we define $\ket{\psi_a}(a\in\Z_-)$ by twisting the region $B$ $|a|$ times in the opposite direction:
\begin{align}\label{eq:psineg}
  \ket{\psi_{-1}}=
    \begin{tikzpicture}[thick,scale=1.5,baseline={([yshift=-.5ex]current bounding box.center)}]
        \begin{scope}[decoration={markings, mark=at position 0.3 with {\arrow{>}}}]
        \draw[fill=lightgray!20!white] (0,0) circle (1);
        \draw[dotted] (0,1) arc (90:-90:0.2 and 1);
        \node at (-0.9,0.9) {$A$};
        \node at (0.9,0.9) {$B$};
        \begin{scope}[shift={(-0.7,0.4)}] 
          \draw[BrickRed] (-0.05,-0.05)--(0.05,0.05);
          \draw[BrickRed] (-0.05,0.05)--(0.05,-0.05);
        \end{scope}
        \begin{scope}[shift={(-0.7,-0.4)}] 
          \draw[BrickRed] (-0.05,-0.05)--(0.05,0.05);
          \draw[BrickRed] (-0.05,0.05)--(0.05,-0.05);
        \end{scope}
        \begin{scope}[shift={(0.7,0.4)}] 
          \draw[BrickRed] (-0.05,-0.05)--(0.05,0.05);
          \draw[BrickRed] (-0.05,0.05)--(0.05,-0.05);
        \end{scope}
        \begin{scope}[shift={(0.7,-0.4)}] 
          \draw[BrickRed] (-0.05,-0.05)--(0.05,0.05);
          \draw[BrickRed] (-0.05,0.05)--(0.05,-0.05);
        \end{scope}
        \begin{knot}[background color=lightgray!20!white,flip crossing=1]
        \strand[BrickRed, postaction={decorate}] (0.7,0.4) -- (-0.7,-0.4);
        \strand[BrickRed, postaction={decorate}] (0.7,-0.4) -- (-0.7,0.4);
        \end{knot}
        \node[BrickRed] at (-0.55,0.6) {\footnotesize$j$} ;
        \node[BrickRed] at (-0.55,-0.6) {\footnotesize$j$} ;
        \node[BrickRed] at (0.55,0.6) {\footnotesize$\bar{j}$} ;
        \node[BrickRed] at (0.55,-0.6) {\footnotesize$\bar{j}$} ;
        
        \draw (0,1) arc (90:270:0.2 and 1);
      \end{scope}
    \end{tikzpicture}
  \ ,\qquad\ket{\psi_{-2}}=
    \begin{tikzpicture}[thick,scale=1.5,baseline={([yshift=-.5ex]current bounding box.center)}]
        \begin{scope}[decoration={markings, mark=at position 0.5 with {\arrow{>}}}]
        \draw[fill=lightgray!20!white] (0,0) circle (1);
        \draw[dotted] (0,1) arc (90:-90:0.2 and 1);
        \node at (-0.9,0.9) {$A$};
        \node at (0.9,0.9) {$B$};
        \begin{scope}[shift={(-0.7,0.4)}] 
          \draw[BrickRed] (-0.05,-0.05)--(0.05,0.05);
          \draw[BrickRed] (-0.05,0.05)--(0.05,-0.05);
        \end{scope}
        \begin{scope}[shift={(-0.7,-0.4)}] 
          \draw[BrickRed] (-0.05,-0.05)--(0.05,0.05);
          \draw[BrickRed] (-0.05,0.05)--(0.05,-0.05);
        \end{scope}
        \begin{scope}[shift={(0.7,0.4)}] 
          \draw[BrickRed] (-0.05,-0.05)--(0.05,0.05);
          \draw[BrickRed] (-0.05,0.05)--(0.05,-0.05);
        \end{scope}
        \begin{scope}[shift={(0.7,-0.4)}] 
          \draw[BrickRed] (-0.05,-0.05)--(0.05,0.05);
          \draw[BrickRed] (-0.05,0.05)--(0.05,-0.05);
        \end{scope}
        \begin{knot}[background color=lightgray!20!white,flip crossing=1]
        \strand[BrickRed, postaction={decorate}] (0.7,0.4) .. controls (0,-0.5) .. (-0.7,0.4);
        \strand[BrickRed, postaction={decorate}] (0.7,-0.4) .. controls (0,0.5) .. (-0.7,-0.4);
        \end{knot}
        \node[BrickRed] at (-0.55,0.6) {\footnotesize$j$} ;
        \node[BrickRed] at (-0.55,-0.6) {\footnotesize$j$} ;
        \node[BrickRed] at (0.55,0.6) {\footnotesize$\bar{j}$} ;
        \node[BrickRed] at (0.55,-0.6) {\footnotesize$\bar{j}$} ;
        
        \draw (0,1) arc (90:270:0.2 and 1);
      \end{scope}
    \end{tikzpicture}
  
  \ ,\qquad\ket{\psi_{-3}}=\cdots
\end{align}
In other words, $\ket{\psi_a}$ is a state which has $|a|$ crossings.
We would like to calculate the pseudo entropy of the transition matrix:
\begin{align}
  \tm{a}{b}\equiv\frac{\ket{\psi_a}\bra{\psi_b}}{\braket{\psi_b|\psi_a}}.
\end{align}
The unnormalized reduced transition matrix $\ttm{a}{b}_A\equiv\Tr_B[\ket{\psi_a}\bra{\psi_b}]$ is
\begin{align}\label{eq:tmab}
  \ttm{a}{b}_A=
    \begin{tikzpicture}[thick,scale=1.5,baseline={([yshift=-.5ex]current bounding box.center)}]
        \begin{scope}[decoration={markings, mark=at position 0.7 with {\arrow{>}}}]
        \draw[fill=lightgray!20!white] (0,0) ellipse (1.5 and 1);
        \draw[dotted] (0,1) arc (90:-90:0.2 and 1);
        \node at (-1.2,0.9) {$A$};
        \node at (1.2,0.9) {$\bar{A}$};
        \begin{scope}[shift={(-1.1,0.4)}] 
          \draw[BrickRed] (-0.05,-0.05)--(0.05,0.05);
          \draw[BrickRed] (-0.05,0.05)--(0.05,-0.05);
        \end{scope}
        \begin{scope}[shift={(-1.1,-0.4)}] 
          \draw[BrickRed] (-0.05,-0.05)--(0.05,0.05);
          \draw[BrickRed] (-0.05,0.05)--(0.05,-0.05);
        \end{scope}
        \begin{scope}[shift={(1.1,0.4)}] 
          \draw[BrickRed] (-0.05,-0.05)--(0.05,0.05);
          \draw[BrickRed] (-0.05,0.05)--(0.05,-0.05);
        \end{scope}
        \begin{scope}[shift={(1.1,-0.4)}] 
          \draw[BrickRed] (-0.05,-0.05)--(0.05,0.05);
          \draw[BrickRed] (-0.05,0.05)--(0.05,-0.05);
        \end{scope}
        \begin{knot}[background color=lightgray!20!white]
            \strand[BrickRed] (1.1,0.4) .. controls (0.9,-0.45) and (0.7,-0.45) .. (0.52,-0.1);
            \strand[BrickRed] (1.1,-0.4) .. controls (0.9,0.45) and (0.7,0.45) .. (0.52,0.1);
        \end{knot}
        \begin{knot}[background color=lightgray!20!white,flip crossing=1]
            \strand[BrickRed] (-1.1,0.4) .. controls (-0.8,-0.5) and (-0.6,-0.5) .. (-0.38,0) .. controls (-0.2,0.42) and (0,0.42) ..(0.15,0.1);
            \strand[BrickRed] (-1.1,-0.4) .. controls (-0.8,0.5) and (-0.6,0.5) .. (-0.38,0) .. controls (-0.2,-0.42) and (0,-0.42) ..(0.15,-0.1);
        \end{knot}
        \node[BrickRed] at (-0.95,0.6) {\footnotesize$j$} ;
        \node[BrickRed] at (-0.95,-0.6) {\footnotesize$j$} ;
        \node[BrickRed] at (0.95,0.6) {\footnotesize$\bar{j}$} ;
        \node[BrickRed] at (0.95,-0.6) {\footnotesize$\bar{j}$} ;
        \node[BrickRed] at (0.4, 0) {\large $\dots$};
        \draw (0,1) arc (90:270:0.2 and 1);
        \node[fill=lightgray!20!white] at (0,-0.6) {\footnotesize $|a-b|$ crossings};
      \end{scope}
    \end{tikzpicture}
  \ ,
\end{align}
which has the crossing number $|a-b|$.
In the figure, $\bar{A}$ means the conjugation of $A$.
Therefore $\Tr_A\left[\left(\ttm{a}{b}_A\right)^n\right]$ has one or two Wilson loops with $n|a-b|$ crossings.\footnote{When $n|a-b|$ is even, there are two Wilson loops while there is one Wilson loop when $n|a-b|$ is odd.}

Here let us pause to compute the partition function on $\BS^3$ with a crossing number $m$.
We call such a manifold as $X_m$.
In this case, we also use a technique introduced in \cite{Witten:1988hf}.
We cut along a two-dimensional submanifold that intersects with Wilson lines for four times, and we perform a twisting transformation on the cross section.
Then we obtain three states with different links.
Since the Hilbert space on the cross section is two-dimensional due to \eqref{eq:invsp}, these three states are linearly dependent, giving the skein relation:
\begin{align}
  \alpha\, Z\left[X_m\right] + \beta\, Z\left[X_{m-1}\right] + \gamma\, Z\left[X_{m-2}\right] = 0\ ,
\end{align}
where we call $\BS^3$ including $m$-crossing Wilson lines $X_m$.
Since now our gauge group is $\SU(N)$ and Wilson loops are in the fundamental representation, the coefficients are\footnote{In fact, those coefficients depend on the choice of the ``framing". The framing of Wilson lines in $\BS^3$ can be chosen to be canonical in the sense that the self-interaction numbers of the links are zero. The result $Z\left[X_m\right]/Z\left[X_1\right]^n$ does not depend on the choice of the framing.}
\begin{align}
  \alpha=-q^{\frac{N}{2}}\ ,\qquad \beta=q^\frac{1}{2}-q^{-\frac{1}{2}}\ ,\qquad \gamma=q^{-\frac{N}{2}}\ ,
\end{align}
where we define $q=e^{2\pi\i/(N+k)}$.
Then we obtain the recursion relation 
\begin{align}
  Z\left[X_m\right] + q^{-\frac{N+1}{2}}\,Z\left[X_{m-1}\right]
    =
    q^{-\frac{N-1}{2}}\,\left(Z\left[X_{m-1}\right]+q^{-\frac{N+1}{2}}\,Z\left[X_{m-2}\right]\right)\ .
\end{align}
Solving this relation with the initial conditions $Z\left[X_0\right] = \CS_0^{~0}\,d_j^2$ and $Z\left[X_1\right] = \CS_0^{~0}\,d_j$, we have 
\begin{align}\label{eq:exc2ZMn}
  \frac{Z\left[X_m\right]}{\Smat{0}{0}}
    =
    \left(q^{-\frac{N-1}{2}}\right)^m\frac{[N+1]\,[N]}{[2]}+\left(-q^{-\frac{N+1}{2}}\right)^m\frac{[N]\,[N-1]}{[2]}\ ,
\end{align}
where
\begin{align}
    [x]\equiv\frac{q^\frac{x}{2}-q^{-\frac{x}{2}}}{q^{\frac{1}{2}}-q^{-\frac{1}{2}}} \ ,
\end{align}
and the quantum dimension is $d_j=[N]$. 
This is what we have wanted to obtain. 

It follows from \eqref{eq:exc2ZMn} 
\begin{align}\label{eq:trtaun}
    \begin{aligned}
  \Tr_A\left[\left(\tau_A^{a|b}\right)^n\right]
    &=
    \frac{Z\left[X_{|(a-b)n|}\right]}{Z\left[X_{|a-b|}\right]^n} \\
    &=
    \left(\Smat{0}{0}\,[N]\right)^{(1-n)}\,\frac{\left(q^{\frac{1}{2}}\right)^{|a-b|\,n}\frac{[N+1]}{[2]}+\left(-q^{-\frac{1}{2}}\right)^{|a-b|\,n}\frac{[N-1]}{[2]}}{\left[\left(q^{\frac{1}{2}}\right)^{|a-b|}\frac{[N+1]}{[2]}+\left(-q^{-\frac{1}{2}}\right)^{|a-b|}\frac{[N-1]}{[2]}\right]^n}\ .
    \end{aligned}
\end{align}
When $a=b$, 
\begin{align}
  \Tr_A\left[\left(\rho^{a}_A\right)^n\right] = \left(\Smat{0}{0}\,[N]^2\right)^{1-n}\ ,
\end{align}
where we have defined $\rho^a_A\equiv\tm{a}{a}_A$ and used the relation $[N]=\frac{[N+1]}{[2]}+\frac{[N-1]}{[2]}$.
Then the entanglement entropy becomes independent of $a$:
\begin{align}
  S\left(\rho^a_A\right) = -\log \CD + 2\log\,[N]\ .
\end{align}

We are now ready to calculate the difference $\Delta S$ of the pseudo entropy from the averaged entanglement entropy, defined by 
\begin{align}\label{eq:deltaSgen1}
  \Delta S
    =
    -\frac{1}{2}\left.\frac{\partial}{\partial n}\log\frac{\Tr_A\left[\left(\tau_A^{a|b}\right)^n\right]\,\Tr_A\left[\left(\tau_A^{a|b}\right)^n\right]^*}{\Tr_A\left[\left(\rho_A^{a}\right)^n\right]\,\Tr_A\left[\left(\rho_A^{b}\right)^n\right]}\right|_{n=1}\ .
\end{align}
Here the argument of the logarithm is
\begin{align}\label{eq:inlog}
\begin{aligned}
  &\frac{\Tr_A\left[\left(\tau_A^{a|b}\right)^n\right]\,\Tr_A\left[\left(\tau_A^{a|b}\right)^n\right]^*}{\Tr_A\left[\left(\rho_A^{a}\right)^n\right]\,\Tr_A\left[\left(\rho_A^{b}\right)^n\right]}\\
  &\quad =
  \left([N]\,[2]\right)^{2(n-1)}\,\frac{[N+1]^2+[N-1]^2 + 2\,(-1)^{|a-b|\,n}\,\cos\left(\frac{2\pi\, |a-b|\,n}{N+k}\right) [N+1]\,[N-1]}{\left[ [N+1]^2+[N-1]^2+2\,(-1)^{|a-b|}\,\cos\left(\frac{2\pi\, |a-b|}{N+k}\right) [N+1]\,[N-1]\right]^n}\ .
  \end{aligned}
\end{align}

Now we analytically continue $n$ in \eqref{eq:trtaun} or \eqref{eq:inlog} to real numbers. 
However we have to be careful because the phase factor $(-1)^{|a-b|n}$ depends on the way of analytic continuation. 
In the followings we compute $S\left(\tm{\psi}{\varphi}_A\right)$ and $\Delta S$ in two different prescriptions of analytic continuations: (1) a naive prescription by deforming $(-1)^{|a-b|n}=e^{\i\pi|a-b|n}$ and (2) restricting $n$ to odd numbers and then analytically continuing to real numbers, which is similar to the replica method for the logarithmic negativity \cite{Calabrese:2012ew}.

\paragraph{(1) A naive prescription}
When $|a-b|$ is even, $(-1)^{|a-b|\,n}=1$ for any integer $n$.
Therefore there is no ambiguity due to the choice of the prescriptions.
Thus the pseudo entropy takes the form:
\begin{align}
    \begin{aligned}
        S\left(\tm{a}{b}_A\right)
        &=
        -\log\mathcal{D}+\log\left[\frac{[N]}{[2]}\right]+\log\left[q^{\frac{|a-b|}{2}}\,[N+1]+q^{-\frac{|a-b|}{2}}\,[N-1]\right] \\
        &\qquad\qquad - \i\,\frac{\pi\,|a-b|}{N+k}\, \frac{q^{\frac{|a-b|}{2}}\,[N+1]-q^{-\frac{|a-b|}{2}}\,[N-1]}{q^{\frac{|a-b|}{2}}\,[N+1]+q^{-\frac{|a-b|}{2}}\,[N-1]}\ ,
    \end{aligned}
\end{align}
and $\Delta S$ becomes 
\begin{align}
  \begin{aligned}
  \Delta S
    &=
        -\log\left( [N]\,[2] \right)+ \frac{1}{2}\,\log\left[[N+1]^2+[N-1]^2 + 2\,\cos\left(\frac{2\pi\, |a-b|}{N+k}\right)[N+1]\,[N-1]\right]\\
    &\qquad 
    + \frac{2\pi\,|a-b|}{N+k}\,\frac{\sin\left(\frac{2\pi\,|a-b|}{N+k}\right)[N+1]\,[N-1]}{[N+1]^2+[N-1]^2 + 2\,\cos\left(\frac{2\pi\, |a-b|}{N+k}\right)[N+1]\,[N-1] }\ .
  \end{aligned}
\end{align}
When $|a-b|$ is odd, the factor $(-1)^{|a-b|\,n}$ remains. Deforming it to $e^{\i\pi\,|a-b|\,n}$, the pseudo entropy results in 
\begin{align}\label{Naive_PE_odd}
    \begin{aligned}
        S\left(\tm{a}{b}_A\right)
        &=
        -\log\mathcal{D}+\log\left[\frac{[N]}{[2]}\right]+\log\left[q^{\frac{|a-b|}{2}}\,[N+1]-q^{-\frac{|a-b|}{2}}\,[N-1]\right] \\
        &\qquad - \i\,\frac{\pi\,|a-b|}{N+k}\,\frac{q^{\frac{|a-b|}{2}}\,[N+1]+(1-N-k)\,q^{-\frac{|a-b|}{2}}\,[N-1]}{q^{\frac{|a-b|}{2}}\,[N+1]-q^{-\frac{|a-b|}{2}}\,[N-1]}\ ,
    \end{aligned}
\end{align}
and $\Delta S$ becomes 
\begin{align}
  \begin{aligned}
  \Delta S
    &=
        -\log\left( [N]\,[2] \right)+ \frac{1}{2}\,\log\left[[N+1]^2+[N-1]^2 - 2\,\cos\left(\frac{2\pi\, |a-b|}{N+k}\right)[N+1]\,[N-1]\right]\\
    &\qquad 
    + \frac{(2-N-k)\,\pi\,|a-b|}{N+k}\,\frac{\sin\left(\frac{2\pi\,|a-b|}{N+k}\right)[N+1]\,[N-1]}{[N+1]^2+[N-1]^2 - 2\,\cos\left(\frac{2\pi\, |a-b|}{N+k}\right)[N+1]\,[N-1] }\ .
  \end{aligned}
\end{align}

In this calculation, we used the relation $-1=e^{\i\pi}$. However, more generally it satisfies $-1=e^{\i(2m+1)\pi}\ (m\in\Z)$, which corresponds to choosing a branch of logarithm such that $(2m-1)\pi<\mathrm{Im}\,[\log z]\le (2m+1)\pi$. 
The pseudo entropy and $\Delta S$ depend on which branch we choose because of differentiating $(-1)^{|a-b|\,n}$ with respect to $n$. 
While the usual entanglement entropy also depends on the branch, it does not affect the real part. 
Therefore, it seems to be unnatural that the real part of the pseudo entropy, and $\Delta S$, depends on the branch.
To avoid this obstruction, we have to use a prescription that does not include the derivative of $(-1)^{|a-b|\,n}$. 

\paragraph{(2) Restricting $\bm{n}$ to odd numbers} 
In the previous calculation, the obstruction is the existence of $(-1)^{|a-b|\, n}$. 
Here we restrict $n$ to odd so that $(-1)^{|a-b|\,n}$ reduces to $(-1)^{|a-b|}$ and after that analytically continue $n$ to real numbers.\footnote{A similar method was used for the calculation of the entanglement entropy for Dirac fields \cite{Azeyanagi:2008}.
Also the logarithmic negativity calculation \cite{Calabrese:2012ew} employs the analytic continuation of even $n$.
Here we simply assume odd $n$ in continuing to $n=1$ as the (pseudo) R\'{e}nyi entropy goes to the (pseudo) entanglement entropy in the limit.
}
In this case, 
\begin{align}\label{Odd_continuation}
    \begin{aligned}
        S\left(\tm{a}{b}_A\right)
        &=
        -\log\mathcal{D}+\log\left[\frac{[N]}{[2]}\right]+\log\left[q^{\frac{|a-b|}{2}}\,[N+1]+(-1)^{|a-b|}\,q^{-\frac{|a-b|}{2}}\,[N-1]\right] \\
        &\qquad\qquad - \i\,\frac{\pi|a-b|}{N+k}\, \frac{q^{\frac{|a-b|}{2}}\,[N+1]-(-1)^{|a-b|}\,q^{-\frac{|a-b|}{2}}\,[N-1]}{q^{\frac{|a-b|}{2}}\,[N+1]+(-1)^{|a-b|}\,q^{-\frac{|a-b|}{2}}\,[N-1]}\ ,
    \end{aligned}
\end{align}
and
\begin{align}\label{eq:deltaSgen}
  \begin{aligned}
  \Delta S
    &=
        -\log\left( [N]\,[2] \right)+ \frac{1}{2}\,\log\left[[N+1]^2+[N-1]^2 + 2\,(-1)^{|a-b|}\,\cos\left(\frac{2\pi |a-b|}{N+k}\right)[N+1]\,[N-1]\right]\\
    &\qquad 
    + (-1)^{|a-b|}\,\frac{2\pi\,|a-b|}{N+k}\,\frac{\sin\left(\frac{2\pi\,|a-b|}{N+k}\right)[N+1]\,[N-1]}{[N+1]^2+[N-1]^2 + 2\,(-1)^{|a-b|}\,\cos\left(\frac{2\pi |a-b|}{N+k}\right)[N+1]\,[N-1] }\ .
  \end{aligned}
\end{align}
\begin{figure}[t]
  \begin{center}
    \begin{minipage}{0.45\hsize}
      \includegraphics[width=\linewidth]{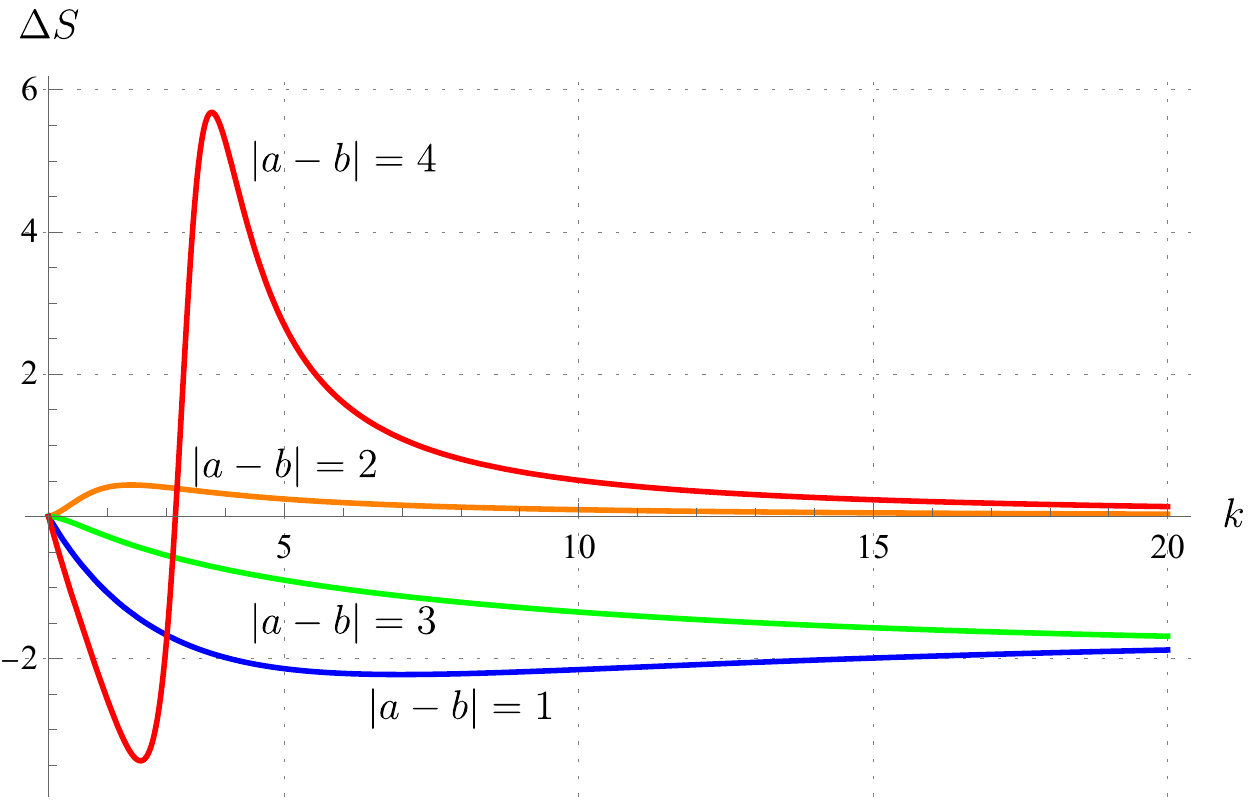}
    \end{minipage}
    \begin{minipage}{0.45\hsize}
      \includegraphics[width=\linewidth]{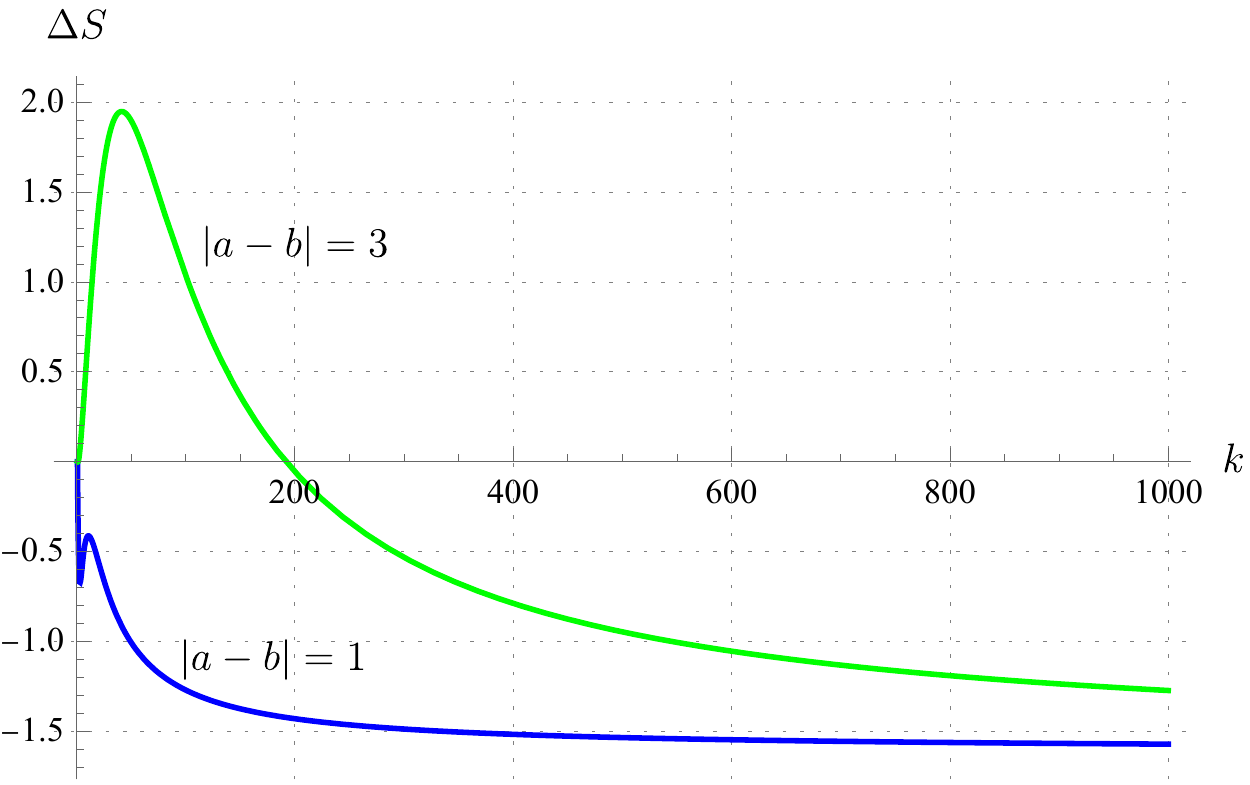}
    \end{minipage}
  \end{center}
  \caption{The difference $\Delta S$ of the pseudo entropy from the averaged entanglement entropy as a function of the levels $k$ when $N=5$. The left panel shows $\Delta S$ of the form \eqref{eq:deltaSgen} by the second prescription (2) of analytic continuation. The blue, orange, green and red curves represent the cases with $|a-b|=1,2,3,4$ respectively. For comparison, the right panel shows $\Delta S$ of the form \eqref{eq:deltaSgen1} by a naive prescription (1) when $|a-b|$ is odd. For even $|a-b|$, $\Delta S$ takes the same values as the left panel.}
      \label{fig:exc4}
\end{figure}

$\Delta S$ depends on $a$ and $b$ only through the difference $|a-b|$ and highly depends on whether $|a-b|$ is even or odd through the sign factor $(-1)^{|a-b|}$.
The left panel of figure \ref{fig:exc4} shows the difference $\Delta S$ for several choices of the level $k$ when $N=5$ (For comparison the right panel shows $\Delta S$ calculated by the previous prescription only for odd $|a-b|$ in the right panel).
The four curves represent the cases of $|a-b|=1,\ldots,4$.
The figure shows that $\Delta S$ can be positive only when $|a-b|$ is even.

\begin{figure}
    \centering
    \begin{tikzpicture}[thick]
        \begin{scope}
            \draw (0, 0) circle (1.5);
            \draw (-1.5, 0) arc (180:360:1.5 and 0.3);
            \draw[dotted] (1.5, 0) arc (0:180:1.5 and 0.3);
            
            \draw[RoyalBlue] (-1, 0) arc (180:360:0.75);
            \draw[RoyalBlue] (-0.5, 0) arc (180:240:0.75);
            \draw[RoyalBlue] (1, 0) arc (0:-100:0.75);
            
            \draw[draw=none, fill=lightgray!50, fill opacity = 0.5] (0, 0) ellipse (1.5 and 0.3);
            
            \draw[RoyalBlue] (1, 0) arc (0:180:1);
            \draw[RoyalBlue] (0.5, 0) arc (0:180:0.5);
            
            \node[circle,draw=BrickRed, fill=BrickRed, inner sep=0pt,minimum size=5pt] at (-1, 0) {};
            \node[circle,draw=BrickRed, fill=BrickRed, inner sep=0pt,minimum size=5pt] at (-0.5, 0) {};
            \node[circle,draw=OliveGreen, fill=OliveGreen, inner sep=0pt,minimum size=5pt] at (1, 0) {};
            \node[circle,draw=OliveGreen, fill=OliveGreen, inner sep=0pt,minimum size=5pt] at (0.5, 0) {};
            
            \node[BrickRed] at (-0.75, 0.2) {$A$};
            \node[OliveGreen] at (0.75, 0.2) {$B$};
            
            \node at (0, 2) {\large $n=$ odd};
            \node at (0, -2) {\large $-\log \CD + \log d_i$};
        \end{scope}
        
        \begin{scope}[xshift=5cm]
            \draw (0, 0) circle (1.5);
            \draw (-1.5, 0) arc (180:360:1.5 and 0.3);
            \draw[dotted] (1.5, 0) arc (0:180:1.5 and 0.3);
            
            \draw[RoyalBlue] (-1, 0) arc (180:360:1);
            \draw[RoyalBlue] (-0.5, 0) arc (180:360:0.5);
            
            \draw[draw=none, fill=lightgray!50, fill opacity = 0.5] (0, 0) ellipse (1.5 and 0.3);
            
            \draw[RoyalBlue] (1, 0) arc (0:180:1);
            \draw[RoyalBlue] (0.5, 0) arc (0:180:0.5);
            
            \node[circle,draw=BrickRed, fill=BrickRed, inner sep=0pt,minimum size=5pt] at (-1, 0) {};
            \node[circle,draw=BrickRed, fill=BrickRed, inner sep=0pt,minimum size=5pt] at (-0.5, 0) {};
            \node[circle,draw=OliveGreen, fill=OliveGreen, inner sep=0pt,minimum size=5pt] at (1, 0) {};
            \node[circle,draw=OliveGreen, fill=OliveGreen, inner sep=0pt,minimum size=5pt] at (0.5, 0) {};
            
            \node[BrickRed] at (-0.75, 0.2) {$A$};
            \node[OliveGreen] at (0.75, 0.2) {$B$};
            
            \node at (0, 2) {\large $n=$ even};
            \node at (0, -2) {\large $-\log \CD + 2\log d_i$};
        \end{scope}
    \end{tikzpicture}
  
    \caption{The partition function for the pseudo entropy with $n$ odd and even  and their values. In the odd case, we can interpret that the two states are related by the entanglement swapping. In the even case, it can be regarded as two copies of entangled pairs.}
    \label{oddevenfigg}
\end{figure}
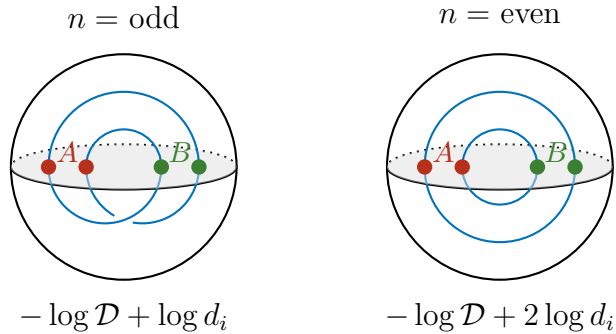

In the classical limit $k\to\infty$, $[x]$ reduces to $x$, so
\begin{align}\label{eq:classicallimit}
  \Delta S\to
  \begin{cases}
    0 & |a-b|:\text{even} \\
    -\log N & |a-b|:\text{odd} 
  \end{cases}
\end{align}
Refer also to appendix B for the SU$(2)$ case.
This can also be seen in figure \ref{fig:exc4}.
We can interpret this behavior as follows. Whether $a$ is even or odd determines the pairs of the excitations connected by Wilson lines in $\ket{\psi_a}$ (see figures in \eqref{eq:psizero}-\eqref{eq:psineg}).
Therefore if $|a-b|$ is even, the pairs of excitations connected in $\ket{\psi_a}$ and those in $\ket{\psi_b}$ are same, but those are different if $|a-b|$ is odd.
\eqref{eq:classicallimit} shows that the links of Wilson lines do not contribute to the pseudo entropy in the classical limit.
When $|a-b|$ is even, $\Delta S$ goes to zero because we can regard $\ket{\psi_a}$ and $\ket{\psi_b}$ as the same states in the classical limit.
When $|a-b|$ is odd, $\Delta S$ has a contribution from the difference of the pairs of excitations.
In \cite{Nakata:2021ubr}, it was shown in multi-qubit systems that if $\ket{\psi}$ and $\ket{\varphi}$ are related by an entanglement swapping, then $\Delta S<0$.
Moreover in the case of odd $|a-b|$, the result \eqref{eq:classicallimit} can be understood as a consequence of entanglement swapping (see figure \ref{oddevenfigg}).

Furthermore, it is also important that $\Delta S$ is non-positive in the classical limit.
We can see that $\Delta S$ can be positive (for $a-b$ even) only when the quantum effect from the links of Wilson loops give a huge contribution to the pseudo entropy.
This is also consistent with the results in the transverse Ising model \cite{Mollabashi:2020yie} and the XY model \cite{Mollabashi:2021xsd}.
In such situations, $\Delta S$ plays a role of the order parameter diagnosing whether the two states $\ket{\psi}$ and $\ket{\varphi}$, used in the definition of the transition matrix, are in the same phase or not.
The transverse Ising model, for example, has the paramagnetic and ferromagnetic phase, which are called quantum phases because those phases are emergent only in quantum systems. Therefore, we may conclude that $\Delta S$ captures the quantum-theoretic difference between the two states $\ket{\psi}$ and $\ket{\varphi}$. 

\subsection{Geometrical interpretation}\label{ss:geometric_interpretation}

Motivated by the geometric formula of holographic entanglement entropy \cite{Ryu:2006bv,Ryu:2006ef,Hubeny:2007xt,Rangamani:2016dms}, 
we explore a possible geometric interpretation of 
topological pseudo entropy in the Chern-Simons gauge theory.
Consider Wilson loops on $\BS^3$ and divide the sphere into two hemispheres.
The surface of each hemisphere is $\BS^2$ and we separate $\BS^2$ into two regions $A$ and $B$ 
along a curve $\Gamma(=\de A=\de B)$.

When there are no Wilson loops, it is clear that the topological entanglement entropy 
is simply given by 
\begin{align}
    S\left(\rho_A\right)
        =-n({\Gamma})\, \log \CD\ ,
\end{align}
where $n({\Gamma})$ is the number of connected components of $\Gamma$.

If $\Gamma$ is connected, i.e., $n(\Gamma)=1$ and the $\Gamma$ intersects with only one Wilson line in the fundamental representation (see the left of figure \ref{interfig}), 
it is easy to evaluate the topological pseudo entropy:
\begin{align}
    S\left(\tau_A\right)
        =\log  d_j-\log \CD\ .
\end{align}
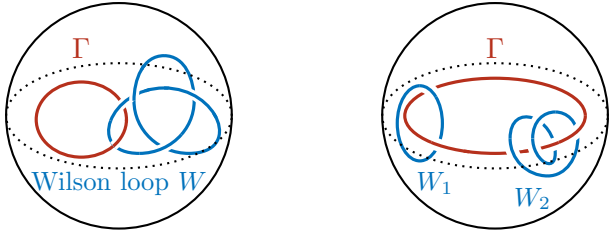
\begin{figure}
  \centering
      \begin{tikzpicture}[thick]
        \begin{scope}
            \draw (0, 0) circle (1.5);
            \begin{knot}[
                consider self intersections,
                end tolerance=.05pt,
                clip width=3,
                flip crossing=1,
                flip crossing=4
                ]
                \strand [very thick, BrickRed] (-0.5, -0.05) ellipse (0.6cm and 0.5cm);
                \strand[very thick, RoyalBlue]
                    (0.6, 0.8) to[out=180,in=90]
                    (0.2, 0.2) to[out=-90,in=180]
                    (0.93, -0.5) to[out=0,in=0,looseness=2]
                    (0.6, 0.37) to[out=180,in=180,looseness=2]
                    (0.27, -0.5) to[out=0,in=-90]
                    (1, 0.2) to[out=90,in=0] (0.6, 0.8);
            \end{knot}
            
            \node[BrickRed] at (-0.5, 0.9) {$\Gamma$};
            \node[RoyalBlue] at (0, -0.9) {\small Wilson loop $W$};
            
            \draw[dotted] (0, 0) ellipse (1.5 and 0.7);
        \end{scope}
        
        \begin{scope}[xshift=5cm]
            \draw (0, 0) circle (1.5);
            
            \begin{knot}[
                consider self intersections,
                end tolerance=.05pt,
                clip width=4,
                flip crossing=2,
                flip crossing=3,
                flip crossing=5,
                flip crossing=7
                ]
                \strand [very thick, RoyalBlue] (-1,-0.1) ellipse (0.3cm and 0.5cm);
                \strand [very thick, BrickRed] (0,0) ellipse (1.2cm and 0.5cm);

                \strand [very thick, RoyalBlue] (0.7,-0.8) to[out=180,in=270] (0.2, -0.3) to[out=90, in=90, looseness=2] (0.8, -0.5) to[out=270, in=-90, looseness=2] (0.5, -0.3) to[out=90, in=90, looseness=2] (1.1, -0.3) to[out=270, in=0, looseness=1] (0.7,-0.8); 
            \end{knot}
            
            \node[BrickRed] at (0, 0.9) {$\Gamma$};
            \node[RoyalBlue] at (-0.8, -0.9) {\small $W_1$};
            \node[RoyalBlue] at (0.5, -1.1) {\small $W_2$};
            
            \draw[dotted] (0, 0) ellipse (1.5 and 0.7);
        \end{scope}

    \end{tikzpicture}
  
  \caption{The intersections between the entangling surface $\Gamma$ and the Wilson loops. The left panel describes the setup with $n(\Gamma\cup W)=1$.  In the right panel, we count it as $n(\Gamma\cup W)=2$.}
\label{interfig}
\end{figure}

However, it is not straightforward to find a simple formula in more general cases.
Thus, we focus on the semi-classical limit $k\to\infty$.
In this limit, if $\Gamma$ is connected, we can find the following simple result:
\begin{align}
    S\left(\tau_A\right)
        =\sum_{j}\,n_i({\Gamma}\cap W)\, \log  d_j-\log \CD\ .
\end{align}
We defined $n_j({\Gamma}\cap W)$ to be the number of the Wilson loops $W$ in the representation $R_j$ which wrap on $\Gamma$, as illustrated in figure \ref{interfig}.
We may regard $n({\Gamma}\cap W)$ as the number of entangled pairs given by the Wilson lines. This is qualitatively similar to the holographic entanglement entropy, where the entanglement entropy is proportional to the area of codimension-two surface like $\Gamma$.
The holographic entanglement entropy suggests a heuristic picture of emergent spacetime from quantum entanglement in that a Bell pair per Planck unit area is expected to be penetrated on the codimension-two surface.
Indeed in our topological entropy, the Wilson loop is linked with $\Gamma$, which gives the contribution proportional to $\log d_j$.
On the other hand, the term proportional to $-\log \CD$ is analogous to the gravity edge mode contribution.

\subsection{Topological pseudo entropy on \texorpdfstring{$\BT^2$}{T2} with Wilson loops}\label{subsec:torus}
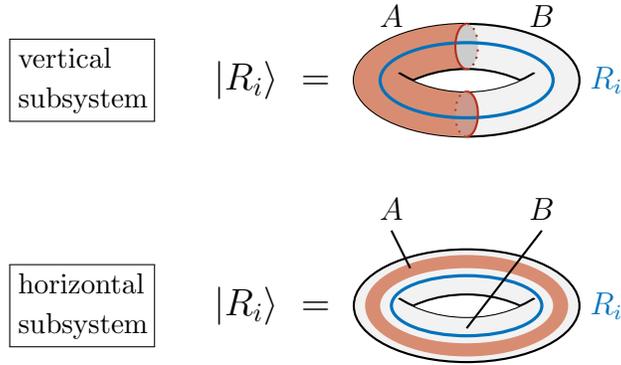
\begin{figure}[h]
  \centering
  \begin{tikzpicture}[thick]
      \begin{scope}
          \begin{scope}[scale=0.5]
            \draw[fill=lightgray!20!white] (0,0) ellipse (3 and 1.5);
            \draw[draw=none, fill=BrickRed!50] (0, 1.5) arc (90:270:3 and 1.5);
            \begin{scope}
            \clip (0,-1.8) ellipse (3 and 2.5);
            \draw (0,2.2) ellipse (3 and 2.5);
            \end{scope}
            \begin{scope}
            \clip (0,2.2) ellipse (3 and 2.5);
            \draw[fill=white] (0,-2.2) ellipse (3 and 2.5);
            \end{scope}
            \draw[draw=none, fill=lightgray!80] (0, 0.9) ellipse (0.3 and 0.6);
            \draw[draw=none, fill=lightgray!80] (0, -0.9) ellipse (0.3 and 0.6);
            \draw[BrickRed, dotted] (0, -0.3) arc (90:270:0.3 and 0.6);
            \draw[BrickRed, dotted] (0, 0.3) arc (-90:90:0.3 and 0.6);
            \draw[draw=none, fill=BrickRed!70, opacity=0.5] (0, -0.9) ellipse (0.3 and 0.6);
            \draw[very thick, RoyalBlue] (0, 0) ellipse (2.3 and 1);
            \draw[BrickRed] (0, 1.5) arc (90:270:0.3 and 0.6);
            \draw[BrickRed] (0, -1.5) arc (-90:90:0.3 and 0.6);

            \node at (-2, 1.7) {\large $A$};
            \node at (2, 1.7) {\large $B$};
            \node[RoyalBlue, right] at (3, 0) {\large $R_i$};
            
            \node[left] at (-8, 0) {\fbox{\parbox{1.7cm}{vertical\\ subsystem}}};
            \node[right] at (-7, 0) {\Large $|R_i\rangle~=$};
        \end{scope}
      \end{scope}
        \begin{scope}[yshift=-3cm]
          \begin{scope}[scale=0.5]
            \draw[fill=lightgray!20!white] (0,0) ellipse (3 and 1.5);
            
            \draw[draw=none, fill=BrickRed!50] (0,0) ellipse (2.7 and 1.35);
            \draw[draw=none, fill=lightgray!20!white] (0,0) ellipse (2.3 and 1);
            \begin{scope}
                \clip (0,-1.8) ellipse (3 and 2.5);
                \draw (0,2.2) ellipse (3 and 2.5);
            \end{scope}
            \begin{scope}
                \clip (0,2.2) ellipse (3 and 2.5);
                \draw[fill=white] (0,-2.2) ellipse (3 and 2.5);
            \end{scope}           
            
            \draw[very thick, RoyalBlue] (0, 0) ellipse (2 and 0.8);

            \draw[-] (-2, 2) node[above] {\large $A$} -- (-1.5, 1);
            \draw[-] (2, 2) node[above] {\large $B$} -- (0, -0.6);
            \node[RoyalBlue, right] at (3, 0) {\large $R_i$};
            
            \node[left] at (-8, 0) {\fbox{\parbox{1.7cm}{horizontal\\ subsystem}}};
            \node[right] at (-7, 0) {\Large $|R_i\rangle~=$};
        \end{scope}
      \end{scope}
  \end{tikzpicture}
  \caption{Topological entanglement entropy of states on a solid torus with a Wilson loop insertion. There are two ways to choose the subsystem $A$ on the surface; the vertical subsystem [Above] and horizontal subsystem [Below].}
\label{torusfig}
\end{figure}

We move onto the case where the subsystem $A$ is a cylinder on a torus $\BT^2$ and 
where there is a Wilson loop in the interior of $\BT^2$ winding handle.
This is depicted as the vertical subsystem in the upper figure \ref{torusfig}.

By the same calculations as \cite{Dong:2008ft}, we have 
\begin{align}
    \begin{aligned}
        \Tr_A\left[\left(\Tr_B\left[\,\ket{R_i}\bra{R_j}\,\right]\right)^n\right]
            &=
            Z\left[\BS^3;R_i\right]^{2(1-n)}\,\delta_{ij}\\
            &=
            \left(\Smat{0}{i}\right)^{2(1-n)}\,\delta_{ij}\ ,
    \end{aligned}
\end{align}
where $\ket{R_j}$ is a state including a Wilson loop in the representation denoted by $j$.
For general unnormalized states 
\begin{align}
  \ket{\psi}=\sum_i\psi_i\ket{R_i}\ ,\qquad\ket{\varphi}=\sum_i\varphi_i\ket{R_i}\ ,
\end{align}
the trace of the $n^{\text{th}}$ power of the reduced transition matrix is
\begin{align}
    \begin{aligned}
    \Tr_A\left[\left(\Tr_B\left[\,\ket{\psi}\bra{\varphi}\,\right]\right)^n\right]
        &=
        \sum_i\varphi_i^*\psi_i\cdots\varphi_i^*\psi_i\,\Tr_A\left[\left(\Tr_B\left[\,\ket{R_i}\bra{R_i}\,\right]\right)^n\right] \\
        &=
        \sum_i(\varphi_i^*\psi_i)^n\left(\Smat{0}{i}\right)^{2(1-n)}\ .
    \end{aligned}
\end{align}
Thus 
\begin{align}
  \Tr_A\left[\left(\tau_A^{\psi|\varphi}\right)^n\right]
    =
        \frac{\sum_i(\varphi_i^*\psi_i)^n\left(\Smat{0}{i}\right)^{2(1-n)}}{\left(\sum_i\varphi_i^*\psi_i\right)^n}\ .
\end{align}

This leads to the pseudo entropy given by
\begin{align}\label{topeotorus}
    S\left(\tau_A^{\psi|\varphi}\right)
        =
        \log \left[\sum_{i}\vp^*_i\psi_i\right]
        -\sum_{i}\vp^*_i\psi_i \log\frac{\vp^*_i\psi_i}{\left(\Smat{0}{i}\right)^2} \ .
\end{align}
When we consider the topological entanglement entropy for the Wilson line $R_i$ we have
\begin{align}
    \begin{aligned}
        S\left(\rho_A\right)
            &=
            2\log \Smat{0}{i}\\
            &=
            -2\log \CD + 2\log  d_i\ .
    \end{aligned}
\end{align}

The difference $\Delta S$ of the pseudo entropy from the average of entanglement entropy is calculated by 
\begin{align}
    \begin{aligned}
        \Delta S
            &=
            \log\frac{|\braket{\varphi|\psi}|^2}{\sqrt{\braket{\psi|\psi}\braket{\varphi|\varphi}}}
            - \frac{1}{\sum_k \varphi_k^*\psi_k} \sum_i\varphi_i^*\psi_i\log\frac{\varphi^*_i\psi_i}{\left(\Smat{0}{i}\right)^2} \\
            &
            \qquad + \frac{1}{2}\left[ \frac{1}{\sum_k|\psi_k|^2}\sum_i|\psi_i|^2\log\frac{|\psi_i|^2}{\left(\Smat{0}{i}\right)^2} + \frac{1}{\sum_k|\varphi_k|^2}\sum_i|\varphi_i|^2\log\frac{|\varphi_i|^2}{\left(\Smat{0}{i}\right)^2}\right] \ . 
    \end{aligned}
\end{align}

On the other hand, if we consider the horizontal subsystem in figure \ref{torusfig}, the topological entanglement entropy is found as follows. 
First, the replica method gives
\begin{align}
    \begin{aligned}
        \Tr_A\left[\left(\rho_A\right)^n\right]
            &=
            \frac{Z_{2n}}{(Z_2)^n}\\
            &=
            \sum_{j_1,\ldots,j_{2n-3}}{N_{\bar{i}\bar{i}}}^{j_1}{N_{ij_1}}^{j_2}
            {N_{\bar{i}j_2}}^{j_3}\ddd {N_{\bar{i}j_{2n-4}}}^{j_{2n-3}} {N_{ij_{2n-3}}}^{\bar{i}} \\
            &=\sum_j\frac{\left|\Smat{i}{j}\right|^{2n}}{\left|\Smat{0}{j}\right|^{2n-2}}\ ,
    \end{aligned}
\end{align}
where $Z_{2n}$ is the partition function on $\BS^1\times \BS^2$ with $n$ Wilson lines $R_i$ and $n$ Wilson lines $R^*_i$ winding around $\BS^1$.
Finally, we find the entanglement entropy
\begin{align}\label{topint}
    S_A
        =
        -\sum_{j} \left|\Smat{i}{j}\right|^2\log \frac{\left|\Smat{i}{j}\right|^2}{\left|\Smat{0}{j}\right|^2}\ .
\end{align}
We will see later that this coincides with a finite term of the topological interface entropy in \eqref{IE_diag_RCFT}.
We can also get \eqref{topint} by setting $\psi_j=\vp_j=\Smat{i}{j}$.

\subsection{Possible definition of boundary states in Chern-Simons theory}\label{subsec:boundary}

Consider a path integral on a three-dimensional hemisphere or a ball in Chern-Simons theory.
We divide its boundary given by $\BS^2$ into $A$ and $B$, such that they are two dimensional 
hemispheres. 

\begin{figure}
  \centering
  \begin{tikzpicture}[thick]
        \begin{scope}
            \draw[fill=lightgray!20!white] (0, 0) circle (1.5);
            \draw[OrangeRed] (0, 1.5) arc (90:270:0.3 and 1.5);
            \draw[dotted, OrangeRed] (0, -1.5) arc (-90:90:0.3 and 1.5);
            
            \draw[RoyalBlue] (-1, 0.5) arc (180:270:1);
            \draw[draw=none, fill=lightgray, fill opacity = 0.5] (0, 0) ellipse (0.3 and 1.5);
            \draw[RoyalBlue] (1, 0.5) arc (0:-90:1);
            
            \node[circle,draw=RoyalBlue, fill=RoyalBlue, inner sep=0pt,minimum size=5pt] at (-1, 0.5) {};
            \node[circle,draw=RoyalBlue, fill=RoyalBlue, inner sep=0pt,minimum size=5pt] at (1, 0.5) {};
            
            \node[RoyalBlue] at (0, 0) {\large $R_i$};
            \node at (-1, -0.5) {$A$};
            \node at (1, -0.5) {$B$};
            
            \node at (-2, 2.5) {\fbox{\large Ishibashi-like state}};
            \node at (-2.5, 0) {\Large $|I_i\rrangle~=$};
        \end{scope}
        
        \begin{scope}[xshift=7cm]
            \begin{scope}
                \draw[fill=Orchid!60, fill opacity=0.5] (0, 0.8) arc (90:270:0.8);
                \draw[draw=none, fill=white] (0, 0) ellipse (0.16 and 0.8);
            \end{scope}
            
            \draw[draw=none, fill=lightgray, fill opacity = 0.5] (0, 0) ellipse (0.3 and 1.5);
            
            \draw[dotted, OrangeRed] (0, -1.5) arc (-90:90:0.3 and 1.5);
            
            \begin{scope}
                \draw[fill=Orchid!60, fill opacity=0.5] (0, -0.8) arc (-90:90:0.8);
                \draw[draw=none, fill=white] (0, 0) ellipse (0.16 and 0.8);
            \end{scope}

            \draw[fill=lightgray!50, fill opacity=0.5] (0, 0) circle (1.5);
            
            \draw[OrangeRed] (0, 1.5) arc (90:270:0.3 and 1.5);
            
            \draw[draw=none, fill=Orchid!60, fill opacity=0.2] (0, 0) ellipse (0.16 and 0.8);
            
            \draw[OrangeRed] (0, 0.8) arc (90:270:0.16 and 0.8);
            
            \draw[dotted, draw=OrangeRed
            ] (0, -0.8) arc (-90:90:0.16 and 0.8);
            
            \node at (-1, -0.5) {$A$};
            \node at (1, -0.5) {$B$};
            
            \draw (0.5, 0) --+ (1, 1) node[above right] {\parbox{5cm}{Inner boundary\\ labeled by $a$}};
            
            \node at (-2.4, 2.5) {\fbox{\large Cardy-like state}};
            \node at (-2.5, 0) {\Large $|B_a\rangle~=$};
        \end{scope}
  \end{tikzpicture}
  \caption{Analogues of boundary states in Chern-Simons theory. The Ishibashi-like state is defined as a state on the surface ($\BS^2$) of a ball where a pair of excitations is located across the common boundary of the two regions $A$ and $B$ [Left]. On the other hand the Cardy-like state is defined as a state on the surface of a ball with the inner boundary surface with a specific boundary condition corresponding to \eqref{Cardy_Ishibashi_rel} [Right].}
\label{boundarystatefig}
\end{figure}
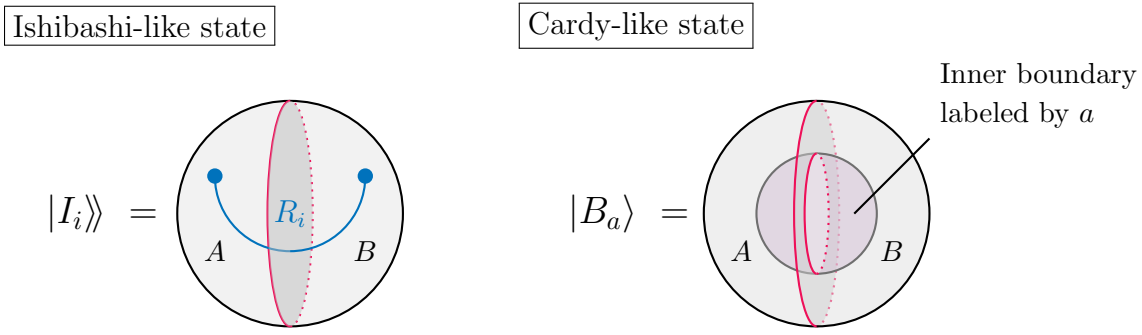

Now we can define the Ishibashi-type state $|I_i\rrangle$ as 
the path integral on the three-dimensional hemisphere $\BB^3$ with an open Wilson line with the representation 
$R_i$ such that one of its end points is on $A$ and the other is on $B$ (see figure \ref{boundarystatefig}).\footnote{The Ishibashi-like state $|I_i\rrangle$ we define in 3d Chern-Simons theory is different from the Ishibashi state $|i\rrangle$ in 2d BCFT used in section \ref{sec:LRPE}.} 
Obviously, they satisfy the same relation as the Ishibashi-type states in boundary CFT$_2$:
\begin{align}\label{CS_Ishibashi}
    \llangle I_i| I_j\rrangle =\delta_{ij}\,\Smat{0}{i}\ .
\end{align} 
It is also straightforward to calculate the entanglement entropy $S_A$ of $|I_i\rrangle$ via the replica trick
and this leads to
\begin{align}
    S_A=\log \Smat{0}{i}= -\log \CD + \log d_i \ .
\end{align}

If we consider the linear combination state
\begin{align}
    \begin{aligned}
    |\psi\lb
        &=\sum_{i}\psi_i\, |I_i\rrangle\ ,\\
    |\vp\lb
        &=\sum_{i}\vp_i\, |I_i\rrangle\ ,
    \end{aligned}
\end{align}
then the transition matrix looks like
\begin{align}
    \tau_A^{\psi|\varphi}
        =
        \frac{\mbox{Tr}_B\left[\,|\psi\lb\la \vp|\,\right]}{\la \vp|\phi\lb}
        =
        \frac{\sum_i \vp^*_i\psi_i\, \mbox{Tr}_B\left[\,|I_i\rrangle\llangle I_i|\,\right]}{\sum_i \vp^*_i\psi_i\, \Smat{i}{0}}\ .
\end{align}
We can calculate the pseudo entropy
\begin{align}
    \Tr_A\left[\left(\tau_A^{\psi|\varphi}\right)^n\right]
        =
        \frac{\sum_i (\vp^*_i\psi_i)^n\, \Smat{i}{0}}{\left(\sum_{i} \left(\vp^*_i\psi_i\right)^n\, \Smat{i}{0}\right)^n}\ ,
\end{align}
leading to the expression
\begin{align}\label{topeebdy}
    S\left(\tm{\psi}{\varphi}_A\right)
        =
        \log \left[\sum_{i}\vp^*_i\psi_i\, \Smat{i}{0}\right]
        -
        \frac{\sum_i \vp^*_i\psi_i\, \Smat{i}{0}\,\log(\vp^*_i\psi_i)}{\sum_i \vp^*_i \psi_i\, \Smat{i}{0}}\ .
\end{align}

Next, we introduce the Cardy-type state by\footnote{While the Cardy-like state $\ket{B_a}$ satisfies the same relation \eqref{Cardy_Ishibashi_relation_BCFT} as the Cardy state $\ket{a}$ in 2d BCFT  they are different states.} (refer to figure \ref{boundarystatefig})
\begin{align}\label{Cardy_Ishibashi_rel}
    |B_a \rangle
        =
        \sum_{i}\frac{\Smat{a}{i}}{\s{\Smat{0}{i}}}\,|I_i\rrangle\ .
\end{align}
It is easy to show that the Cardy-type states are orthogonal to each other 
\begin{align}
    \langle B_a| B_b\rangle =\delta_{ab}\ .
\end{align}
Even though in physical two-dimensional CFTs, the Cardy-type state satisfied the open-closed duality,
the above result corresponds to the truncation to the lowest energy mode of open string. 
If we calculate the topological entanglement entropy for $|B_a\rangle$ we get from \eqref{topeebdy}
\begin{align}\label{CS_Cardy_S2}
    S_A
        =
        -\sum_{i} \left(\Smat{a}{i}\right)^2\,\log \frac{\left(\Smat{a}{i}\right)^2}{\Smat{0}{i}}\ .
\end{align}
This is the same as the finite part of the left-right entanglement entropy \eqref{LREE_Cardy} of the Cardy state characterized by the boundary condition $a$ in 2d boundary CFT examined in section \ref{sec:LRPE}.

We can also evaluate the partition function on the hemisphere $\BB^3$ with the boundary condition of 
$|B_a\rangle$:
\begin{align}
    Z\left[\BB^3; B_a\right]
        =
        \llangle I_0|B_a \rangle
        =
        \Smat{a}{0}\sqrt{\Smat{0}{0}}\ .
\end{align}
We define the $g$-function by
\begin{align}
    g_a=\llangle \tilde I_0| B_a \rangle=\Smat{a}{0}\ ,
\end{align}
where $|\tilde I_0\rrangle =\frac{1}{\s{\Smat{0}{0}}}\,|I_0\rrangle $ is the normalized vacuum state.
Then we have
\begin{align}\label{eq:ballba}
    Z\left[\BB^3; B_a\right]
        =\sqrt{\Smat{0}{0}}\ g_a\ .
\end{align}
Note that the partition functions we have obtained above satisfy 
\begin{align}
    Z\left[\BS^3\right]=\sum_a Z\left[\BB^3; B_a\right]^*Z\left[\BB^3; B_a\right]\ ,
\end{align}
which can be regarded as a completeness relation for $\ket{B_a}$.

Finally, we calculate the entanglement entropy of the vacuum state on a disk $\BD^2$ with the Cardy-type boundary condition.
By \eqref{eq:ballba}, the resulting entanglement entropy takes the form
\begin{align}\label{eq:completeness}
    \begin{aligned}
        \log Z\left[\BB^3; B_a\right]=\frac{1}{2}\log Z\left[\BS^3\right]+\log g_a\ .
    \end{aligned}
\end{align}
The second term is analogous to the boundary entropy in BCFT. 
Indeed, the form
\begin{align}\label{boundary_entropy_PF}
    \log Z\left[\BB^d; B_a\right]-\frac{1}{2}\log Z\left[\BS^d\right]
\end{align}
is proposed to be a candidate for a $C$-function in 3d \cite{Nozaki:2012qd} and 4d \cite{Gaiotto:2014gha}, and it was shown in $d$-dimensional BCFT that the boundary entropy is defined by $S_{\text{bdy}}=S^{\text{(BCFT)}}-S^{\text{(CFT)}}/2$ equals to \eqref{boundary_entropy_PF} up to a UV divergence \cite{Kobayashi:2018lil}.

\section{Pseudo entropy in CFT}
\label{sec:CHM}

We switch gears and move to examining the pseudo entropy for a simple choice of the entangling region in CFT.
In section \ref{ss:CHM} we review the Casini-Huerta-Myers (CHM) map for a spherical entangling surface on $\BR^d$, and describe the pseudo entropy as the path integral on $\BS^1\times \BH^{d-1}$.
In section \ref{ss:CS_revisit} we illustrate the application of the CHM map by showing the calculation of the pseudo entropy in the three-dimensional Chern-Simons theory, reproducing the results in section \ref{subsec:excitation} from a slightly different viewpoint.
We then expand on the relation between the pseudo entropy and interface entropy in CFT$_2$, which allows us to read off the pseudo entropies of non-topological theories from their interface entropies in section \ref{ss:Relation_PE_IE}.

\subsection{Conformal map}\label{ss:CHM}
We begin with reviewing the CHM map \cite{Casini:2011kv} that equates the entanglement entropy across a sphere in CFT$_d$ to the calculation of the partition function on $\BS^d$ or $\BS^1\times \BH^{d-1}$.

\begin{figure}[ht]
    \centering
    \begin{tikzpicture}[thick]
        \draw[fill=lightgray!10] (0,0) -- (4, 0) -- (3, -2) -- (-1, -2) -- cycle;
        \draw[-{Stealth[length=3mm, width=2mm]}] (0,0) node[left] {$0$} -- (0, 2) node [left] {\large $t$};
        \draw[fill=YellowOrange] (1.5, -1) node {\large $A$} ellipse (1.2cm and 0.6cm);
        \draw[-stealth] (2, -1) -- (3, -1) node[right] {$r$};
        \draw[-stealth] (0.7,-1.2) arc
            [start angle=0,
             end angle=270,
             x radius=0.3cm,
             y radius=0.3cm
            ] node [below] {$\tau$};
        \node at (3.4, -0.3) {\large $B$};
        \node at (1.5, 1) {\large $|\varphi\rangle$};
        \node at (1.5, -3) {\large $|\psi\rangle$};
    \end{tikzpicture}
    \caption{The spherical entangling surface $\Sigma=\partial A$ at a time slice ($t=0$) in flat space $\BR^{d}$.}
    \label{fig:PE_setup}
\end{figure}
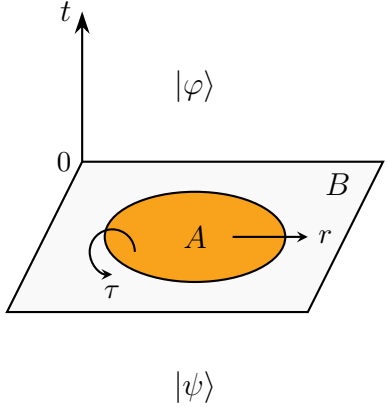

In what follows, we bipartite a constant time slice into two regions $A$ and its complement $B$ in flat space $\BR^d$ with the metric:
\begin{align}
    \d s^2_\text{Flat} = \d t^2 + \d r^2 + r^2\,\d \Omega_{d-2}^2 \ ,
\end{align}
and let the entangling surface $\Gamma= \partial A$ be spherical (see figure \ref{fig:PE_setup}):
\begin{align}
    \Gamma = \{ t=0,\, r = R \} \ .
\end{align}

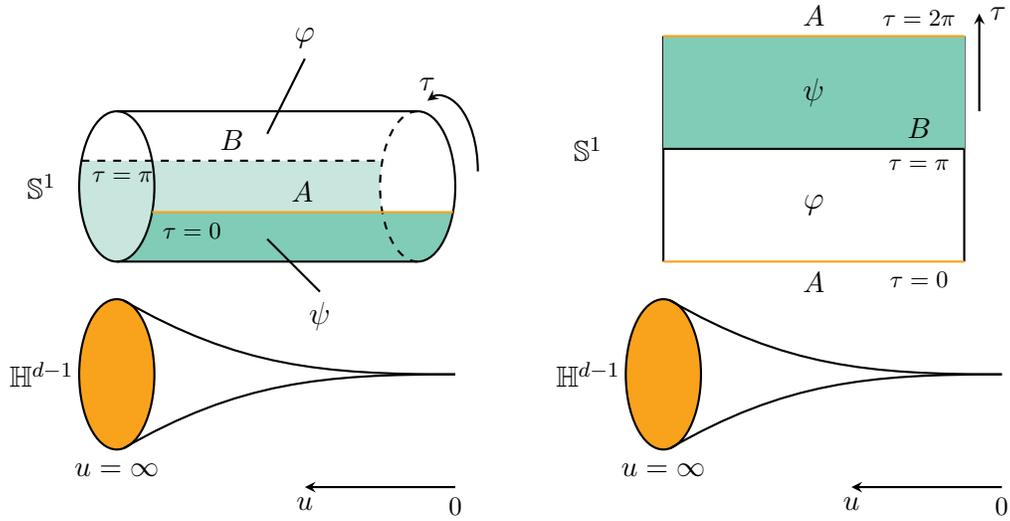
\begin{figure}[t]
    \centering
    \begin{tikzpicture}[thick]
        \draw[draw=none, fill=SeaGreen!30] ($(4,0) + (160:0.5 and 1)$) arc (160:270:0.5 and 1) -- (0,-1) -- (4,-1) -- ($(0, 0) + (270:0.5 and 1)$) arc (270:160:0.5 and 1) -- cycle;
        \draw[draw=none, fill=SeaGreen!70] ($(4,0) + (-20:0.5 and 1)$) arc (-20:-90:0.5 and 1) -- (0,-1) -- (4,-1) -- ($(0, 0) + (270:0.5 and 1)$) arc (-90:-20:0.5 and 1) -- cycle;
        \draw (0.5,0) arc[
        start angle=0,
        end angle=360,
        x radius=0.5,
        y radius =1] ;
        \draw[dashed] (4,1) arc (90:270:0.5 and 1);
        \draw (4,1) arc (90:270:-0.5 and 1);
        \draw (0,1) -- (4,1);
        \draw (0,-1) -- (4,-1);
        \draw[-stealth] (4.8,0.2) arc[
        start angle=0,
        end angle=110,
        x radius=0.5,
        y radius =1] node[above] {$\tau$};
        \draw[YellowOrange] (-20:0.5 and 1) node[below right, black] {\footnotesize $\tau=0$} -- node[black, above] {$A$} +  (4,0) ;
        \draw[dashed] (160:0.5 and 1) node[below right] {\footnotesize $\tau=\pi$} -- node[black, above] {$B$} +  (4,0);
        \node at (-1,0) {$\BS^1$};
        \draw (2, 0.7) --+ (0.5,1) node[above] {\large $\varphi$};
        \draw (2, -0.7) --+ (0.7,-0.7) node[below] {\large $\psi$};
        
        \draw[fill=YellowOrange] (0.5,-2.5) arc[
        start angle=0,
        end angle=360,
        x radius=0.5,
        y radius =1] ;
        \draw (0,-2.5)+(80:0.5 and 1) to [out=-30, in=180] (4.5,-2.5);
        \draw (0,-2.5)+(-80:0.5 and 1) to [out=30, in=180] (4.5,-2.5);
        \draw[-stealth] (4.5,-4) node[below] {\small $0$} --+ (-2,0) node[below] {$u$};
        \node at (-1,-2.5) {$\BH^{d-1}$};
        \node[above] at (0, -4) {$u=\infty$};
    \end{tikzpicture}
    \qquad
    \begin{tikzpicture}[thick]
        \draw (0, 2) -- (0, -1);
        \draw (4, 2) -- (4, -1);
        \draw[draw=none, fill=SeaGreen!70] (0, 0.5) -- (4, 0.5) -- (4, 2) -- (0, 2) --cycle;
        \draw[YellowOrange] (0,2) -- node[black, above] {$A$} (4, 2);
        \draw[YellowOrange] (0, -1) --  node[black, below] {$A$}(4,-1);
        \draw (0, 0.5) -- (4, 0.5);

        \draw[-stealth] (4.2, 1) --+ (0, 1.3) node[right] {$\tau$};
        \node[below] at (3.4, -1) {\footnotesize $\tau = 0$};
        \node[below] at (3.4, 0.5) {\footnotesize $\tau = \pi$};
        \node[above] at (3.4, 2) {\footnotesize $\tau = 2\pi$};
        \node[above] at (3.4, 0.5){$B$};
        \node at (2, 1.25) {\large $\psi$};
        \node at (2, -0.25) {\large $\varphi$};
        \node at (-1, 0.5) {$\BS^1$};
        
        \draw[fill=YellowOrange] (0.5,-2.5) arc[
        start angle=0,
        end angle=360,
        x radius=0.5,
        y radius =1] ;
        \draw (0,-2.5)+(80:0.5 and 1) to [out=-30, in=180] (4.5,-2.5);
        \draw (0,-2.5)+(-80:0.5 and 1) to [out=30, in=180] (4.5,-2.5);
        \draw[-stealth] (4.5,-4) node[below] {\small $0$} --+ (-2,0) node[below] {$u$};
        \node at (-1,-2.5) {$\BH^{d-1}$};
        \node[above] at (0, -4) {$u=\infty$};
    \end{tikzpicture}
    \caption{The Euclidean configuration for the pseudo entropy across the spherical entangling surface after the CHM map [Left] and the path integral representation of the transition matrix $\tau^{\psi|\varphi}_A$ [Right].}
    \label{fig:CHM_map}
\end{figure}

The flat space is conformally equivalent to $\BS^1 \times \BH^{d-1}$ with the metric 
\begin{align}
    \d s_\text{Hyp}^2 = \d \tau^2 + \d u^2 + \sinh^2 u\,\d \Omega_{d-2}^2 \ , \qquad (0\le \tau < 2\pi, \, 0\le u < \infty) \ ,
\end{align}
by the CHM map \cite{Casini:2011kv}
\begin{align}
    t = R\, \frac{\sin\tau}{\cosh u + \cos \tau} \ , \qquad r = R\, \frac{\sinh u}{\cosh u + \cos \tau} \ .
\end{align}
Indeed, the two spaces are related by
\begin{align}
    \d s^2_\text{Flat} = \Omega_\text{Hyp}^2\,\d s^2_\text{Hyp} \ , \qquad \Omega_\text{Hyp} = \frac{R}{\cosh u + \cos \tau} \ .
\end{align}
In the latter geometry, the replica geometry can be given simply by scaling the periodicity of $\tau$ by $n$.
The entangling region $A$ (and its complement $B$) is mapped to the time slice at $\tau = 0$ (and at $\tau = \pi$) and $\Gamma$ is pushed to the infinity of the hyperbolic space (see figure \ref{fig:CHM_map}):
\begin{align}
    \Gamma = \{ \tau = 0, \, u = \infty\} \ .
\end{align}

Now let us turn to the pseudo entropy between two states $|\psi\rangle$ and $|\varphi\rangle$.
To prepare the transition matrix $\tau_A^{\psi|\varphi}$ we use the Euclidean path integral where the ket state $|\psi\rangle$ is represented as a path integral from $t=-\infty$ to $t=0$ while the bra state $\langle \varphi|$ represented as a path integral from $t=\infty$ to $t=0$ in the flat space.
After the CHM map, each state covers half of the cylinder as in figure \ref{fig:CHM_map}:
\begin{align}
        \langle \varphi|  :~ \tau \in [0, \pi] \ , \qquad
        |\psi\rangle :~  \tau \in [\pi, 2\pi] \ .
\end{align}
With this in mind the $n^{\text{th}}$ pseudo R\'enyi entropy defined by \eqref{Pseudo_Renyi} is calculable from the path integral representation of the replica partition function:
\begin{align}
    \Tr_A\left[ \left(\tau_A^{\psi|\varphi}\right)^n\right] \equiv \frac{Z(n)}{(Z(1))^n}\ , \qquad Z(1) \equiv \langle \varphi | \psi\rangle  \ .
\end{align}

\subsection{Chern-Simons calculation revisited}\label{ss:CS_revisit}

\begin{figure}[t]
    \centering
    
    \begin{tikzpicture}[thick]
        \begin{scope}
            \draw (0, 0) -- (4, 0) -- (5, 2) -- (1, 2) -- cycle;
            \draw[fill=YellowOrange] (2.5, 1) node[above] {\large $A$} ellipse (1.2cm and 0.6cm);
            
            \node[circle,draw=RoyalBlue, fill=RoyalBlue, inner sep=0pt,minimum size=2pt] at (2, 1) {};
            \node[below] at (2, 1) {\footnotesize $j$};
            \node[circle,draw=RoyalBlue, fill=RoyalBlue, inner sep=0pt,minimum size=2pt] at (3, 1) {};
            \node[below] at (3, 1) {\footnotesize $\bar j$};
            
            \node[circle,draw=RoyalBlue, fill=RoyalBlue, inner sep=0pt,minimum size=2pt] at (1, 0.5) {};
            \node[below] at (1, 0.5) {\footnotesize $j$};
            \node[circle,draw=RoyalBlue, fill=RoyalBlue, inner sep=0pt,minimum size=2pt] at (3.7, 0.5) {};
            \node[below] at (3.7, 0.5) {\footnotesize $\bar j$};
        \end{scope}

        \begin{scope}[xshift=6cm]
            \draw (0, 0) -- (4, 0) -- (5, 2) -- (1, 2) -- cycle;
            \draw[fill=YellowOrange] (2.5, 1) node[above] {\large $A$} ellipse (1.2cm and 0.6cm);
            
            \node[circle,draw=RoyalBlue, fill=RoyalBlue, inner sep=0pt,minimum size=2pt] at (2, 1) {};
            \node[below] at (2, 1) {\footnotesize $j$};
            \node[circle,draw=RoyalBlue, fill=RoyalBlue, inner sep=0pt,minimum size=2pt] at (3, 1) {};
            \node[below] at (3, 1) {\footnotesize $j$};
            
            \node[circle,draw=RoyalBlue, fill=RoyalBlue, inner sep=0pt,minimum size=2pt] at (1, 0.5) {};
            \node[below] at (1, 0.5) {\footnotesize $\bar j$};
            \node[circle,draw=RoyalBlue, fill=RoyalBlue, inner sep=0pt,minimum size=2pt] at (3.7, 0.5) {};
            \node[below] at (3.7, 0.5) {\footnotesize $\bar j$};
        \end{scope}
    \end{tikzpicture}
    \caption{The spherical entangling region with two excitations inside and the other two outside. There are two configurations: (1) two excitations in $A$ are in the same representation [Left], (2) two excitations in $A$ are in the conjugate representation to each other [Right].}
    \label{fig:CS_setup}
\end{figure}
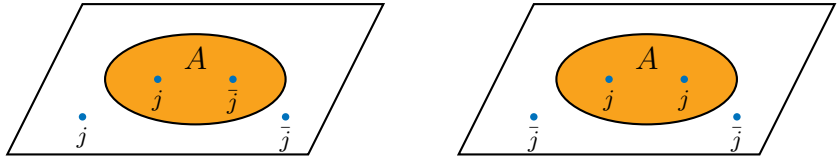

To illustrate the application of the CHM map, we revisit the pseudo entropy in the three-dimensional Chern-Simons theory considered in section \ref{subsec:excitation}.
Hence, we focus on states with two excitations in the entangling region $A$ and the other two excitations in the complement in the Chern-Simons theory.
Corresponding to the two cases studied in section \ref{subsec:excitation} there are two configurations depending on whether the two excitations in $A$ are in the same representation or not as shown in figure \ref{fig:CS_setup}.

\subsubsection{Case 1}
\begin{figure}[h]
    \centering
    
        \begin{tikzpicture}[thick]
            \draw (0, 2) -- (0, -2);
            \draw (4, 2) -- (4, -2);
            \draw[draw=none, fill=SeaGreen!70] (0, 0) -- (4, 0) -- (4, 2) -- (0, 2) --cycle;
            \draw[very thick, YellowOrange] (0,2) -- node[black, above] {$A$} (4, 2);
            \draw[very thick, YellowOrange] (0, -2) --  node[black, below] {$A$}(4,-2);
            \draw (0, 0) -- (4, 0);
    
            \draw[-stealth] (-0.2, 1) --+ (0, 1.3) node[above] {$\tau$};
            \node[left] at (-0.2, -2) {\footnotesize $0$};
            \node[left] at (-0.2, 0) {\footnotesize $\pi$};
            \node[left] at (-0.2, 2) {\footnotesize $2\pi$};
            \node[above] at (3.4, 0){$B$};
            \node[right] at (3, 1) {\large $\psi$};
            \node[right] at (3, -1) {\large $\varphi$};
            
            \draw[RoyalBlue] (1, 2) arc (180:360:1 and 0.5);
            \node[circle,draw=RoyalBlue, fill=RoyalBlue, inner sep=0pt,minimum size=2pt] at (1, 2) {};
            \node[above] at (1, 2) {\footnotesize $j$};
            \node[circle,draw=RoyalBlue, fill=RoyalBlue, inner sep=0pt,minimum size=2pt] at (3, 2) {};
            \node[above] at (3, 2) {\footnotesize $\bar j$};
            
            \draw[RoyalBlue] (1, 0) arc (180:0:1 and 0.5);
            \node[circle,draw=RoyalBlue, fill=RoyalBlue, inner sep=0pt,minimum size=2pt] at (1, 0) {};
            \node[below left] at (1, 0) {\footnotesize $j$};
            \node[circle,draw=RoyalBlue, fill=RoyalBlue, inner sep=0pt,minimum size=2pt] at (3, 0) {};
            \node[below right] at (3, 0) {\footnotesize $\bar j$};
            
            \draw[RoyalBlue] (1, 0) to [out=270,in=90] (3, -2);
            \draw[RoyalBlue] (3, 0) to [out=270,in=30] (2.1, -0.95);
            \draw[RoyalBlue] (1.9, -1.05) to [out=210,in=90] (1, -2);
            \node[circle,draw=RoyalBlue, fill=RoyalBlue, inner sep=0pt,minimum size=2pt] at (1, -2) {};
            \node[below] at (1, -2) {\footnotesize $j$};
            \node[circle,draw=RoyalBlue, fill=RoyalBlue, inner sep=0pt,minimum size=2pt] at (3, -2) {};
            \node[below] at (3, -2) {\footnotesize $\bar j$};
            
        \end{tikzpicture}
    \caption{The path integral representation of the transition matrix $\tau_A^{\psi|\varphi}$ for the case with two excitations in the fundamental representation inside $A$. We draw only the two-dimensional space parametrized by the coordinates $\tau$ and $u$.}
    \label{fig:CS_Config1}
\end{figure}
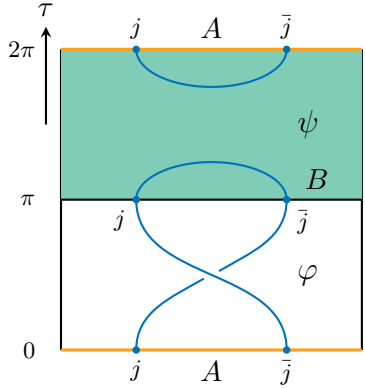

We begin with the case with two excitations in the fundamental representation $R_j$ inside $A$ and two excitations in the anti-fundamental representation $R_{\bar j}$.
The two states $|\psi\rangle$, $|\varphi\rangle$ given by \eqref{eq:exc1} are conformally equivalent to the configuration in figure \ref{fig:CS_Config1}.

The replica partition function can be given by gluing $n$ copies of the transition matrix cyclically along $\tau$, resulting in a cylinder of circumference $2\pi n$ (times $\BH^2$) with $n$ disjoint Wilson loops inserted.
The replica manifold is topologically $\BS^3$, so we find
\begin{align}
    Z(n) = Z\left[\BS^3; R_j^{\otimes n}\right] \ .
\end{align}
One can simplify the right hand side to a product of the partition functions on $\BS^3$ with one Wilson loop insertion using the relation \eqref{eq:cutting}:
\begin{align}
    Z(n) = \frac{Z\left[\BS^3; R_j\right]^n}{Z\left[\BS^3\right]^{n-1}} \ .
\end{align}
Hence we find the pseudo entropy of this configuration:
\begin{align}
    \begin{aligned}
    S\left( \tau_A^{\psi|\varphi}\right) 
        &= 
            \log Z\left[\BS^3\right] \\
        &= 
            - \log \CD \ ,
    \end{aligned}
\end{align}
which reproduces the previous result \eqref{exc1pseudo}.

\subsubsection{Case 2}

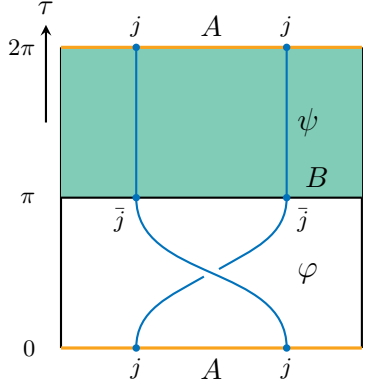
\begin{figure}[th]
    \centering
    
        \begin{tikzpicture}[thick]
            \draw (0, 2) -- (0, -2);
            \draw (4, 2) -- (4, -2);
            \draw[draw=none, fill=SeaGreen!70] (0, 0) -- (4, 0) -- (4, 2) -- (0, 2) --cycle;
            \draw[very thick, YellowOrange] (0,2) -- node[black, above] {$A$} (4, 2);
            \draw[very thick, YellowOrange] (0, -2) --  node[black, below] {$A$}(4,-2);
            \draw (0, 0) -- (4, 0);
    
            \draw[-stealth] (-0.2, 1) --+ (0, 1.3) node[above] {$\tau$};
            \node[left] at (-0.2, -2) {\footnotesize $0$};
            \node[left] at (-0.2, 0) {\footnotesize $\pi$};
            \node[left] at (-0.2, 2) {\footnotesize $2\pi$};
            \node[above] at (3.4, 0){$B$};
            \node[right] at (3, 1) {\large $\psi$};
            \node[right] at (3, -1) {\large $\varphi$};
            
            \draw[RoyalBlue] (1, 2) -- (1, 0);
            \draw[RoyalBlue] (3, 2) -- (3, 0);
            \node[circle,draw=RoyalBlue, fill=RoyalBlue, inner sep=0pt,minimum size=2pt] at (1, 2) {};
            \node[above] at (1, 2) {\footnotesize $j$};
            \node[circle,draw=RoyalBlue, fill=RoyalBlue, inner sep=0pt,minimum size=2pt] at (3, 2) {};
            \node[above] at (3, 2) {\footnotesize $j$};
            
            \node[circle,draw=RoyalBlue, fill=RoyalBlue, inner sep=0pt,minimum size=2pt] at (1, 0) {};
            \node[below left] at (1, 0) {\footnotesize $\bar j$};
            \node[circle,draw=RoyalBlue, fill=RoyalBlue, inner sep=0pt,minimum size=2pt] at (3, 0) {};
            \node[below right] at (3, 0) {\footnotesize $\bar j$};
            
            \draw[RoyalBlue] (1, 0) to [out=270,in=90] (3, -2);
            \draw[RoyalBlue] (3, 0) to [out=270,in=30] (2.1, -0.95);
            \draw[RoyalBlue] (1.9, -1.05) to [out=210,in=90] (1, -2);
            \node[circle,draw=RoyalBlue, fill=RoyalBlue, inner sep=0pt,minimum size=2pt] at (1, -2) {};
            \node[below] at (1, -2) {\footnotesize $j$};
            \node[circle,draw=RoyalBlue, fill=RoyalBlue, inner sep=0pt,minimum size=2pt] at (3, -2) {};
            \node[below] at (3, -2) {\footnotesize $j$};
            
        \end{tikzpicture}
    \caption{The path integral representation of the transition matrix $\tau_A^{\psi|\varphi}$ for the case with two excitations, one in the fundamental representation and the other in the anti-fundamental representation inside $A$.}
    \label{fig:CS_Config2}
\end{figure}

Next we move to the second case with two excitations, one in the fundamental representation and the other in the anti-fundamental representation inside $A$, and take the two states $|\psi\rangle, |\varphi\rangle$ as in figure \ref{fig:CS_Config2}.
This configuration corresponds to the choice of the states $|\psi\rangle = |\psi_0\rangle$ and $|\phi\rangle = |\psi_1\rangle$ in section \ref{sec:CS_case2}.

In this case, the replica partition function falls into two classes depending on whether $n$ is even or not.

\paragraph{$n$: even}
When $n$ is even the replica manifold is topologically equivalent to $\BS^3$ with two Wilson loops inserted with linking number $n/2$:
\begin{align}
    \begin{aligned}
        Z(n) 
        &= 
        Z\left[\BS^3; R_j^{\otimes 2}\big|_{n/2\, \text{link}}\right] \\
        &=
            \frac{[N]}{\CD\,[2]}\left[ q^{-\frac{(N-1)n}{2}}\,[N+1] + q^{-\frac{(N+1)n}{2}}\,[N-1] \right] \ ,
    \end{aligned}
\end{align}
where we use \eqref{eq:exc2ZMn} in the last line.
If we analytically continue even $n$ to a real number and calculate the pseudo entropy from the above partition function, we find
\begin{align}
    \begin{aligned}
            S\left( \tau_A^{\psi|\varphi}\right) 
                &=
                - \log\CD + \log \left[ \frac{[N]}{[2]}\right] 
                + \log \left[ q^\frac{1}{2}\,[N+1] + q^{-\frac{1}{2}}\,[N-1]\right] \\
                &\qquad 
                    -\i\,\frac{\pi}{N+k}\, \frac{q^\frac{1}{2}\,[N+1] - q^{-\frac{1}{2}}\,[N-1]}{q^\frac{1}{2}\,[N+1] + q^{-\frac{1}{2}}\,[N-1]} \ .
    \end{aligned}
\end{align}

\paragraph{$n$: odd}
When $n$ is odd the replica manifold is topologically equivalent to $\BS^3$ with one Wilson loop inserted with $n$ crossings (self-intersection at $n$ points):
\begin{align}
    \begin{aligned}
        Z(n) 
            &=
            Z\left[\BS^3; R_j\big|_{n\, \text{crossing}}\right] \\
            &=
            \frac{[N]}{\CD\,[2]}\left[ q^{-\frac{(N-1)n}{2}}\,[N+1] - q^{-\frac{(N+1)n}{2}}\,[N-1] \right] \ ,
    \end{aligned}
\end{align}
where we use again \eqref{eq:exc2ZMn} in the last line.
By analytically continuing $n$ to a real number, the pseudo entropy is calculated to be 
\begin{align}
    \begin{aligned}
            S\left( \tau_A^{\psi|\varphi}\right) 
                &=
                - \log\CD + \log \left[ \frac{[N]}{[2]}\right] 
                + \log \left[ q^\frac{1}{2}\,[N+1] - q^{-\frac{1}{2}}\,[N-1]\right] \\
                &\qquad 
                    -\i\,\frac{\pi}{N+k}\, \frac{q^\frac{1}{2}\,[N+1] + q^{-\frac{1}{2}}\,[N-1]}{q^\frac{1}{2}\,[N+1] - q^{-\frac{1}{2}}\,[N-1]} \ ,
    \end{aligned}
\end{align}
which reproduces \eqref{Odd_continuation} for $|a-b|=1$ derived with the odd $n$ analytic continuation.

\subsection{Relation to interface entropy in two dimensions}\label{ss:Relation_PE_IE}
The argument for the CHM map slightly differs in two dimensions from the one given in the previous section.
To be concrete, we consider an interval $A=[u,v]$ on $\BR$ at time slice $t=0$ and prepare two different states $\langle \varphi|$ and $|\psi\rangle$ in the Euclidean path integral.
Using a transformation 
\begin{align}
    e^w = \frac{z-u}{v-z}\ ,
\end{align}
the original space (parametrized by the complex coordinates $z$) can be mapped to a cylinder of circumference $\tau \sim \tau + 2\pi$ with the coordinates $w\equiv \sigma + \i\,\tau$ as in figure \ref{fig:PE_2d}.
This is indeed a conformal map as seen from the transformation of the metric:
\begin{align}
    \d z\,\d \bar z = \left( \frac{v-u}{4\,\sinh(w/2)\sinh(\bar w/2)}\right)^2\,\d w\, \d \bar w\ .
\end{align}

\begin{figure}[th]
    \centering
    \begin{tikzpicture}[thick]
        \draw[draw=none, fill=SeaGreen!70]  (0, 3) -- (3, 3) -- (3, 2) -- (0, 2) --cycle;
        \node at (1.5, 2.5) {$\psi$};
        \draw[draw=none, fill=SeaGreen!70]  (0, -1) -- (3, -1) -- (3, -2) -- (0, -2) --cycle;
        \node at (1.5, 1.5) {$\varphi$};
        \draw (0, 3) -- (0, -3);
        \draw (3, 3) -- (3, -3);
        \draw[very thick, orange] (0, 3) -- (3, 3) node[above left] {\footnotesize $A_-^{(1)}$} ; 
        \draw (0, 2) -- (3, 2);
        \draw[very thick, orange] (0, 1) -- (3, 1) node[above left] {\footnotesize $A_+^{(1)}$} ; 
        \node[below left, orange] at (3, 1) {\footnotesize $A_-^{(2)}$};
        \draw[very thick, orange] (0, -1) -- (3, -1) node[above left] {\footnotesize $A_-^{(n)}$} ;
        \draw (0, -2) -- (3, -2);
        \draw[very thick, orange] (0, -3) -- (3, -3) node[below left] {\footnotesize $A_+^{(n)}$} ;
        \draw[dotted] (1.5, 0.3) -- (1.5, -0.3);
        \draw[-stealth] (-0.1, 2.5) -- (-0.1, 3.25) node[above] {\footnotesize $\tau$};
        \node[left] at (-0.1, 3) {\footnotesize $2\pi n$};
        \node[left] at (-0.1, 1) {\footnotesize $2\pi (n-1)$};
        \node[left] at (-0.1, -1) {\footnotesize $2\pi$};
        \node[left] at (-0.1, -3) {\footnotesize $0$};
    \end{tikzpicture}
    \caption{The replica partition function $Z(n)$ for the pseudo entropy.}
    \label{fig:replica_PF}
\end{figure}
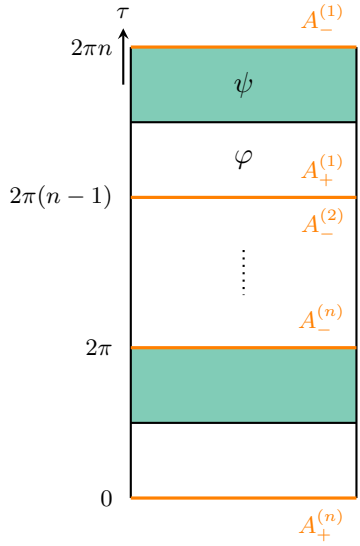

Note that the left and right boundaries correspond to the boundaries of the disks around the endpoints of $A$ which play a role of the UV cutoff in calculation of the partition function.
Then $Z(n)$ is given by the partition function on the replica manifold which can be constructed straightforwardly by gluing $n$ copies of the strip as in figure \ref{fig:replica_PF}.

\begin{figure}[th]
    \centering
    \begin{tikzpicture}[thick]
        \draw[draw=none, fill=SeaGreen!70] (0, 0) -- (4, 0) -- (4, -2) -- (0, -2) --cycle;
        \draw (0, 2) -- (4, 2) -- (4, -2) -- (0, -2) --cycle;
        \draw[RoyalBlue, fill=white] (1.3, 0) circle (0.2) ;
        \draw[RoyalBlue, fill=white] (2.7, 0) circle (0.2);
        \draw (0, 0) -- (1.1, 0);
        \draw (2.9, 0) -- (4, 0);
        \draw[very thick, Orange] (1.5, 0) -- node[above] {$A_+$} (2.5, 0);
        \node[below, Orange] at (2,0) {$A_-$};
        \node[above left, RoyalBlue] at (1.2, 0) {$c_L$};
        \node[above right, RoyalBlue] at (2.8, 0) {$c_R$};
        \node at (2, 1) {\large $\varphi$};
        \node at (2, -1) {\large $\psi$};
        
        \draw[decoration={markings,mark=at position 1 with
        {\arrow[scale=2,>=stealth]{>}}},postaction={decorate}] (4.5, 0) -- node[below] {conformal map} (7, 0); 
        
        \begin{scope}{xshift=1cm}
            \draw[draw=none, fill=SeaGreen!70]  (8, 2) -- (11, 2) -- (11, 0) -- (8, 0) --cycle;
            \draw[RoyalBlue] (8, 2) -- node[left] {$c_L$} (8, -2);
            \draw[RoyalBlue] (11, 2) -- node[right] {$c_R$} (11, -2);
            \draw (8, 0) -- (11, 0);
            \draw[very thick, orange] (8, 2) -- node[above right] {$A_-$} (11, 2); 
            \draw[very thick, orange] (8, -2) -- node[below right] {$A_+$} (11, -2);
            \node[left] at (7.8, -2) {\footnotesize $0$};
            \node[left] at (7.8, 2) {\footnotesize $2\pi$};
            \draw[-stealth] (7.8, 1.5) --  (7.8, 2.25) node[above] {\small $\tau$};
            \node at (9.5, 1) {\large $\psi$};
            \node at (9.5, -1) {\large $\varphi$};
        \end{scope}
    \end{tikzpicture}
    \caption{The pseudo entropy for an interval in CFT$_2$ [Left] and the conformal transformation to the cylinder [Right].}
    \label{fig:PE_2d}
\end{figure}
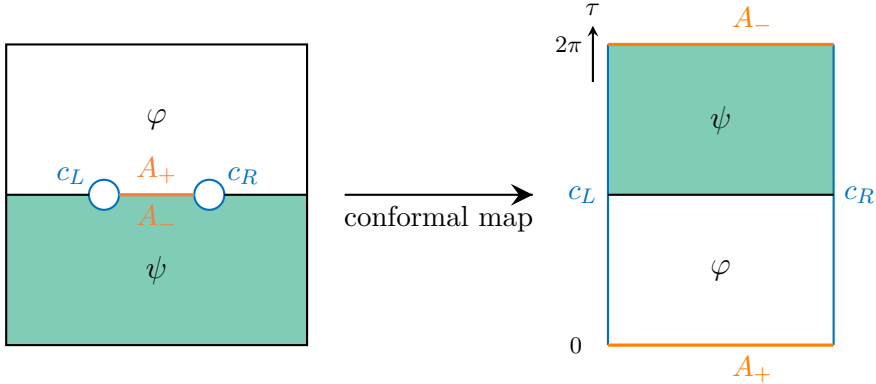

It is sometimes useful to make a further map from the cylinder to a sphere by the following coordinate transformation:
\begin{align}
    \d w\, \d \bar w = \d \sigma^2 + \d \tau^2 = \frac{1}{\sin^2\phi} \left[ \d\phi^2 + \sin^2\phi\,\d\tau^2\right] \ ,
\end{align}
where $\sigma = \log \tan(\phi/2)$ with $\phi \in [0,\pi]$.
Combining the two transformations, we find the map from the original space to the sphere:
\begin{align}\label{Map-to-2sphere}
    \d z\,\d \bar z = \Omega(\phi,\tau)^2 \left[ \d\phi^2 + \sin^2\phi\,\d\tau^2\right] \ , \qquad \Omega \equiv \frac{v-u}{2(1 - \cos\tau \sin \phi)} \ .
\end{align}

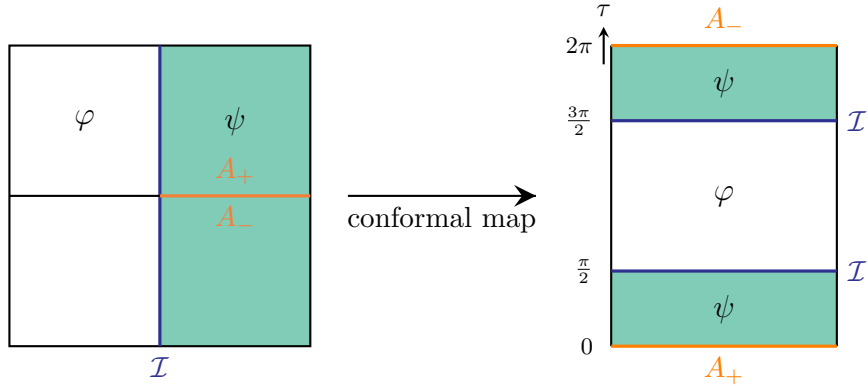
\begin{figure}[th]
    \centering
    \begin{tikzpicture}[thick]
        \draw[draw=none, fill=SeaGreen!70] (2, 2) -- (4, 2) -- (4, -2) -- (2, -2) --cycle;
        \draw (0, 2) -- (4, 2) -- (4, -2) -- (0, -2) --cycle;
        \draw[very thick, Blue] (2, 2) -- (2, -2) node[below] {$\CI$};
        \draw (0, 0) -- (2, 0);
        \draw[very thick, Orange] (2, 0) -- node[above] {$A_+$} (4, 0);
        \node[below, Orange] at (3,0) {$A_-$};
        \node at (1, 1) {\large $\varphi$};
        \node at (3, 1) {\large $\psi$};
        
        \draw[decoration={markings,mark=at position 1 with
        {\arrow[scale=2,>=stealth]{>}}},postaction={decorate}] (4.5, 0) -- node[below] {conformal map} (7, 0); 

        \draw[draw=none, fill=SeaGreen!70]  (8, 2) -- (11, 2) -- (11, 1) -- (8, 1) --cycle;
        \draw[draw=none, fill=SeaGreen!70]  (8, -1) -- (11, -1) -- (11, -2) -- (8, -2) --cycle;
        \draw (8, 2) -- (8, -2);
        \draw (11, 2) -- (11, -2);
        \draw[very thick, orange] (8, 2) -- node[above] {$A_-$} (11, 2);
        \draw[very thick, Blue] (8, 1) -- (11, 1) node[right] {$\CI$} ;
        \draw[very thick, Blue] (8, -1) -- (11, -1) node[right] {$\CI$};
        \draw[very thick, orange] (8, -2) -- node[below] {$A_+$} (11, -2);

        \node at (9.5, 1.5) {\large $\psi$};
        \node at (9.5, 0) {\large $\varphi$};
        \node at (9.5, -1.5) {\large $\psi$};
        
        \draw[-stealth] (7.9, 1.75) -- (7.9, 2.25) node[above] {\small $\tau$};
        \node[left] at (7.9, 2) {\footnotesize $2\pi$};
        \node[left] at (7.9, 1) {\footnotesize $\frac{3\pi}{2}$};
        \node[left] at (7.9, -1) {\footnotesize $\frac{\pi}{2}$};
        \node[left] at (7.9, -2) {\footnotesize $0$};
    \end{tikzpicture}
    
    \caption{The entanglement entropy across a conformal interface $\CI$ (interface entropy) in two dimensions [Left].
    The entangling region $A$ is taken to be a half line right to the interface $\CI$.
    The configuration can be mapped to the cylinder by a conformal transformation [Right].
    }
    \label{fig:Interface_map}
\end{figure}

A closely related measure to the pseudo entropy is the entanglement entropy across a conformal interface, also known as interface entropy.
We here consider a restricted case where two different states $|\psi\rangle$ and $|\varphi\rangle$ in CFT$_2$ are glued along an interface $\CI$ at the origin of a time slice as in figure \ref{fig:Interface_map}.
The interface entropy is the entanglement entropy for the entangling region $A$ extending from the origin to the right, which quantifies the difference between the two states \cite{Sakai:2008tt}.

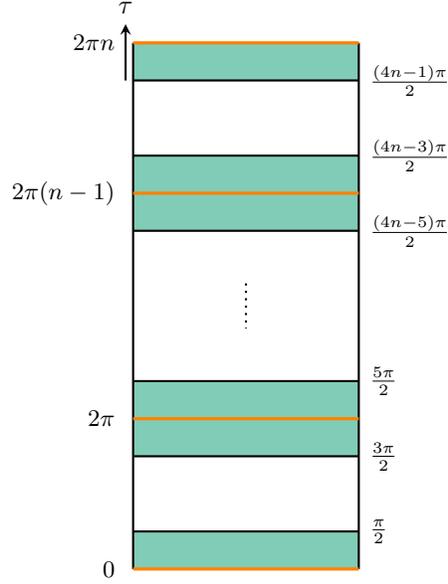
\begin{figure}[ht]
    \centering
    \begin{tikzpicture}[thick]
        \draw[draw=none, fill=SeaGreen!70]  (0, 3.5) -- (3, 3.5) -- (3, 3) -- (0, 3) --cycle;
        \draw[draw=none, fill=SeaGreen!70]  (0, 2) -- (3, 2) -- (3, 1) -- (0, 1) --cycle;
        \draw[draw=none, fill=SeaGreen!70]  (0, -2) -- (3, -2) -- (3, -1) -- (0, -1) --cycle;
        \draw[draw=none, fill=SeaGreen!70]  (0, -3) -- (3, -3) -- (3, -3.5) -- (0, -3.5) --cycle;
        \draw (0, 3.5) -- (0, -3.5);
        \draw (3, 3.5) -- (3, -3.5);
        \draw[very thick, orange] (0, 3.5) -- (3, 3.5);
        \draw (0, 3) -- (3, 3);
        \draw (0, 2) -- (3, 2);
        \draw[very thick, orange] (0, 1.5) -- (3, 1.5);
        \draw (0, 1) -- (3, 1);
        \draw (0, -1) -- (3, -1);
        \draw[very thick, orange] (0, -1.5) -- (3, -1.5);
        \draw (0, -2) -- (3, -2);
        \draw (0, -3) -- (3, -3);
        \draw[very thick, orange] (0, -3.5) -- (3, -3.5);
        \draw[dotted] (1.5, 0.3) -- (1.5, -0.3);
        \draw[-stealth] (-0.1, 3) -- (-0.1, 3.75) node[above] {\small $\tau$};
        \node[left] at (-0.1, 3.5) {\footnotesize $2\pi n$};
        \node[left] at (-0.1, 1.5) {\footnotesize $2\pi (n-1)$};
        \node[left] at (-0.1, -1.5) {\footnotesize $2\pi$};
        \node[left] at (-0.1, -3.5) {\footnotesize $0$};
        
        \node[right] at (3, 3) {\footnotesize $\frac{(4n-1)\pi}{2}$};
        \node[right] at (3, 2) {\footnotesize $\frac{(4n-3)\pi}{2}$};
        \node[right] at (3, 1) {\footnotesize $\frac{(4n-5)\pi}{2}$};
        \node[right] at (3, -1) {\footnotesize $\frac{5\pi}{2}$};
        \node[right] at (3, -2) {\footnotesize $\frac{3\pi}{2}$};
        \node[right] at (3, -3) {\footnotesize $\frac{\pi}{2}$};
    \end{tikzpicture}
    \caption{The replica partition function for the interface entropy.}
    \label{fig:Interface_replica}
\end{figure}

Using a canonical conformal map from flat space to a cylinder and gluing $n$ copies along the entangling surface one obtains the replica manifold as a cylinder
of circumference $\tau ~\tau + 2\pi n$ with $n$ interfaces inserted at specific locations (see figure \ref{fig:Interface_replica}):
\begin{align}
    \CI: \left\{\tau = \frac{(2i -1)\pi}{2} \ , \quad (i=1,2,\cdots, 2n)\right\} \ . 
\end{align}
Compared with figure \ref{fig:replica_PF} this is the same replica manifold as the pseudo entropy in the previous subsection with $\tau$ shifted by $\pi/2$.
Hence, we establish the equality between the pseudo entropy and interface entropy in CFT$_2$:
\begin{align}
    S\left( \tm{\psi}{\varphi}_{A:\, \text{interval}}\right) = S^\CI_{A:\, \text{half-line}}  \qquad \text{in CFT$_2$} \ .
\end{align}

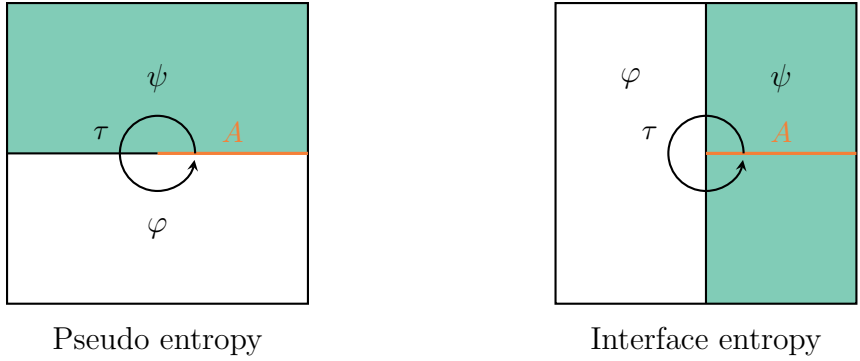
\begin{figure}[ht]
    \centering
    \begin{tikzpicture}[thick]
        \draw[draw=none, fill=SeaGreen!70] (0, 2) -- (4, 2) -- (4, 0) -- (0, 0) --cycle;
        \draw (0, 2) -- (4, 2) -- (4, -2) -- (0, -2) --cycle;
        \draw[draw=none] (0, -2) -- (4, -2) -- (4, 0) -- (0, 0) --cycle;
        \draw[very thick, Orange] (2, 0) -- node[above] {$A$} (4, 0);
        \draw (0, 0) -- (2, 0);
        \draw[-stealth] (2.5, 0) arc[start angle=0, end angle=350,radius=0.5] node[midway, above left] {$\tau$};
        \node at (2, -1) {\large $\varphi$};
        \node at (2, 1) {\large $\psi$};
        \node at (2, -2.5) {\large Pseudo entropy};
    \end{tikzpicture}
    \hspace{3cm}
    \begin{tikzpicture}[thick]
        \draw[draw=none, fill=SeaGreen!70] (2, 2) -- (4, 2) -- (4, -2) -- (2, -2) --cycle;
        \draw (0, 2) -- (4, 2) -- (4, -2) -- (0, -2) --cycle;
        \draw (2, 2) -- (2, -2);
        \draw[very thick, Orange] (2, 0) -- node[above] {$A$} (4, 0);
        \draw[-stealth] (2.5, 0) arc[start angle=0, end angle=350,radius=0.5] node[midway, above left] {$\tau$};
        \node at (1, 1) {\large $\varphi$};
        \node at (3, 1) {\large $\psi$};
        \node at (2, -2.5) {\large Interface entropy};
    \end{tikzpicture}
    \caption{The pseudo entropy and interface entropy across a hyperplane. Their replica manifolds are the same up to the $\tau$ shift by $\pi/2$.}
    \label{fig:PE_IE}
\end{figure}

More generally, taking the entangling surface to be a hyperplane in flat space the pseudo entropy equals to the interface entropy in \emph{any QFT in $d\ge 2$ dimensions}:
\begin{align}
    S\left( \tm{\psi}{\varphi}_{A}\right) = S^\CI_{A} \qquad \text{for} \quad \Gamma = \partial A : \{ x^0=x^1 =0 \} \ .
\end{align}
This is clear from figure \ref{fig:PE_IE} where one sees the replica manifold of the former is obtained by rotating that of the latter by $\pi/2$ degree.

\subsubsection{Compact scalar theory}

We calculate the pseudo entropy between ground states of massless compact bosons of different radii $R_1$ and $R_2$ in two dimensions.
When the entangling region is an interval it amounts to the entanglement entropy across an interface between the two compact boson theories as we saw in section \ref{ss:Relation_PE_IE}.

There are four types of conformal interfaces $\CI^\pm_{k_1k_2}$ labeled by $\pm$ and their conjugates, which act as intertwiners from one side of a free boson theory to the other.
The pair of relatively prime numbers $(k_1, k_2)$ can be interpreted as the winding numbers of D-brane along the two cycles of the torus parametrized by two compact bosons.

We focus on the case with $\CI^+_{k_1k_2}$ and read off the pseudo entropy from the result of \cite{Sakai:2008tt} for the interface entropy by translating the parameters appropriately:
\begin{align}
    S^{\CI^+_{k_1k_2}}\left( \tau_A^{\psi|\varphi}\right) = \sigma_\text{s}  \left(|\sin (2\theta_+)|\right)\, \log\left( \frac{L}{\epsilon}\right) - \log |k_1 k_2| \ ,
\end{align}
where $L\equiv v-u$, $\epsilon$ are the lengths of the interval and the UV cutoff for the pseudo entropy, respectively.
The parameter $\theta_+$ defined through the relation
\begin{align}
    \tan \theta_+ \equiv \frac{k_2 R_2}{k_1 R_1} \ ,
\end{align}
controls the transmittance of the interface.
The function $\sigma_\text{s} (x)$ defined by
\begin{align}
    \sigma_\text{s} (x ) \equiv \frac{1}{6} + \frac{x}{3} + \frac{1}{\pi^2}\left[ (x+1)\log(x+1)\,\log x + (x-1)\,\text{Li}_2( 1-x) + (x+1)\,\text{Li}_2 (-x)\right] \ ,
\end{align}
interpolates between $\sigma_s(0) = 0$ and $\sigma_s(1)=1/3$ monotonically.

\subsubsection{Free fermion}
The entanglement entropy across a conformal interface in the Ising model in two dimensions was investigated in \cite{Brehm:2015lja}.
Interfaces in the Ising model can be described as boundary conditions in either the $\BZ_2$-orbifold theory of a free boson or the real Majorana fermion theory.
In the latter description, there is an independent set of interfaces: NS, R, and neutral interfaces, labeled by a parameter $\phi$ controlling their transmittance.
It follows from the result of \cite{Brehm:2015lja} the pseudo entropy for the NS interface becomes
\begin{align}
    S^\text{NS}\left( \tau_A^{\psi|\varphi}\right) = \sigma_\text{f}  \left(|\sin (2\phi)|\right)\, \log\left( \frac{L}{\epsilon}\right)  \ ,
\end{align}
where the function $\sigma_\text{f} (x)$ defined by
\begin{align}
    \sigma_\text{f} (x ) \equiv \frac{x-1}{6} - \frac{1}{\pi^2}\left[ (x+1)\log(x+1)\,\log x + (x-1)\,\text{Li}_2( 1-x) + (x+1)\,\text{Li}_2 (-x)\right] \ ,
\end{align}
interpolates between $\sigma_f(0) = 0$ and $\sigma_f(1)=1/6$ monotonically.
The pseudo entropies for the R and neutral interfaces are also given by
\begin{align}
    \begin{aligned}
        S^\text{R}\left( \tau_A^{\psi|\varphi}\right) 
            &= 
                S^\text{NS}\left( \tau_A^{\psi|\varphi}\right) \ , \\
        S^\text{neutral}\left( \tau_A^{\psi|\varphi}\right) 
            &= 
                S^\text{NS}\left( \tau_A^{\psi|\varphi}\right) - \log 2 \ .
    \end{aligned}
\end{align}

\subsubsection{Topological interface}

Suppose we are given two CFTs glued along a straight line $\CC$.
To preserve a part of the conformal symmetry, the energy flow perpendicular to $\CC$ has to be continuous:
\begin{align}\label{Conformal_Defect}
    T^{(1)} - \bar T^{(1)}|_{\CC} = T^{(2)} - \bar T^{(2)}|_{\CC} \ .
\end{align}
The gluing line $\CC$ may be seen as an defect operator $\CI$ intertwining the Hilbert space $\CH^{(1)}$ of one theory with the other $\CH^{(2)}$.
The condition \eqref{Conformal_Defect} implies that such an operator satisfies the commutation relations:
\begin{align}\label{Defect_Virasoro_Commutation_Relation}
    \left(L^{(1)}_n - \bar L^{(1)}_n\right)\,\CI = \CI\,\left(L^{(2)}_n - \bar L^{(2)}_n\right) \ .
\end{align}
Finding solutions to these relations is a hard problem, but it simplifies if the defects satisfy stronger conditions:
\begin{align}\label{Topological_condition}
    L^{(1)}_n\,\CI = \CI\,L^{(2)}_n \ , \qquad \bar L^{(1)}_n\,\CI = \CI\,
    \bar L^{(2)}_n \ .
\end{align}
Since they commute with all the Virasoro generators, their locations can be moved freely, and hence $\CI$ becomes topological.

Topological defects have been extensively studied in a rational CFT (RCFT) whose Hilbert space takes the form:
\begin{align}
    \CH = \bigoplus_{i, \bar j}\,M_{i \bar j}\, \CV_i \otimes \bar\CV_{\bar j} \ ,
\end{align}
where $i$ and $\bar j$ label (a finite number of) irreducible representations $\CV_i$ and $\bar\CV_{\bar j}$ of the Virasoro algebra in chiral and anti-chiral sectors respectively and $M_{i\bar j}$ is the multiplicity of the pair representation $(i, \bar j)$ appearing in the spectrum of the theory.
In this case, the topological defects between two RCFTs with multiplicity matrices $M_{i\bar j}^{(1)}$ and $M_{i\bar j}^{(2)}$  can be written as \cite{Petkova:2000ip,Brehm:2015plf}
\begin{align}
    \CI_K = \sum_{\bf i}\,d_{K\bf i}\,P^{\bf i} \ ,
\end{align}
where $K$ labels types of topological interface and ${\bf i} \equiv (i, \bar j; \alpha,\beta)$ is the index for the projector $P^{\bf i}$ intertwining between a representation $\left(\CV_i\otimes \bar \CV_{\bar j}\right)^{(\alpha)}$ with $\alpha = 1, \cdots, M_{i\bar j}^{(1)}$ of one RCFT to another representation $\left(\CV_i\otimes \bar \CV_{\bar j}\right)^{(\beta)}$ with $\beta = 1, \cdots, M_{i\bar j}^{(2)}$ of the other RCFT.

When the two CFTs are isomorphic and hence $M_{i\bar j}^{(1)} = M_{i\bar j}^{(2)}$, hence for conformal defects between the same CFT, the projector $P^{\bf i}$ is given a realization as
\begin{align}
    P^{(i, \bar j; \alpha, \beta)} \equiv \sum_{ {\bf n}, \bar {\bf n}}\, \left(|i,  {\bf n}\rangle\otimes |\bar j, \bar  {\bf n}\rangle\right)^{(\alpha)}\, \left(\langle i,  {\bf n}|\otimes \langle \bar j, \bar  {\bf n}|\right)^{(\beta)} \ ,
\end{align}
where $|i, {\bf n}\rangle\otimes |\bar j, \bar {\bf n}\rangle$
is an orthogonal basis for the representation $\CV_i\otimes \bar\CV_{\bar j}$.
Clearly $\CI_K$ satisfies the conditions \eqref{Topological_condition} which now reduce to the commutation relations with the Virasoro generators
\begin{align}\label{Interface_commutator}
    [L_n, \CI_K] = [\bar L_n, \CI_K] = 0 \ ,
\end{align}
and hence correlation functions depend only on the homotopy class of the contour of $\CI_K$ \cite{Petkova:2000ip}.
Defects intertwining between the same theory are called interfaces, so topological interfaces are the solutions to the conditions \eqref{Interface_commutator}.

The replica partition function for calculating the entanglement entropy across a topological interface is given by a torus partition function with $2n$ insertion of interface operators:
\begin{align}
    \begin{aligned}
        Z_{K}(n) 
            &= \tr\left[ \left( \CI_K\, e^{-t H}\, \CI_K^\dagger\,e^{-t H}\right)^n\right]\\
            &= \tr\left[ \left(\CI_K\,\CI_K^\dagger\right)^n\,e^{-2t H}\right] \\
            &= \sum_{(i,\bar j)}\, \Tr\left[\left(d_{K {\bf i}}\,d_{K^\ast {\bf i}}\right)^{n}\right]\, \chi_i \left(e^{-2nt}\right)\, \chi_{\bar j} \left(e^{-2nt}\right) \ ,
    \end{aligned}
\end{align}
where $H$ is the Hamiltonian on a cylinder
\begin{align}
    H = L_0 + \bar L_0 - \frac{c}{12} \ , 
\end{align}
and we used the commutation relation \eqref{Interface_commutator} in the second equality.
$\chi_i$ is the character in the $\CV_i$ representation and
the parameter $t$ is related to the UV and IR cutoffs $\epsilon$, $L$ as
\begin{align}
    t = \frac{2\pi^2}{\log ( L/\epsilon)} \ .
\end{align}
We took a trace $\text{Tr}$ with respect to the multiplicity indices $\alpha, \beta$ by regarding $d_{K(i,\bar j;\alpha, \beta)}$ as a matrix with the notation $d_{K^\ast (i,\bar j;\alpha, \beta)} \equiv d_{K(i,\bar j;\beta, \alpha)}^\ast$.
Using the modular property of the character, the entanglement entropy becomes \cite{Brehm:2015plf}
\begin{align}
    S_{\CI_K} 
        =
        \frac{c}{6}\log \left(\frac{L}{\epsilon}\right) -\sum_{(i, \bar j)}\,\Tr\left[ p^K_{\bf i}\,\log  \frac{p^K_{\bf i}}{p^\text{Id}_{\bf i}}\right] \ ,
\end{align}
where $p^K_{\bf i}$ is a probability distribution characterized by the modular $\CS$-matrix as
\begin{align}
    p^K_{\bf i} \equiv \frac{\Smat{0}{i}\,\left(\Smat{0}{\bar{j}}\right)^\ast}{\sum_{(i,\bar j)}\,\Smat{0}{i}\,\left(\Smat{0}{\bar{j}}\right)^\ast\,\Tr \left[ d_{K {\bf i}}\,d_{K^\ast {\bf i}}\right]}\,d_{K {\bf i}}\,d_{K^\ast {\bf i}} \ , \qquad p^\text{Id}_{\bf i} \equiv \Smat{0}{i}\,\left(\Smat{0}{\bar{j}}\right)^\ast\,\delta_{\alpha\beta} \ .
\end{align}

For diagonal theories with multiplicity $M_{i\bar j} = \delta_{i\bar j}$, the CFTs on both sides are the same theory and topological interfaces are one-to-one correspondence to the primary operators labeled by $a$:
\begin{align}\label{Topological_Interface}
    \CI_a = \sum_i\,\frac{\Smat{a}{i}}{\Smat{0}{i}}\,P^{i} \ ,
\end{align}
where $P^{i}$ is the projector acting on the representation $\CV_i\otimes \bar\CV_{i}$:
\begin{align}
    P^i \equiv \sum_{{\bf n}, \bar {\bf n}}\, |i, {\bf n}\rangle\otimes |i, \bar {\bf n}\rangle\, \langle i, {\bf n}|\otimes \langle i, \bar {\bf n}| \ .
\end{align}
Then the probability distribution simplifies to
\begin{align}
    p^a_i = \left|\Smat{a}{i}\right|^2 \ ,\qquad p^\text{Id}_i = \left|\Smat{0}{i}\right|^2 \ ,
\end{align}
and we find \cite{Gutperle:2015kmw,Brehm:2015plf}
\begin{align}\label{IE_diag_RCFT}
     S_{\CI_a} = \frac{c}{6}\log \left(\frac{L}{\epsilon}\right) -2\sum_i\,\left|\Smat{a}{i}\right|^2\,\log \left| \frac{\Smat{a}{i}}{\Smat{0}{i}}\right| \ .
\end{align}

It would be worthwhile to note that the constant term contributed from the interface takes the same form as the topological entanglement entropy for any state $|\psi\rangle$ in the Chern-Simons theory on a torus we considered in section \ref{subsec:torus}
(see also section 2.5.2 in \cite{Dong:2008ft}):
\begin{align}
    S_\psi = -2 \sum_{i}\, |\psi_i|^2\, \log \frac{|\psi_i|}{\Smat{0}{i}} \ ,
\end{align}
where $\psi_i$ are coefficients for the state $|\psi\rangle$ which is the superposition of a single Wilson loop in the irreducible representation $R_i$:
\begin{align}
    |\psi\rangle = \sum_i\,\psi_i\,|R_i\rangle \ .
\end{align}
Indeed one can reproduce the finite part of the interface entropy \eqref{IE_diag_RCFT} by setting $\psi_i$ to a specific value $\psi_i = \Smat{a}{i}$ (see \eqref{topint}).
This coincidence may not be so surprising given the well-known correspondence between the WZW model and the Chern-Simons theory, but we are not aware of any direct link between them.

\section{Left-right pseudo entanglement entropy}\label{sec:LRPE}

A closely related measure to the interface entropy is the left-right entanglement entropy (LREE) in BCFT$_2$ \cite{PandoZayas:2014wsa,Das:2015oha}.
For a boundary state $|B\rangle$ subject to the gluing condition
\begin{align}
    \left( L_n - \bar L_{-n}\right)\,|B\rangle = 0 \ ,
\end{align}
there exists an orthogonal basis spanned by the Ishibashi states $|i\rrangle$ \cite{Ishibashi:1988kg}:
\begin{align}
    |i\rrangle = \sum_{{\bf n}} |i, {\bf n}\rangle \otimes \overline{|i, {\bf n}\rangle} \ .
\end{align}
Note that they are non-normalizable states, but their inner product can be regularized as\footnote{We can equally normalize the Ishibashi state 
as  $\llangle i | j\rrangle = \delta_{ij}$. This is the normalization 
employed in \cite{Wen:2016snr}, which is equivalent to the rescaling of the coefficient such that $\psi_i\to \psi_i/\s{\Smat{0}{i}}$.} (see e.g.\,\cite{Behrend:1999bn}) 
\begin{align}
    \llangle i | j\rrangle = \delta_{ij}\,\Smat{0}{i} \ .
\end{align}
Hence one can expand any boundary state $|\psi\rangle$ by the Ishibashi state as follows:
\begin{align}
    |\psi\rangle = \sum_{i}\, \psi_i\,|i\rrangle \ .
\end{align}
The LREE is the von Neumann entropy of the reduced density matrix for the left (holomorphic) sector:
\begin{align}
    \rho_L^{\psi} \equiv \frac{1}{\langle\psi | \psi\rangle}\,\Tr_R\left[ |\psi\rangle\,\langle \psi| \right] \ .
\end{align}

Now we introduce the transition matrix for two boundary states $|\psi\rangle$ and $|\varphi\rangle$ by
\begin{align}
    \tau_L^{\psi|\varphi} \equiv \frac{1}{\langle\varphi | \psi\rangle}\,\Tr_R\left[ |\psi\rangle\,\langle \varphi| \right] \ ,
\end{align}
and define the left-right pseudo entropy (LRPE) as the von Neumann entropy of the transition matrix.
Since the boundary states are non-normalizable we regularize them by slightly evolving them in the imaginary time direction:
\begin{align}
    |\psi\rangle \to e^{-\epsilon H}\,|\psi\rangle\ .
\end{align}
For a theory on a cylinder of circumference $\ell$ the Hamiltonian becomes
\begin{align}
    H = \frac{2\pi}{\ell} \left( L_0 + \bar L_0 - \frac{c}{12}\right) \ ,
\end{align}
which yields the following expressions:
\begin{align}
    \begin{aligned}
        \langle \varphi|\, e^{-2\epsilon H} \,| \psi\rangle 
            &= \sum_i\,\psi_i\,\varphi_i^\ast\,\chi_i \left( e^{- \frac{8\pi \epsilon}{\ell}}\right) \ , \\
        \Tr_R\left[ e^{-\epsilon H}|\psi\rangle\,\langle \varphi|\,e^{-\epsilon H} \right] 
            &=
            \sum_{i, {\bf n}}\,\psi_i\,\varphi_i^\ast\,\,e^{- \frac{8\pi \epsilon}{\ell}\left( h_i + N_{\bf n} - \frac{c}{24} \right)}\,|i, {\bf n}\rangle\, \langle i, {\bf n}| \ ,\\
        \Tr_L \left[ \left(\tau_L^{\psi|\varphi}\right)^n \right]
            &=
            \frac{1}{\left[ \sum_i\,\psi_i\,\varphi_i^\ast\,\chi_i \left( e^{- \frac{8\pi \epsilon}{\ell}}\right) \right]^n} \, \sum_i\,\left(\psi_i\,\varphi_i^\ast\right)^n\,\chi_i \left( e^{- \frac{8\pi \epsilon n}{\ell}}\right)\ .
    \end{aligned}
\end{align}
Here $h_i$ and $N_{\bf n}$ are the conformal dimension and the level of the descendant state $|i, {\bf n}\rangle$ respectively, and $\chi_i(q)\equiv \tr_{\CV_i}\,q^{L_0 - \frac{c}{24}}$ is the character for the representation $i$.
Using the modular transformation
\begin{align}
    \chi_i \left( e^{-\frac{8\pi \epsilon n}{\ell}}\right) = \sum_{j}\Smat{i}{j}\,\chi_j\left( e^{-\frac{\pi\ell }{2\epsilon n}}\right)  \ ,
\end{align}
and taking the $\epsilon \to 0$ limit we find the LRPE:
\begin{align}\label{LRPE_formula}
    \begin{aligned}
        S\left(\tau_L^{\psi|\varphi}\right)
            &=
            - \partial_n\,\Tr_L \left[ \left(\tau_L^{\psi|\varphi}\right)^n \right]\big|_{n=1} \\
            &=
            \frac{\pi c\ell}{24\,\epsilon} - \frac{\sum_i\,\Smat{i}{0}\,\psi_i\,\varphi_i^\ast\,\log (\psi_i\,\varphi_i^\ast\,)}{\sum_i\,\Smat{i}{0}\,\psi_i\,\varphi_i^\ast}
            +
            \log \left[\sum_i\,\Smat{i}{0}\,\psi_i\,\varphi_i^\ast\right] \ .
    \end{aligned}    
\end{align}
This expression is not necessarily real, but when $\psi = \varphi$ it reduces to the LREE derived in \cite{Das:2015oha}:
\begin{align}\label{LREE}
    S^{(\text{LR})}\left( |\psi\rangle \right) 
        =
            \frac{\pi c\ell}{24\,\epsilon} - \frac{\sum_i\,\Smat{i}{0}\,|\psi_i|^2\,\log |\psi_i|^2}{\sum_i\,\Smat{i}{0}\,|\psi_i|^2}
            +
            \log \left[\sum_i\,\Smat{i}{0}\,|\psi_i|^2\right] \ .
\end{align}
which is real as expected.

In a diagonal theory, the Cardy states $|a\rangle$ can be written as a superposition of the Ishibashi states:\footnote{Note that this is the same relation as the expression of the interface operator \eqref{Topological_Interface} if one replaces $\Smat{0}{i}$ with $\left(\Smat{0}{i}\right)^2$.}
\begin{align}\label{Cardy_Ishibashi_relation_BCFT}
     |a\rangle = \sum_i\,\frac{\Smat{a}{i}}{\sqrt{\Smat{0}{i}}}\,|i\rrangle \ .
\end{align}
As a simple example let us take 
\begin{align}
    |\psi\rangle = |a\rangle \ , \qquad |\varphi\rangle = |i\rrangle \ ,
\end{align}
then the LRPE becomes
\begin{align}
    S\left(\tau_L^{\psi|\varphi}\right)
        =
        \frac{\pi c\ell}{24\,\epsilon} + \log\,\Smat{i}{0} \ .
\end{align}
Interestingly, the LRPE does not depend on the choice of the Cardy state $|a\rangle$ as long as it overlaps with the Ishibashi state $|i\rrangle$

Next, consider the LREE for the Ishibashi state $|\psi\rangle = |i\rrangle$.
It follows from \eqref{LREE} with $\psi_i=\varphi_i =1$, $\psi_{k\neq i} = \varphi_{k\neq i} = 0$ that
\begin{align}
    S^\text{(LR)}\left(|i\rrangle\right)
        =
        \frac{\pi c\ell}{24\,\epsilon} + \log\,\Smat{i}{0} \ .
\end{align}
This is the same as the LRPE considered above.

Another example is the LREE for the Cardy state $|\psi\rangle = |a\rangle$, which is given by substituting $\psi_i = \varphi_i = \Smat{a}{i}/\sqrt{\Smat{0}{i}}$ to \eqref{LRPE_formula} \cite{Das:2015oha,Brehm:2015plf} 
\begin{align}\label{LREE_Cardy}
    S^\text{(LR)}\left(|a\rangle\right) = \frac{\pi c\ell}{24\,\epsilon} - \sum_i\,\left(\Smat{a}{i}\right)^2\,\log \left[\frac{(\Smat{a}{i})^2}{\Smat{0}{i}}\right] \ ,
\end{align}
which agrees with the topological entanglement entropy \eqref{CS_Cardy_S2} of the Chern-Simons theory for the Cardy state.
Note that \eqref{LREE_Cardy} is close to but differs from \eqref{IE_diag_RCFT} by the denominator in the logarithm.
This difference can be accounted for by the fact that both holomorphic and anti-holomorphic sectors contribute to the interface entropy while there is only the holomorphic sector in the unfolded theory of BCFT in the LREE.
In the latter case the correspondence between the topological entanglement entropy in the Chern-Simons theory and the LREE of a boundary state is clear as the left and right moving CFTs appear as the edge modes of the Chern-Simons theory on either side of the entangling surface \cite{Das:2015oha}.
It would be interesting to understand the above coincidence between the interface entropy in 2d and the topological entropy in 3d along the same lines of thought.

\section{Conclusions}\label{sec:conclusion}
In this paper, we studied pseudo entropy in quantum field theory, mainly concentrating on its topological properties.

In section \ref{sec:CS}, we focused on the excited states in Chern-Simons theory.
This provides a class of important examples where pseudo entropy can be analytically computed in 
quantum field theory.
We found non-trivial behavior of the pseudo entropy in the presence of four excitations on $\BS^2$.
Such excited states are prepared by path integrals with inserting appropriate Wilson lines.
In contrast to topological entanglement entropy, we have seen that topological (R\'enyi) pseudo entropies are directly related to partition functions with knotted Wilson loops.
In other words, generic partition functions with Wilson loops should be interpreted as 
topological pseudo entropy rather than topological entropy because the initial state and final state are different. Since the dependence on the crossings vanishes in the classical limit, we can interpret that the crossings give purely quantum contributions to pseudo entropy.
In particular, it is remarkable that the pseudo entropy may be larger than the entanglement entropy, i.e., the difference $\Delta S$ defined in \eqref{diff_pseudo_entropy_limit} can be positive.
This contrasts with a standard quantum many-body system or quantum field theory 
within the same quantum phase, where $\Delta S$ is always non-positive \cite{Mollabashi:2020yie,Mollabashi:2021xsd}. This is consistent with the known fact that the anyonic states created by Wilson loops in Chern-Simons theory belong to different quantum phases in general. 
Note that this is the first example in dimensions higher than two, where pseudo entropy was explicitly evaluated in non-trivial topological phases.

We also explored a geometric interpretation of topological pseudo entropy in 
Chern-Simons theory. We found a universal result when a single Wilson loop is linked with 
the surface $\Gamma=\de A$ once.
Although universal results are not available in more general cases, we noted  that in the semiclassical limit $k\to \infty$, the topological pseudo entropy captures the number of Wilson loops which link with the surface $\Gamma=\partial A$.
This is analogous to the holographic entanglement entropy in the sense that it also measures entanglement or the number of Bell pairs around an extremal surface $\Gamma$.
The geometrical interpretation including full quantum effects remains as a future problem.  

In section \ref{sec:CHM}, we investigated the properties of the pseudo entropy in CFT.
In particular, we found the close relation between the pseudo entropy and the interface entropy in 2d CFT, which can be generalized to any QFT in $d\ge2$ with the restriction to the case where the subsystem $A$ is a half space, i.e., $\partial A:\{x^0=x^1=0\}$.
The extension of the relation to a more general $A$ would be challenging and is left as a future problem.
We used this relation to calculate the pseudo entropies in several interface CFTs. The finite term in the resulting interface entropy coincides with that in Chern-Simons theory on a torus. 

The CHM map in section \ref{ss:CHM} can be concatenated by a further conformal map $\tanh \frac{u}{2} = \tan \frac{\theta}{2}$ to $\BS^d$ with the metric:
\begin{align}
    \d s_\text{Sph}^2 = \d \theta^2 + \cos^2 \theta\, \d\tau^2 + \sin^2\theta\,\d \Omega_{d-2}^2 \ , \qquad \left(0\le \theta \le \frac{\pi}{2}\right) \ .
\end{align}
The resulting map may open the way to evaluate the pseudo entropy in CFT through the sphere partition function with two states glued alternately along the $\tau$ coordinate.
The same map was applied to the interface entropy to derive a universal relation between the entropy and sphere free energy \cite{Kobayashi:2018lil,Goto:2020per}, but there is a crucial difference between the pseudo entropy and interface entropy as the number of the interfaces between the two states depends on the replica parameter in the former while it is  independent in the latter.
Thus, the calculation of the replica partition function is a highly daunting task.
While the exact results of such a partition function are far from our reach in general, it would become tractable for supersymmetric field theories.
Supersymmetries are broken on the replica manifold due to the conical singularity, but may be maintained by introducing a sort of chemical potential to the pseudo entropy in a similar manner to the supersymmetric R\'enyi entropy \cite{Nishioka:2013haa}, resulting in being calculable due to the supersymmetric localization (see also \cite{Nishioka:2014mwa,Huang:2014gca,Hama:2014iea,Alday:2014fsa,Huang:2014pda,Zhou:2015cpa,Zhou:2015kaj,Nian:2015xky,Giveon:2015cgs,Mori:2015bro,Nishioka:2016guu,Yankielowicz:2017xkf,Hosseini:2019and} for the generalizations in various dimensions).
In particular, it would be worthwhile to see if the supersymmetric extended pseudo entropy could probe two difference phases related by duality such as the $S$-duality wall in $\CN =4$ supersymmetric Yang-Mills theory in four dimensions
\cite{Gaiotto:2008ak,Gaiotto:2008sa,Gaiotto:2008sd}.

\acknowledgments
We are grateful to Yasuaki Hikida, Ali Mollabashi, Kotaro Tamaoka, and Zixia Wei 
for valuable discussions.
The work of T.\,N. was supported in part by the JSPS Grant-in-Aid for Scientific Research (C) No.19K03863.
The work of T.\,N. and T.\,T. is supported by the JSPS Grant-in-Aid for Scientific Research (A) No.21H04469.
T.\,T. is supported by the Simons Foundation through the ``It from Qubit'' collaboration, Inamori Research Institute for Science and World Premier International Research Center Initiative (WPI Initiative) from the Japan Ministry of Education, Culture, Sports, Science and Technology (MEXT).
T.\,T. is also supported by JSPS Grant-in-Aid for Challenging Research (Exploratory) 18K18766.


\appendix

\section{Modular properties in \texorpdfstring{$\SU(2)$}{SU(2)} case}\label{ap:sutwo}

Here we summarize the explicit partition functions for the $\SU(2)$ i.e. $N=2$ case.

The $\CS$-matrices are
\begin{align}
    \Smat{i}{j}
        =\s{\frac{2}{k+2}}\sin\left[\frac{\pi\,(2i+1)(2j+1)}{k+2}\right]\ .
\end{align}
In particular we have
\begin{align}
    \begin{aligned}
    \Smat{0}{0}
        &=
        \s{\frac{2}{k+2}}\,\sin\left[\frac{\pi}{k+2}\right]\ , \\
    \Smat{\frac{1}{2}}{0}
        &=
        \s{\frac{2}{k+2}}\,\sin\left[\frac{2\pi}{k+2}\right]\ , \\
    \Smat{\frac{1}{2}}{\frac{1}{2}}
        &=
        \s{\frac{2}{k+2}}\,\sin\left[\frac{4\pi}{k+2}\right]\ .
    \end{aligned}
\end{align}
The quantum dimension reads
\begin{align}
    d_j=[2j+1]=\frac{\sin\left[\frac{\pi(2j+1)}{k+2}\right]}{\sin\left[\frac{\pi}{k+2}\right]}\ .
\end{align}

The partition function $Z\left[ X_n\right]$ for Wilson loop with $n$ crossing is in general given by 
\begin{align}
    \frac{Z\left[X_n\right]}{\Smat{0}{0}}=\left(q^{-\frac{N-1}{2}}\right)^n\frac{[N+1][N]}{[2]}+\left(-q^{-\frac{N+1}{2}}\right)^n\frac{[N][N-1]}{[2]}\ ,
\end{align}
where $q=e^{2\pi\i/(k+N)}$.

For $N=2$ we get explicitly 
\begin{align}
    Z\left[ X_n\right]
        =
        \s{\frac{2}{k+2}}\cdot\left(\frac{-i}{2}\right)\cdot
        \left[e^{\frac{\pi\i(3-n)}{k+2}}-e^{-\frac{\pi\i(3+n)}{k+2}}+(-1)^n\, e^{\frac{\pi\i(1-3n)}{k+2}}
        -(-1)^n\, e^{-\frac{\pi\i(1+3n)}{k+2}}\right]\ .
\end{align}

In particular, we find
\begin{align}
    \begin{aligned}
    Z\left[X_0\right]
        &=
        \frac{\left(\Smat{0}{\frac{1}{2}}\right)^2}{\Smat{0}{0}}\ ,\\
    Z\left[X_1\right]
        &=
        \Smat{0}{\frac{1}{2}}\ ,\\
    Z\left[X_2\right]
        &=
        \Smat{\frac{1}{2}}{\frac{1}{2}}\ .
    \end{aligned}
\end{align}

In the $k\to \infty$ limit we obtain the simple result:
\begin{align}
    Z\left[ X_n\right]~ \longrightarrow ~ \frac{\pi\,\sqrt{2}}{k^\frac{3}{2}}\,\left[3+(-1)^n\right]\ .
\end{align}

\section{Multi-boundary states in Chern-Simons theory}\label{ap:multi_bdy}
In this section we consider the multi-boundary states of spatial regions $\Sigma=\bigsqcup \BT^2$ consisting of several tori without any Wilson loops.
The calculation of the entanglement entropies of these states is performed in \cite{Balasubramanian:2016sro}, which is easily generalized to pseudo entropy.
These states can be prepared by drilling out the internal region of a subregion $\Sigma=\bigsqcup \BT^2$ from $\BS^3$. The resulting state can be expanded by the states $\ket{R_j}$ in figure \ref{torusfig}:
\begin{align}
  \ket{\psi}=\sum_{j_1,\ldots j_n}c_{j_1,\ldots,j_n}\ket{R_{j_1},\ldots,R_{j_n}},
\end{align}
where $n$ is the number of tori and $\ket{R_{j_1},\ldots,R_{j_n}}\equiv\ket{R_{j_1}}\otimes\cdots\otimes\ket{R_{j_n}}$. The coefficients
\begin{align}
  c_{j_1,\ldots,j_n}&=\braket{R_{j_1},\ldots,R_{j_n}|\psi}
\end{align}
are partition functions on $\BS^3$ with $n$ Wilson loops of representations $R_{j_1},\ldots,{R_{j_n}}$.

Here we only consider a simple example of the two tori states $\Sigma=\BT^2\sqcup \BT^2$ in $\U(1)$ Chern-Simons theory. In this case, the coefficients turn out to be 
\begin{align}
  c_{j_1,j_2}=Z\left[\BS^3;R_{j_1},R_{j_2}\right]=\exp\left(\frac{2\pi \i}{k}\,q_1\,q_2\,\ell_{12}\right),
\end{align}
where $q_1,q_2$ are the $\U(1)$ charges of the two loops and $\ell_{12}$ is the linking number between the two loops.
We define two states 
\begin{align}
    \begin{aligned}
    \ket{\psi}
        &=
            \frac{1}{k}\sum_{q_1,q_2}\exp\left(\frac{2\pi\i}{k}\,q_1\,q_2\,\ell_{12}\right)\ket{R_{j_1}}\otimes\ket{R_{j_2}}, \\
    \ket{\varphi}
        &=
            \frac{1}{k}\sum_{q_1,q_2}\exp\left(\frac{2\pi\i}{k}\,q_1\,q_2\,\ell_{12}'\right)\ket{R_{j_1}}\otimes\ket{R_{j_2}} \ ,
    \end{aligned}
\end{align}
with different linking numbers.

We call one of the two tori $A$, which corresponds to the region including a Wilson loop with $R_{j_1}$, and the other $B$. Let us calculate the entanglement entropies between $A$ and $B$ following \cite{Balasubramanian:2016sro}. The reduced density matrix for $\ket{\psi}$ is
\begin{align}
    \begin{aligned}
    \rho_A^{\psi}\equiv\Tr_B\left[\ket{\psi}\bra{\psi}\right]
        &=
            \frac{1}{k^2}\sum_{q_1,q_1',q_2}e^{\frac{2\pi\i}{k}(q_1-q_1')\,\ell_{12}\,q_2}\ket{R_{j_1}}\bra{R_{j_1'}} \\
        &=
            \frac{1}{k}\sum_{q_1,q_1'}\eta_{q_1,q_2}(k,\ell_{12})\ket{R_{j_1}}\bra{R_{j_1'}}\ ,
    \end{aligned}
\end{align}
where 
\begin{align}
  \eta_{q_1,q_2}(k,\ell_{12})\equiv
  \begin{cases}
    1 \qquad \ell_{12}(q_1-q_2)=0\mod{k}\\
    0 \qquad \ell_{12}(q_1-q_2)\neq0\mod{k}
  \end{cases}.
\end{align}
The R\'{e}nyi entropy for $\ket{\psi}$ is
\begin{align}\label{eq:multiRE}
    S^{(n)}\left(\rho_A^{\psi}\right)
        &=
            \frac{1}{1-n}\log\left[\frac{1}{k^n}\sum_{q_1,\ldots,q_n}\eta_{q_1,q_2}(k,\ell_{12})\cdots\eta_{q_{n},q_1}(k,\ell_{12})\right]\ .
\end{align}
When we fix $0\le q_1\le k-1$, there are $\mathrm{gcd}(k,\ell_{12})$ values of $0\le q_2\le k-1$ satisfying the relation $\ell_{12}(q_1-q_2)=0\mod{k}$.
Similarly, $q_3,\ldots,q_n$ also take $\mathrm{gcd}(k,\ell_{12})$ values. Therefore the summation in the logarithm in \eqref{eq:multiRE} will be $k\,(\mathrm{gcd}(k,\ell_{12}))^{n-1}$, so 
\begin{align}
  S^{(n)}\left(\rho_A^{\psi}\right)=\log\left[\frac{k}{\mathrm{gcd}(k,\ell_{12})}\right]\ ,
\end{align}
and the entanglement entropy is clearly 
\begin{align}
  S\left(\rho_A^{\psi}\right)=\log\left[\frac{k}{\mathrm{gcd}(k,\ell_{12})}\right]\ .
\end{align}
Similarly, the entanglement entropy for $\ket{\varphi}$ is 
\begin{align}
  S\left(\rho_{A}^{\varphi}\right)=\log\left[\frac{k}{\mathrm{gcd}(k,\ell_{12}')}\right]\ .
\end{align}

\begin{figure}[t]
  \begin{center}
    \includegraphics[width=0.6\linewidth,pagebox=cropbox,clip]{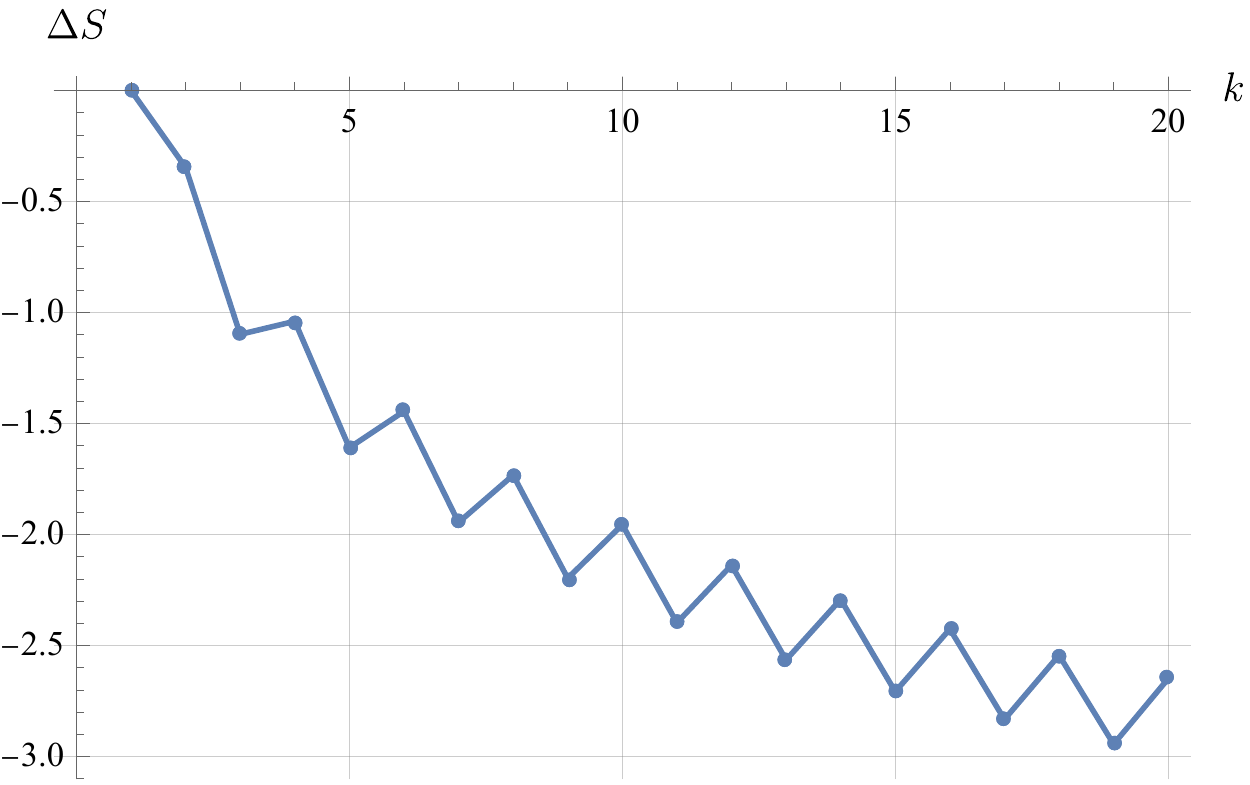}
    \caption{The difference $\Delta S$ of the pseudo entropy from the average of the entanglement entropies varying the level $k$. We set $\ell_{12}=2,\ell_{12}'=1$. } 
    \label{fig:multiU1}
  \end{center}
\end{figure}

Next, let us calculate the pseudo entropy. The inner product is 
\begin{align}
    \begin{aligned}
        \braket{\varphi|\psi}
            &=
                \frac{1}{k^2}\sum_{q_1,q_2}\exp\left(\frac{2\pi\i}{k}\,q_1\,q_2\,(\ell_{12}-\ell'_{12})\right) \\
            &=
                \frac{1}{k}\sum_{q_1}\eta_{\ell_{12},\ell_{12}'}(q_1,k) \\
            &=
                \frac{1}{k}\,\mathrm{gcd}(\ell_{12}-\ell_{12}',k)\ .
    \end{aligned}
\end{align}
To obtain the third line, we have used the fact that there are $\mathrm{gcd}(\ell_{12}-\ell_{12}',k)$ values of $0\le q_1\le k-1$ satisfying $q_1(\ell_{12}-\ell_{12}')=0\mod{k}$.
When $\ell_{12}=\ell'_{12}$, i.e. $\ket{\psi}=\ket{\varphi}$, the inner product is $1$ since $\mathrm{gcd}(0,k)=k$.
The reduced transition matrix is
\begin{align}
    \begin{aligned}
        \tau_A^{\psi|\varphi}
            &=
                \frac{1}{k^2\braket{\varphi|\psi}}\sum_{q_1,q_1',q_2}e^{\frac{2\pi\i}{k}(q_1\,\ell_{12}-q_1'\,\ell_{12}')\,q_2}\ket{R_{j_1}}\bra{R_{j_1}'} \\
            &=
                \frac{1}{\mathrm{gcd}(\ell_{12}-\ell_{12}',k)}\sum_{q_1,q_2}\tilde{\eta}_{q_1,q_1'}(\ell_{12},\ell_{12}',k)\ket{R_{j_1}}\bra{R_{j_1}'}\ ,
    \end{aligned}
\end{align}
where 
\begin{align}
  \tilde{\eta}_{q_1,q_2}(\ell_{12},\ell_{12}', k)\equiv
  \begin{cases}
    1 \qquad q_1\,\ell_{12}-q_1'\,\ell_{12}'=0\mod{k}\\
    0 \qquad q_1\,\ell_{12}-q_1'\,\ell_{12}'\neq0\mod{k}
  \end{cases}.
\end{align}
Let $N(\ell_{12},\ell'_{12},k)$ be the number of $0\le q_1'\le k-1$ satisfying $q_1\,\ell_{12}-q_1'\,\ell_{12}'=0\mod{k}$ when fixing $q_1$, which in fact does not depend on $q_1$. Then the pseudo entropy is
\begin{align}
    \begin{aligned}
        S\left(\tau_A^{\psi|\varphi}\right)
            &=
            \lim_{n\to1}\frac{1}{1-n}\log\left[\frac{N(\ell_{12},\ell'_{12},k)}{\gcd(\ell_{12}-\ell_{12}',k)}\right]^{n-1} \\
            &=
                \log\left[\frac{\gcd(\ell_{12}-\ell_{12}',k)}{N(\ell_{12},\ell'_{12},k)}\right]\ .
    \end{aligned}
\end{align}

Figure \ref{fig:multiU1} shows the difference $\Delta S$ when the linking numbers are $\ell_{12}=2,\ell_{12}'=1$.
We can see that $\Delta S<0$ similar to the other typical examples.


\bibliographystyle{JHEP}
\bibliography{PE}
\end{document}